\documentclass[preprint,preprintnumbers,showpacs,floatfix,amsmath,amssymb]{revtex4}
\usepackage{times}
\usepackage{epsfig}
\usepackage{amssymb}
\usepackage{amsmath}
\usepackage[english]{babel}
\usepackage{graphics}

\begin{document}

\bibliographystyle{apsrev}

\title{Multiple complex networks emerging from individual interactions}

\author{L. E. C. da Rocha and L. da F. Costa}
\email{luciano@if.sc.usp.br}

\affiliation{Grupo de Pesquisa em Vis\~ao Cibern\'etica,
Instituto de F\'isica de S\~ao Carlos,
Universidade de S\~ao Paulo, Av. Trabalhador S\~ao Carlense 400,
Caixa Postal 369, 13560-970, S\~ao Carlos, SP, Brazil}

\date{5th August 2007}

\begin{abstract}

Systems composed of distinct complex networks are present in many
real-world environments, from society to ecological systems. In the
present paper, we propose a network model obtained as a consequence of
interactions between two species (\textit{e.g.} predator and prey).  Fields are
produced and sensed by the individuals, defining spatio-temporal
patterns which are strongly affected by the attraction intensity
between individuals from the same species. The dynamical evolution of
the system, including the change of individuals between different
clusters, is investigated by building two complex networks having the
individuals as nodes. In the first network, the edge weight is given
by the Euclidean distance between every two individuals and, in the
case of the second network, by the amount of time two individuals stay
close one another. A third network is obtained from the two previous
networks whose nodes correspond to the spatially congruent groups.
The system evolves to an organized state where Gaussian and
scale-free-like strength distributions emerge, respectively, in the
predator and prey networks. Such a different connectivity is mainly a
consequence of preys elimination.  Some configurations favor the
survival of preys or higher efficiency of predator activity.

\end{abstract}


\maketitle

\section{Introduction}

Despite the impressive development undergone by complex networks along
the recent years, relatively little attention has been focused on the
investigation of systems incorporating more than one network, and even
lesser attention has been given to the study of specific dynamics (at
the nodes or of the own network topology) taking place in such
networks.  Yet, several real-world systems are inherently composed by
multiple networks.  In biological sciences, for instance,
protein-protein interaction networks are immediately related to
transcriptional and metabolic networks~\cite{OltvaiPyramid02}.  Similar
entanglements are found in many other areas, such as in the Internet
or in the www, which are directly related to social and cultural
networks, as well as economical constraints~\cite{CostaMultiple05}. Strictly speaking, it is
actually hard to think of a real-world system which can be completely
represented in terms of a single network.

Some works considering multiple networks have been reported in the
literature.  The ``Solomon Network'' system was introduced by Erez and
collaborators~\cite{Erez05}. It consists in a multi-layered system
with a set of nodes (agents) common to all layers. In each layer,
different rules according to social and/or economical ties define the
interaction between the nodes. On the other hand, the coupling between
layers is obtained by interaction between variables associated to the
respective nodes in different layers. Another layered-based formalism
was proposed independently by Kurant \textit{et
al.}~\cite{Kurant06a}. Using transportation networks, they constructed
a logical network representing the traffic flows which were mapped
onto another network representing the physical
infrastructure~\cite{Kurant06a,Kurant06b}.  Park \textit{et al.}
constructed two networks of musicians in which the edges of one
network represented the similarity while, in the other, they
represented the collaboration between the musicians~\cite{Park06}.
The intersection of both networks generated a new dataset where the
structural properties showed significant differences between both
types of networks.  We have been particularly interested in addressing
the dynamics of multiple networks systems.  In a previous
work~\cite{Rocha07}, we considered a system composed of regular and
complex networks. While a diffusion pattern evolved in the regular
network, the complex network were expected to self-organize to control
and if possible, eliminate the pattern. An interaction rule between
both networks was responsible to activate the complex network nodes.

Of special relevance in which concerns multi-layered system are
ecological environments where the evolution is constrained by the
interactions among its components, besides climate conditions and
other external influences. Modelling all such constrains might be
difficult and often unnecessary while investigating specific phenomena
of interest. Considering an ideal environment (favorable climate and
abundant supplies), the interaction between animals could be described
as an interaction function representing the odor
intensity~\cite{Mingfeng:odor}.  Conversely, the interaction rule
could be avoided to some extent as in the famous Lotka-Volterra system
(predator-prey model)~\cite{Lotka_book,Volterra_book}, which considers
only the amount of individuals of one species as a constrain to the
evolution of the other species. Adding a diffusive term generalizes
the system in which concerns spatial constraints, but does not
properly address the question of single
interactions~\cite{Murray_book}. Although collective behavior may
emerge from such simple interaction
rules~\cite{Waldau:book,Sznajd:opinion,Farkas:mexicanwave}, there are
open questions related to emergence of spatio-temporal patterns on
complex adaptive systems.  Since the predator prey dynamics involves
two types of agents (or individuals) in constant interaction, we
considered a multi networks view of the system evolution. In this
case, each network with moveable nodes represents one species, while
the interaction rule between the nodes of both networks is defined in
terms of sensitive fields.

More specifically, in the present paper we propose a system composed of two interacting
species, with emphasis on the representation of the spatial position
of each individual. One of the species (henceforth called predator)
relies on the other species (prey) as the only source of food. In this
scarce food environment, predators move by sensing the presence of
preys while the preys are expected to sense the predators proximity
and move away. Mutual attraction between same species individuals
tends to generate spatial clusters which imply decrease in the
mobility of single individuals inside those clusters. Two weighted
complex networks are constructed along all time steps in terms of the
dynamics between predators and preys. One of them expresses the
Euclidean distance between any two animals as weights. This network is
geographical, in the sense that the each node incorporate information
about the position of the respective individual.  In the other
network, the weights reflect the history of proximity between pairs of
animals.  In other words, the weights in the second network are
proportional to the total time each pair of animals spend together.  A
third network is obtained so that each of its nodes corresponds to one
spatial cluster and the weights provide information about the number
of exchanges of animals between pairs of respective spatial clusters.

The current paper begins by reviewing basic concepts related to
complex networks and follows stating the interaction rules of the
proposed model and the parameters of the system. The results section
discusses the emergence of spatial clusters patterns and then analyse
the evolution of structural properties of the resulting complex
network and their relation with the proposed dynamics.

\section{Complex Networks Concepts}

A complex network $\Gamma$ with $N$ nodes can be defined by a set
$V(\Gamma)$ of nodes $\mu_i$ ($i=1,2,\ldots,N$) and a set of edges
$E(\Gamma)$ connecting a pair of nodes $(\mu_i,\mu_j)$. Given a set
$W(\Gamma)$ of real values, a weight $\omega_{\mu_i,\mu_j}$ is
assigned to an edge by mapping one element of $W(\Gamma)$ to one
element of $E(\Gamma)$. A sub-network $\kappa$ is defined by a set of
nodes $V(\kappa) \subset V(\Gamma)$ and a set of edges
\mbox{$E(\kappa) \subseteq \{ (\mu_i, \mu_j):(\mu_i, \mu_j) \in
E(\Gamma) \text{ and } \mu_i, \mu_j \in V(\kappa) \}$}. The
sub-network $\kappa$ is connected if and only if, any node of
$V(\kappa)$ can be reached by any other node of
$V(\kappa)$~\cite{Barabasi:review, Newman:review, Dorog:review,
Boccaletti:review, Costa:review}.

The local connectivity of a node $\mu_i$ can be quantified by its
degree $k_{i}$, which provides the number of nodes directly connected
to $\mu_i$, \textit{i.e.} the neighbours of $\mu_i$. When the
connections have weights, the strength $s_{i}$ of node $\mu_i$ is
obtained by summing all weights assigned to connections involving
$\mu_i$~\cite{Costa:review}. The number of closed triangles in the
neighbourhood of node $\mu_i$ is given by the generalized clustering
coefficient $cc_i$ (eq.~\ref{eq:01}) which considers the weights
$\omega_{\mu_i,\mu_j}$ of each connection established between a node
$\mu_i$ and its neighbours $\mu_j$~\cite{Barrat:weight}.

\begin{equation}
\label{eq:01}
  cc_i = \dfrac{1}{s_i(k_i-1)}\sum_{j,k=1}^{N,N}\dfrac{\omega_{\mu_i,\mu_j} + \omega_{\mu_i,\mu_k}}{2}a_{ij}a_{ik}a_{jk}
\end{equation}
where, $a_{ij}$ represents the connection between $\mu_i$ and $\mu_j$
such that $a_{ij}a_{ik}a_{jk}=1$ indicates that a closed triangle
exists in the neighbourhood of node $\mu_i$. Finally, the overall
structure of a network can be summarized by the averaged value of each
considered measurement.

\section{Interaction Model}

\subsection{Movement and Interaction Dynamics}

Consider a squared bi-dimensional region $O$ of size $L=512$ in the
continuous space $\Omega$. We randomly distribute $N=300$ animals (or
individuals $I_i$, where: $i=1,2,\ldots,N$) of each species (predator
and prey) inside $O$ such that any two species-independent individuals
$I_i$ and $I_j$ are at a minimum Euclidean distance $r_{i,j} = R_{min}
= 4$ from each other. At every time step $\Delta t=0.1$, each
individual $I_i$ updates its respective position $P^t_{x_i,y_i}$
according to $P^{t+1}_{x_i,y_i}=P^t_{x_i,y_i} + \mathbf{v_i}\Delta
t$. The resulting \textit{sensitive field} is understood to provide
the velocity $\mathbf{v_i}$ of each individual at any position and
time. In such a way, an animal only senses other individuals within
its \textit{perception radius}, which is fixed over time as
\mbox{$R_{per}=128$.}

The interactions between predators and preys are directly related to
their species. In case they belong to the same species, the sensitive
field generated by one over the other is given by a dimensionless
function (eq.~\ref{eq:02a}), where the parameters $\lambda=4.0$ and
$\mu=0.04$ have respectively units of $[L]$ and $[L]^{-2}$. The
alternate characteristic of the function $g(r_{i,j})$ is chosen
because it provides a balance of attraction and repulsion between same
species. On the other hand, when the animals belong to different
species, the individual $I_i$ of one species senses $I_j$ of the other
species in such a way that its reaction intensity is inversely
proportional to the proximity between them.  This is expressed in
equation~\ref{eq:02b}, where the parameter $\sigma = 1.0$ has units of
area $[L]^2$.

\begin{subequations}
\label{eq:02}
\begin{align}
\textbf{g}_{j}(r_{i,j}) &= \Bigl[4e^{-\mu{(r_{i,j}-\lambda)}^2}-e^{-\mu{(r_{i,j}-3\lambda)}^2}\Bigr]\text{\^r}_{i,j} \label{eq:02a} \\
\textbf{f}_{j}(r_{i,j}) &= \dfrac{\sigma}{r_{i,j}^2}\text{\^r}_{i,j} \qquad  \text{ where: } \;\; \|\textbf{r}_{i,j}\| < R_{per} \label{eq:02b}
\end{align}
\end{subequations}

The sum over all contributions $I_j$ determines the direction and
intensity of an individual $I_i$ final velocity (eq.~\ref{eq:03}). The
signs and values of the parameters $\alpha$ and $\beta$ in
equation~\ref{eq:03} should be selected so as to correctly represent
the system of interest, and they have units of velocity
$[L][T]^{-2}$. In the present paper, we will investigate different
combinations of the parameter $\beta$ for predators and preys
(eq.~\ref{eq:04}).

\begin{equation}
\label{eq:03}
  \textbf{v}_{i} = \alpha\sum_{j=1}^{N-1}\textbf{f}_{j}(r_{i,j}) + \beta \sum_{j=1}^{N-1} \textbf{g}_{j} (r_{i,j})
\end{equation}

\begin{equation}
\label{eq:04}
\left\{
  \begin{array}{llll}
    \text{Predator:} \; & \alpha_{predator} \; < 0 \; & \text{ and } \; & \beta_{predator} \; \geqslant 0 \\
    \text{Prey:}     \; & \alpha_{prey}     \; > 0 \; & \text{ and } \; & \beta_{prey}    \; \geqslant 0
  \end{array}
\right.
\end{equation}

Since the model is motivated by an ecological environment, we suppose
that preys have plenty of food. Consequently, a prey $I_j$ only dies
when close enough $r_{i,j} \leqslant R_{eli}=4$ to a predator
$I_i$. The conservation of the number of animals is provided by a
feedback process where each eliminated prey is promptly replaced by a
new one. The feedback process consists in randomly arranging the new
preys within the region $O$, while respecting the minimum distance
$R_{min}=4$ between two individuals as in the initial conditions. As
the abundance of preys provides enough resources to predators, we
suppose that they never die along the simulation interval.

Although periodic boundary conditions are usually adopted in order to
minimize finite size effects, an ecological system is often better
described by other types of boundary conditions. In fact, animals tend
to live inside a specific region $O$ of the ecosystem $\Omega$ called
\textit{habitat} (Similarly, people use to restrict their everyday
life within cities or even within home, work and school places). The
habitat contains not only one species, but also many animal and
vegetal species, which constitute a rich environment. Consequently,
the animals are able to interact with a number of species enclosed in
such a region. The boundaries of the habitat can be related to
landmarks such as mountains and rivers, or even islands. In order to
reproduce such real boundary constrains, we adopt elastic boundaries
conditions (eq.~\ref{eq:05}) which restrict the movement region but
also allow preys to escape while close to the edges.

\begin{equation}
\label{eq:05}
\left\{
  \begin{array}{llll}
  \text{If } & x(y) > L & \text{then } & x(y) \leftarrow 2L - x(y)       \\
  \text{If } & x(y) < 0 & \text{then } & x(y) \leftarrow \vert x(y)\vert \\
  \end{array}
\right.
\end{equation}

\subsection{Complex Networks Construction}

\begin{figure*}[ht]
  \begin{center}
  \includegraphics[scale=0.27]{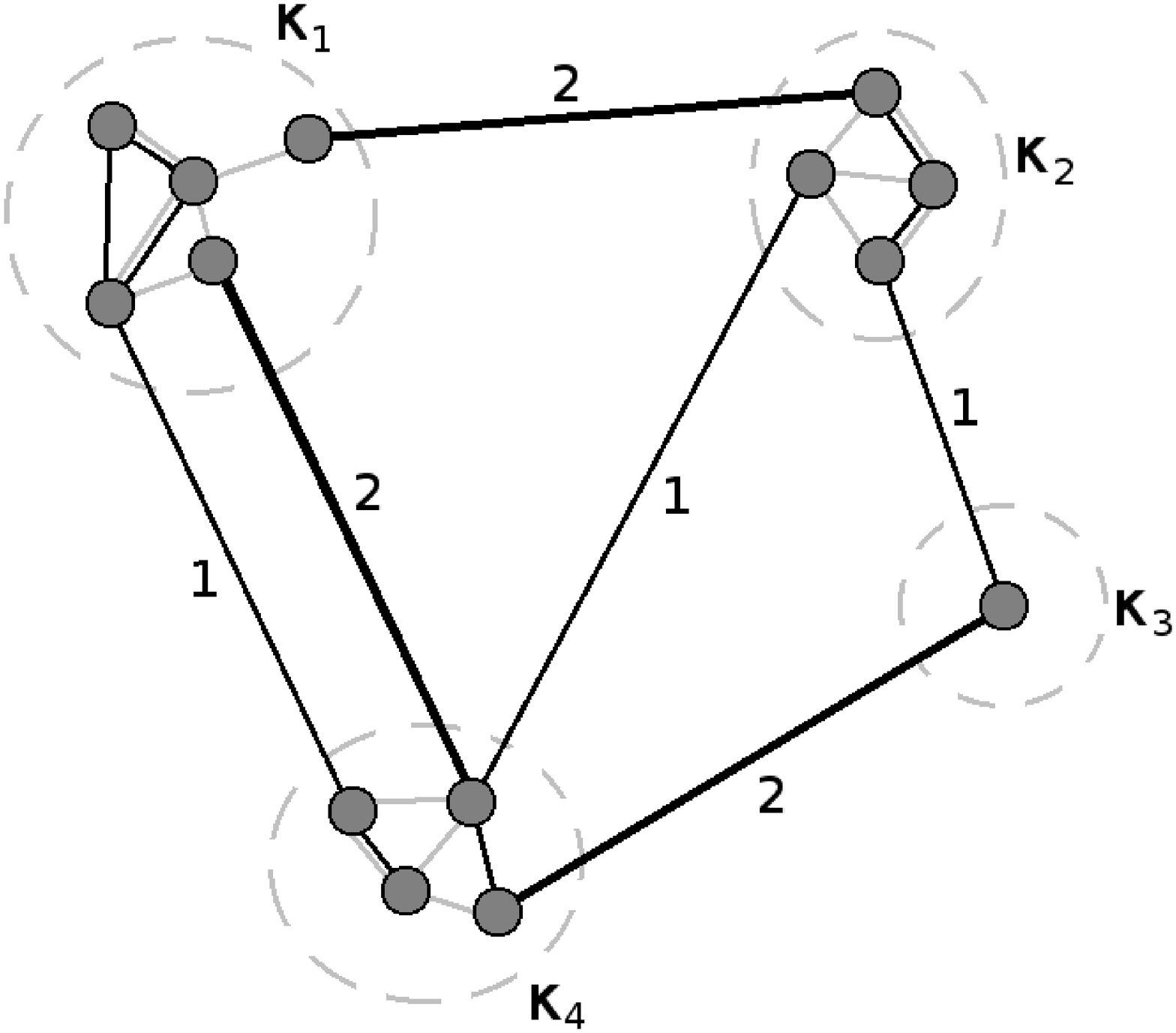}
  \includegraphics[scale=0.27]{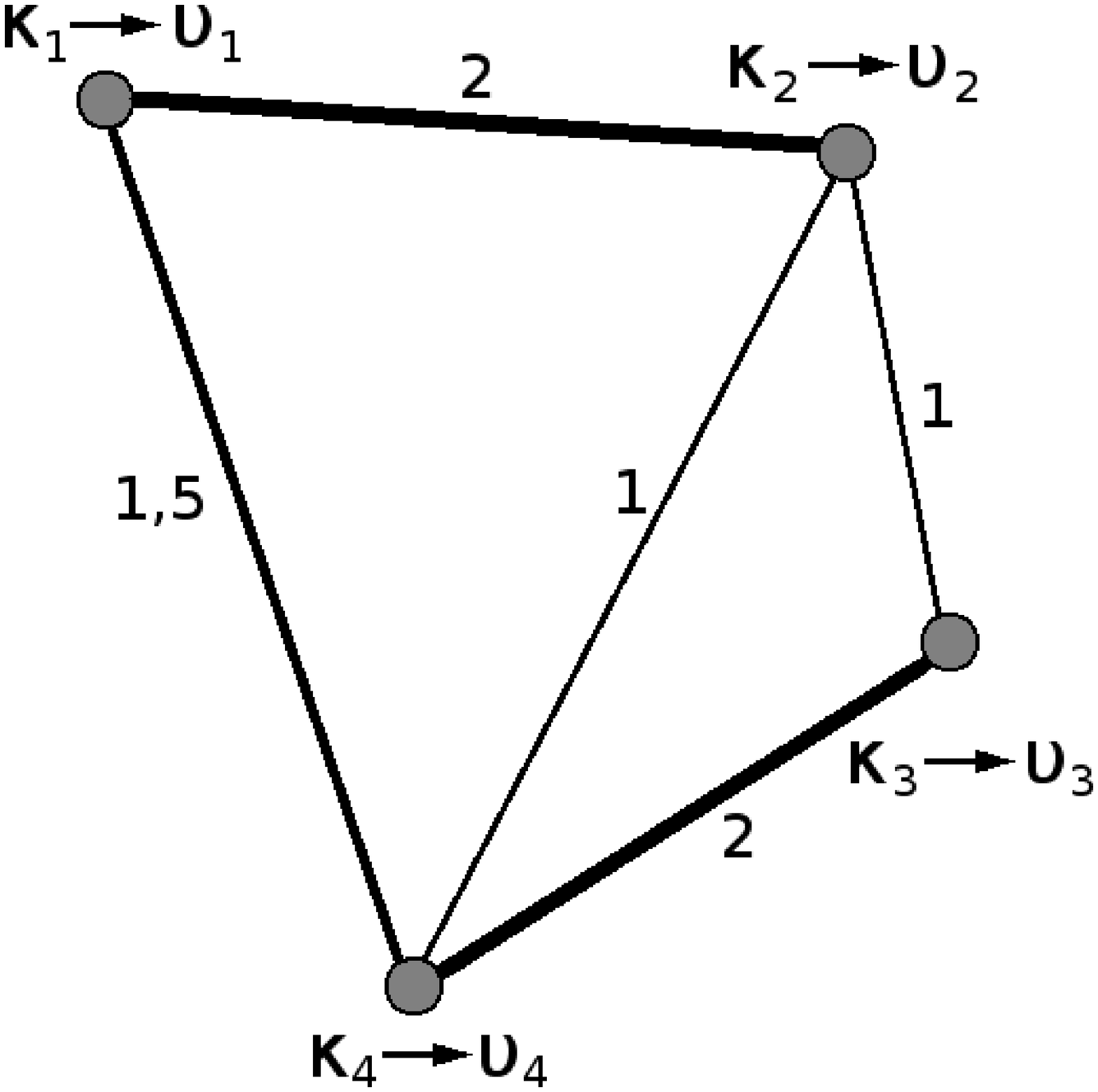} \\
  (a) \hspace{6cm} (b)
  \end{center}
  \caption{(a) A sample of both networks where the nodes represent the
  individuals. Light-gray edges belong to $\Gamma'_{geo}$ and black
  edges to network $\Gamma_{his}$. Dashed lines emphasize the
  sub-networks $\kappa_k$. (b) The resulting network $\Gamma_{res}$ is
  obtained by merging $\Gamma'_{geo}$ and $\Gamma_{his}$. Each node
  $\upsilon_k$ of $\Gamma_{res}$ is related to one sub-network
  $\kappa_k$ in Figure~\ref{fig:01}-a. In both Figures, the edge
  thickness is proportional to the respective weight.}  \label{fig:01}
\end{figure*}

At each time step, a geographical and weighted network fully connected
$\Gamma_{geo}$ is obtained by associating a node $\mu_i$ to each
individual $I_i$ of both species and setting the weight
$\omega_{\mu_i,\mu_j}$ of each edge as the Euclidean distance
$r_{i,j}$ between the respective two individuals. Since the animals
are constantly moving throughout the sub-space $O$, the network
$\Gamma_{geo}$ has a dynamical structure which changes at every
step. By eliminating the edges above a threshold $T=30$ and setting
$\omega_{\mu_i,\mu_j}=1$ for the others, we obtain a new complex
network $\Gamma'_{geo}$ fragmented in many connected sub-networks
$\kappa_k$ which we call \textit{spatial groups}~(Fig.~\ref{fig:01}-a).

The second complex network $\Gamma_{his}$ is constructed by
considering the history of proximity between every two nodes of
$\Gamma_{geo}$ (Fig.~\ref{fig:01}-a). The nodes $V(\Gamma_{his})
\Leftrightarrow V(\Gamma_{geo})$ are initially fully connected so that each
edge is set with $w_{\mu_i,\mu_j}=0$. Whenever two individuals $I_i$
and $I_j$ fall close enough \mbox{($r_{i,j}
\leqslant T=30$)}, the weight
$w^{t+1}_{\mu_i,\mu_j}=w^{t}_{\mu_i,\mu_j}+1$ is updated. The
connections involving dead preys are immediately eliminated.

The resulting network $\Gamma_{res}$ is obtained by a merging
mechanism which is done by associating a node $\upsilon_k$ to each
sub-network $\kappa_k$ of $\Gamma'_{geo}$ such that a new set of nodes
$V(\Gamma_{res})$ is obtained. If $M$ is the number of edges between
the sub-network $\kappa_k$ and $\kappa_l$, the weight $\Xi_{k,l}$
(eq.~\ref{eq:06}) of the corresponding edge $(\upsilon_k,\upsilon_l)$
in $\Gamma_{res}$ is given by the average value of $w_{\mu_i,\mu_j}$
established between the members $\mu_i$ and $\mu_j$ of different
spatial groups $\kappa_k$ (Fig.~\ref{fig:01}-b).

\begin{equation}
\label{eq:06}
  \Xi_{\upsilon_k,\upsilon_l}= \dfrac{1}{M} \sum_{\mu_i,\mu_j} w_{\mu_i,\mu_j}\quad \text{where: } \mu_i \subseteq V(\kappa_k)\text{ and }\mu_j \subseteq V(\kappa_l)
\end{equation}

\section{Cluster Emergence}

\subsection{Spatio-temporal patterns}

\begin{figure}
  \begin{center}
    \includegraphics[scale=0.16]{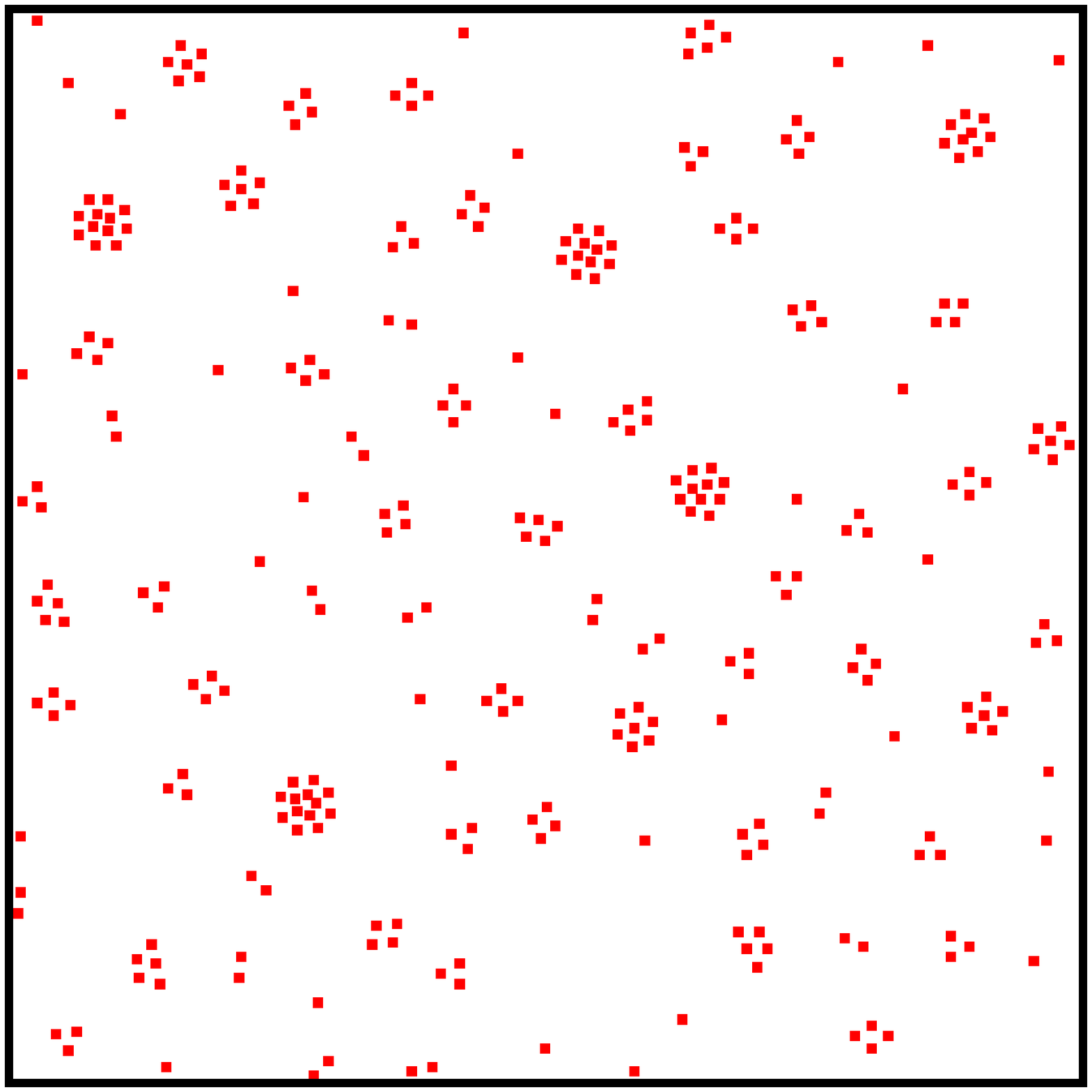}
    \includegraphics[scale=0.16]{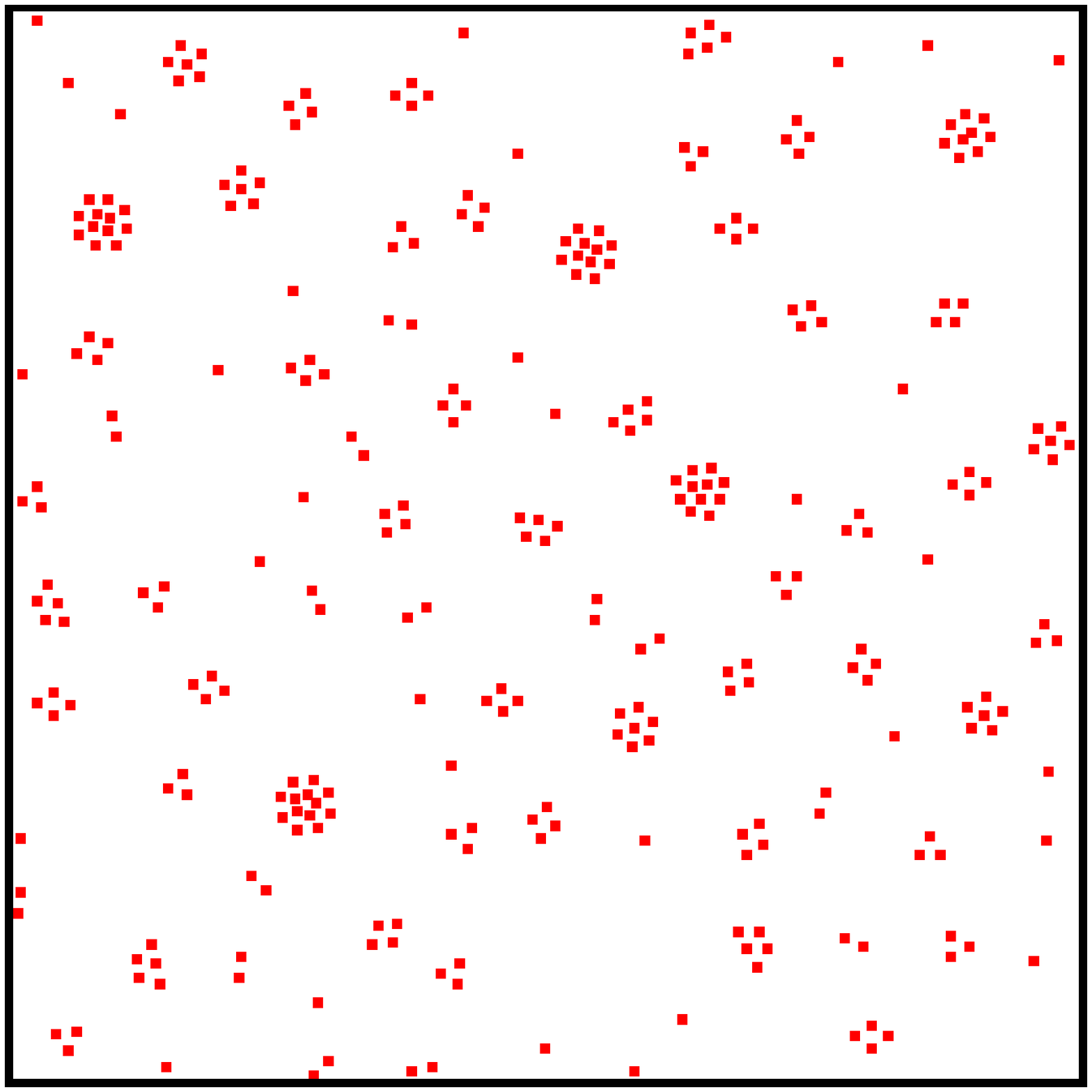}  \\
    (a) $t=250000$ \hspace{0.3cm} (b) $t=500000$        \\
    \caption{Snapshots of the emergent spatio-temporal pattern for one
    single species with $\beta=0.125$.}  \label{fig:02}
  \end{center}
\end{figure}

\begin{figure}
  \begin{center}
    \includegraphics[scale=0.16]{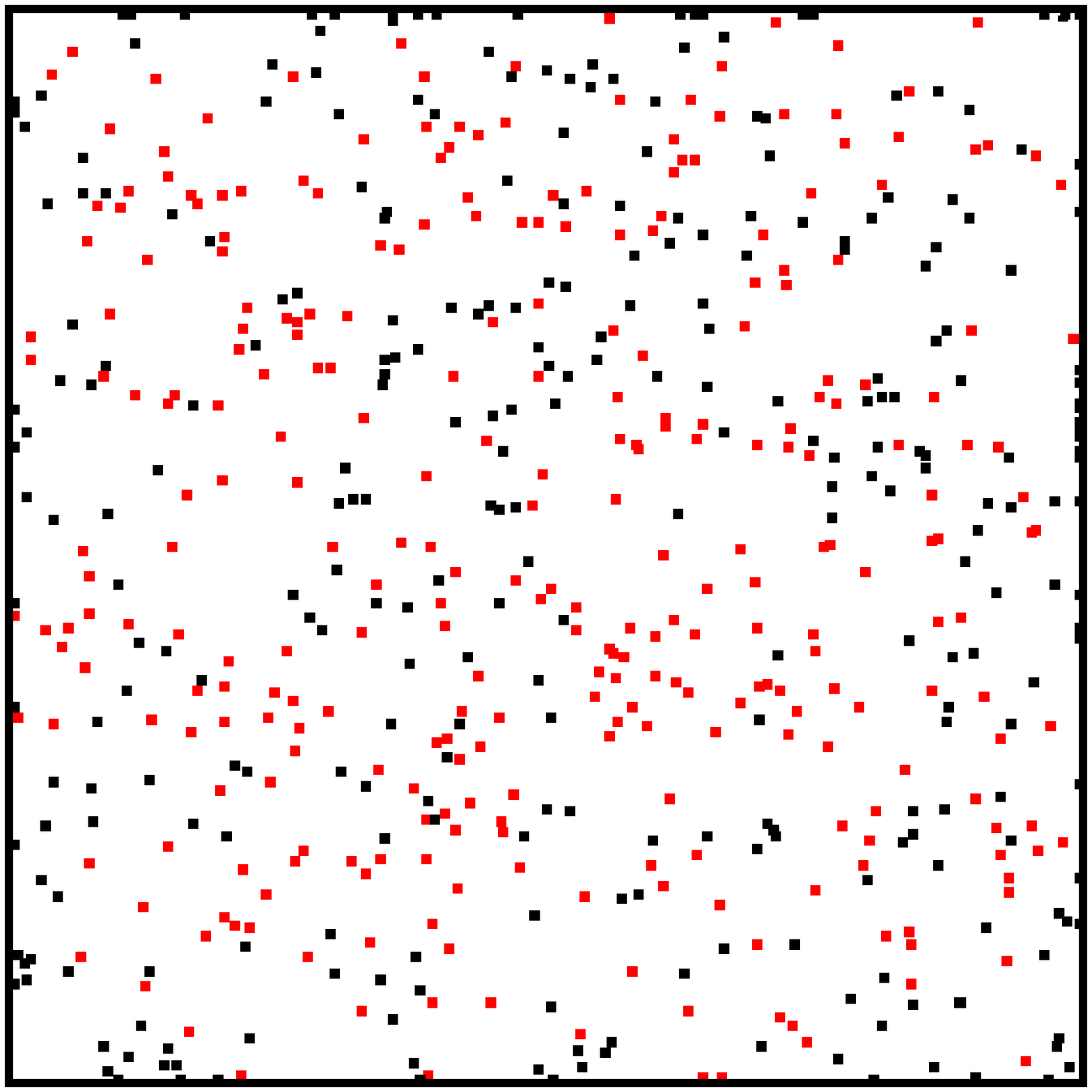}
    \includegraphics[scale=0.16]{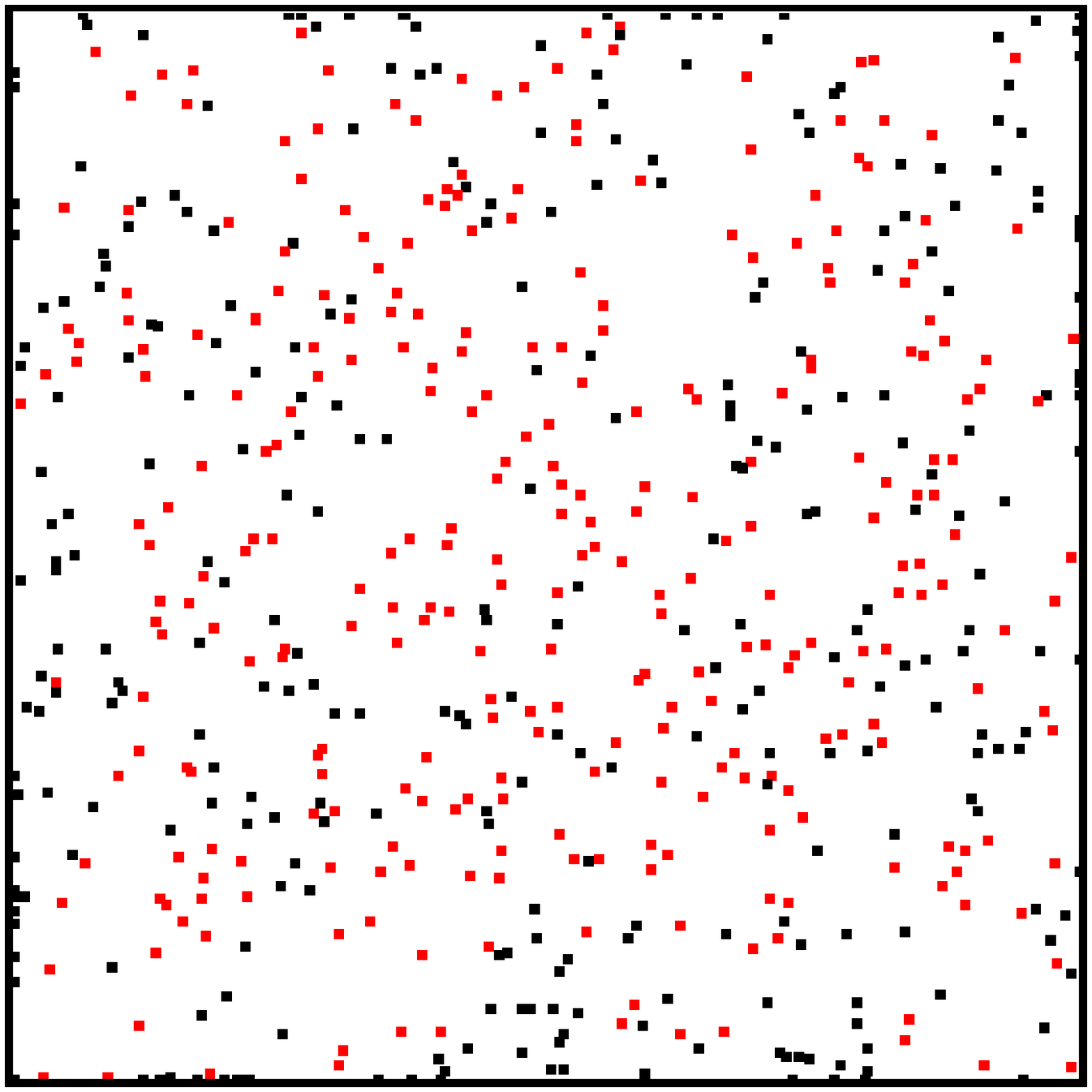}  \\
    (a) $t=250000$ \hspace{0.3cm} (b) $t=500000$    \\
    \caption{Snapshots of predators (red -- on-line) and preys (black
    -- on-line) spatio-temporal patterns with null attraction
    ($\beta=0$) between same species individuals. } \label{fig:03}
  \end{center}
\end{figure}

The interaction between different species, along with the prey
replacement mechanism, are fundamental for providing a non-stationary
state. Spatio-temporal cluster patterns\footnote{A set of movies related to the configurations investigated in the present paper is available at \textit{http://cyvision.ifsc.usp.br/$\sim$ luisrocha/paper2/}} emerge from the dynamics due
to interaction between same specie individuals. In the absence of one
species the system evolves to a stable state with many small static
clusters (Fig.~\ref{fig:02}). Contrariwise, in the presence of
interacting preys and predators, but null attraction between same
specie individuals, the system evolves to a non-stable state, without
emergence of clusters (Fig.~\ref{fig:03}).

\begin{figure*}
  \begin{center}
    (i) \\
    \includegraphics[scale=0.17]{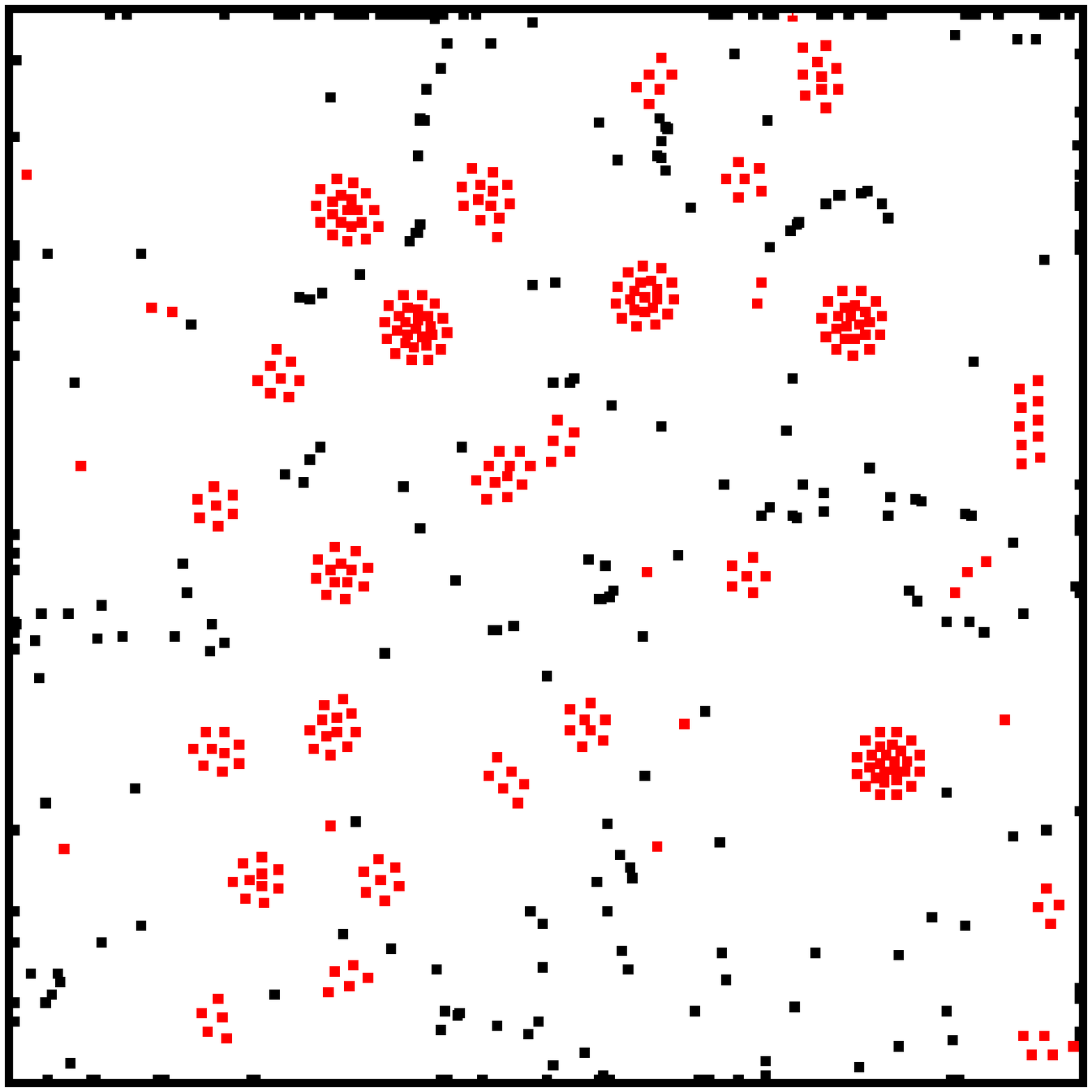}
    \includegraphics[scale=0.17]{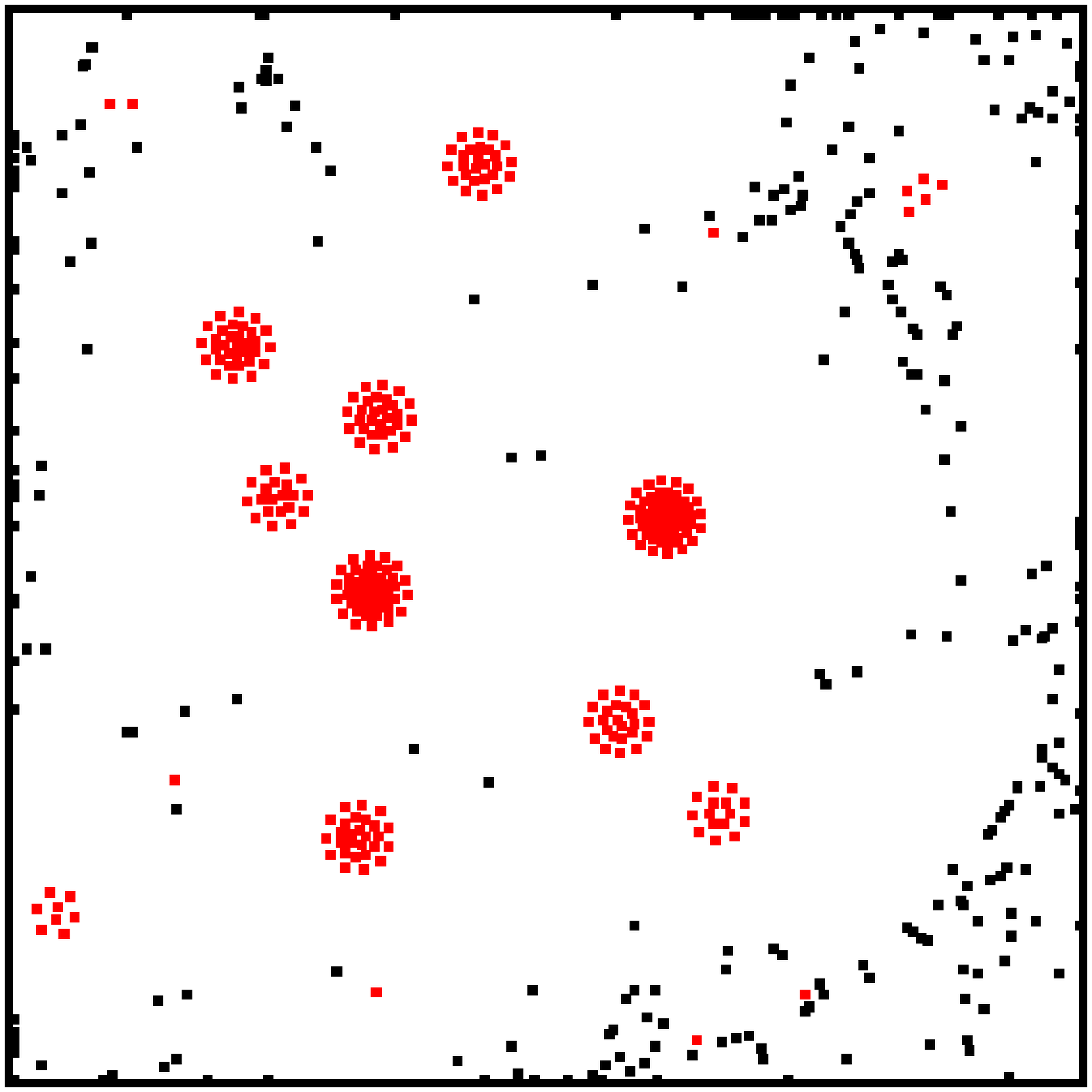}
    \includegraphics[scale=0.17]{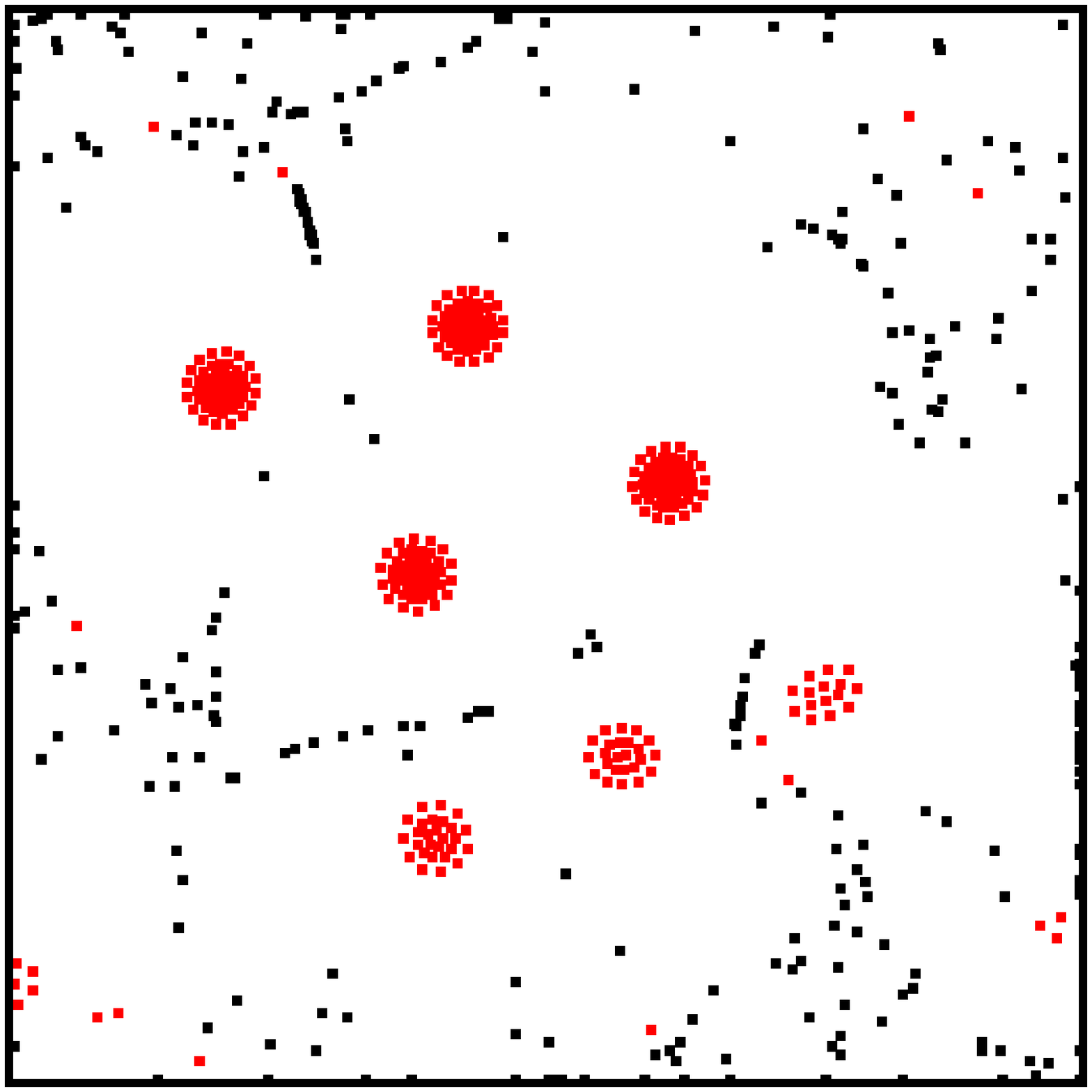}
    \includegraphics[scale=0.17]{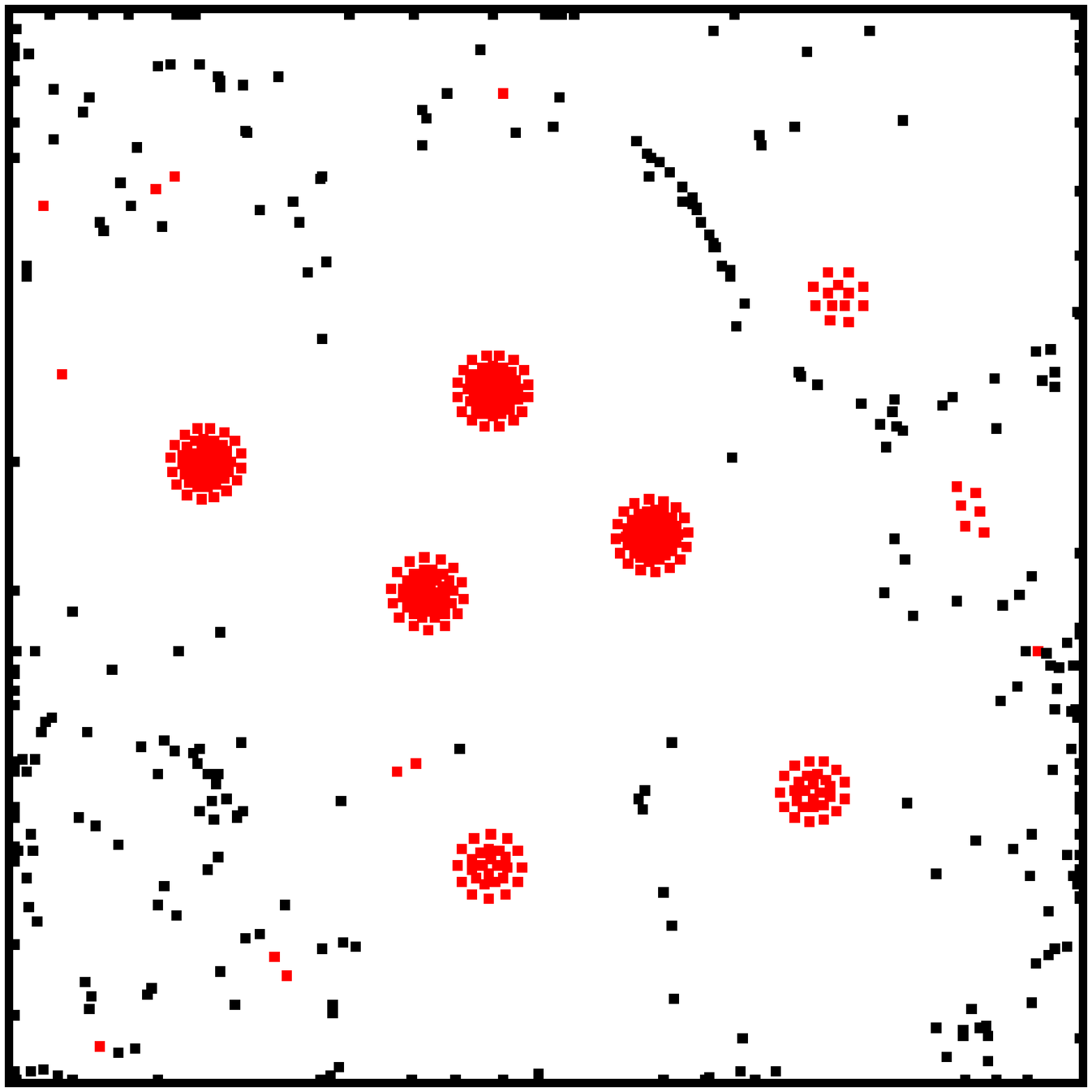}
    \includegraphics[scale=0.17]{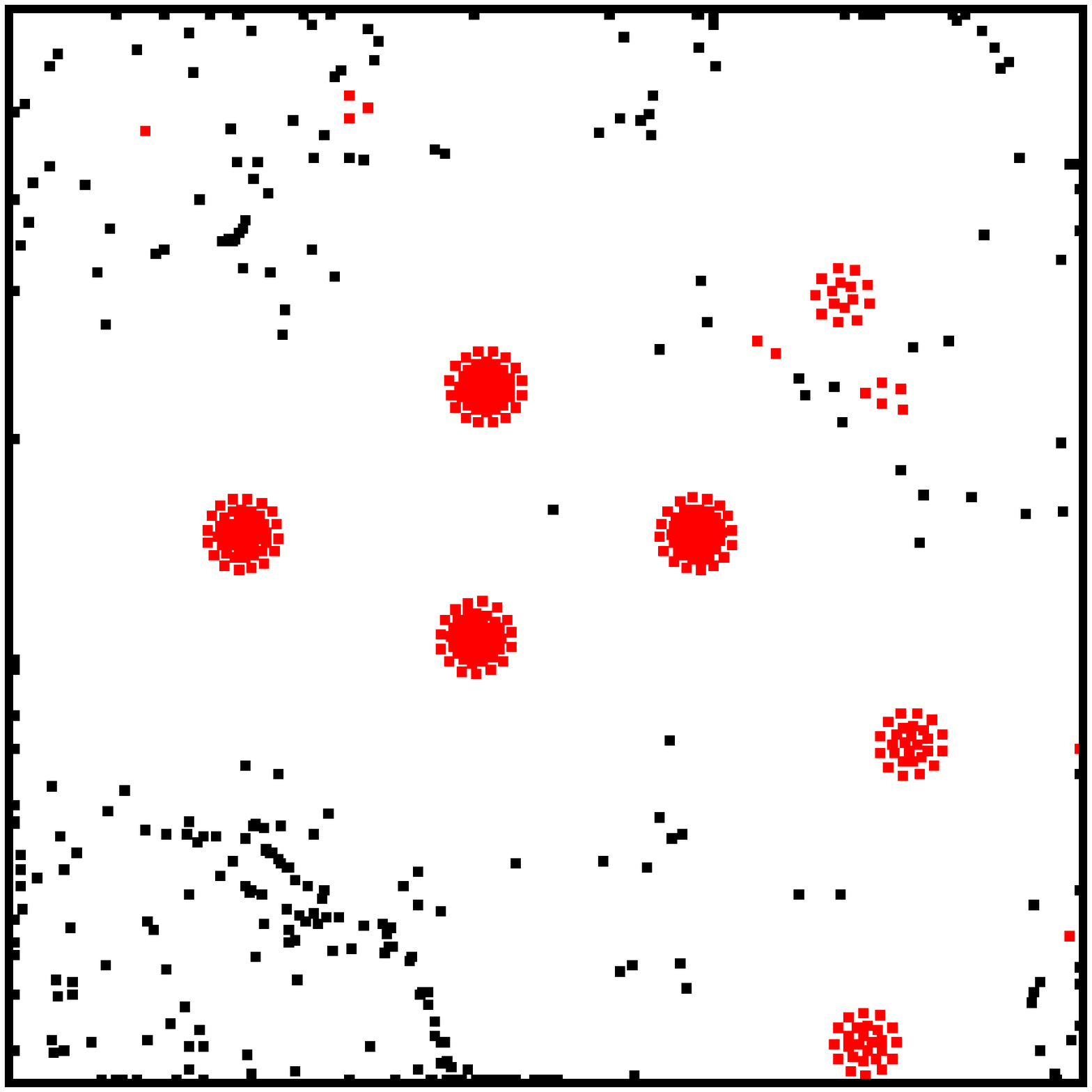}   \\
    (a) $t=10000$ \hspace{0.8cm} (b) $t=75000$ \hspace{0.8cm} (c) $t=150000$ \hspace{0.8cm} (d) $t=300000$ \hspace{0.8cm} (e) $t=500000$ \\
    (ii) \\
    \includegraphics[scale=0.17]{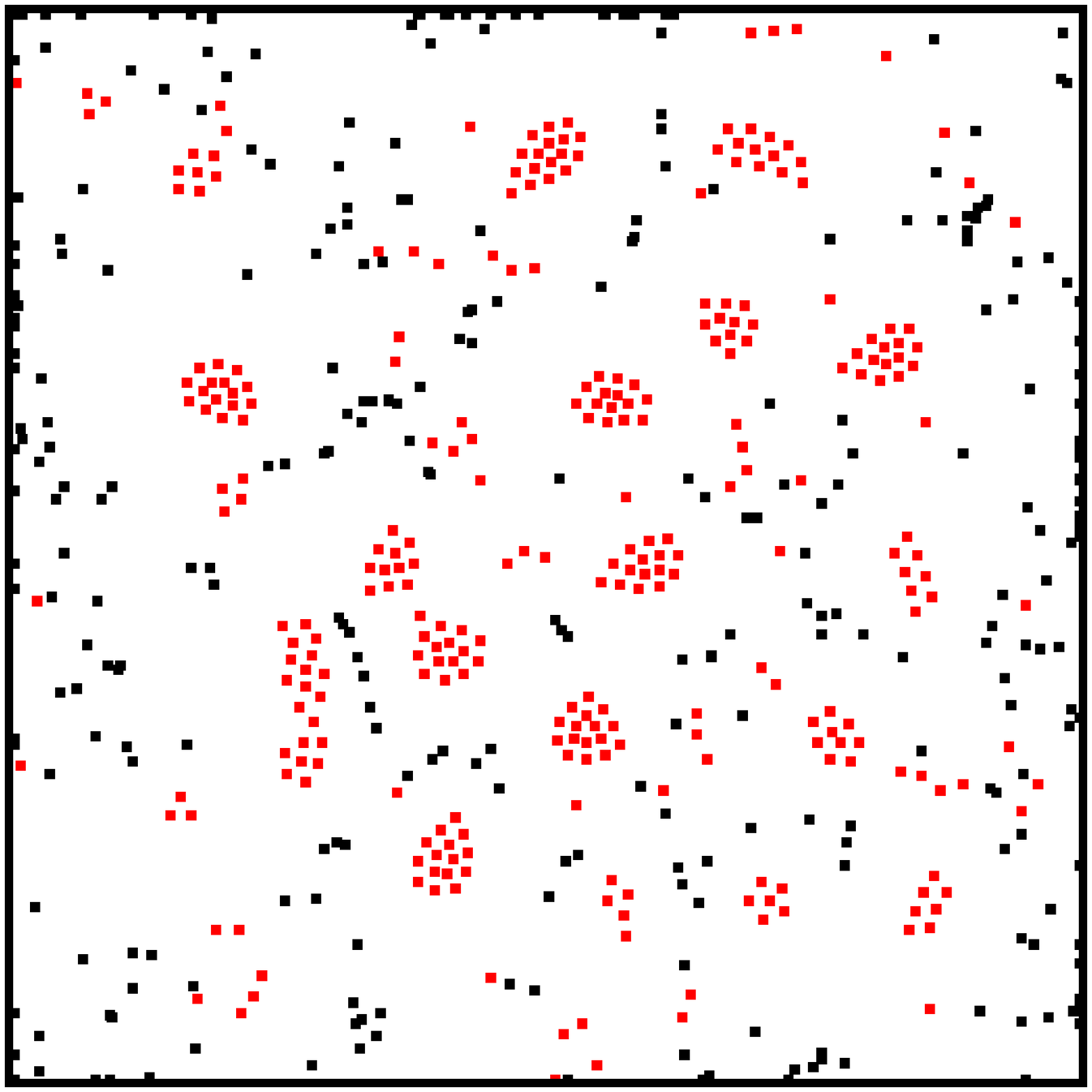}
    \includegraphics[scale=0.17]{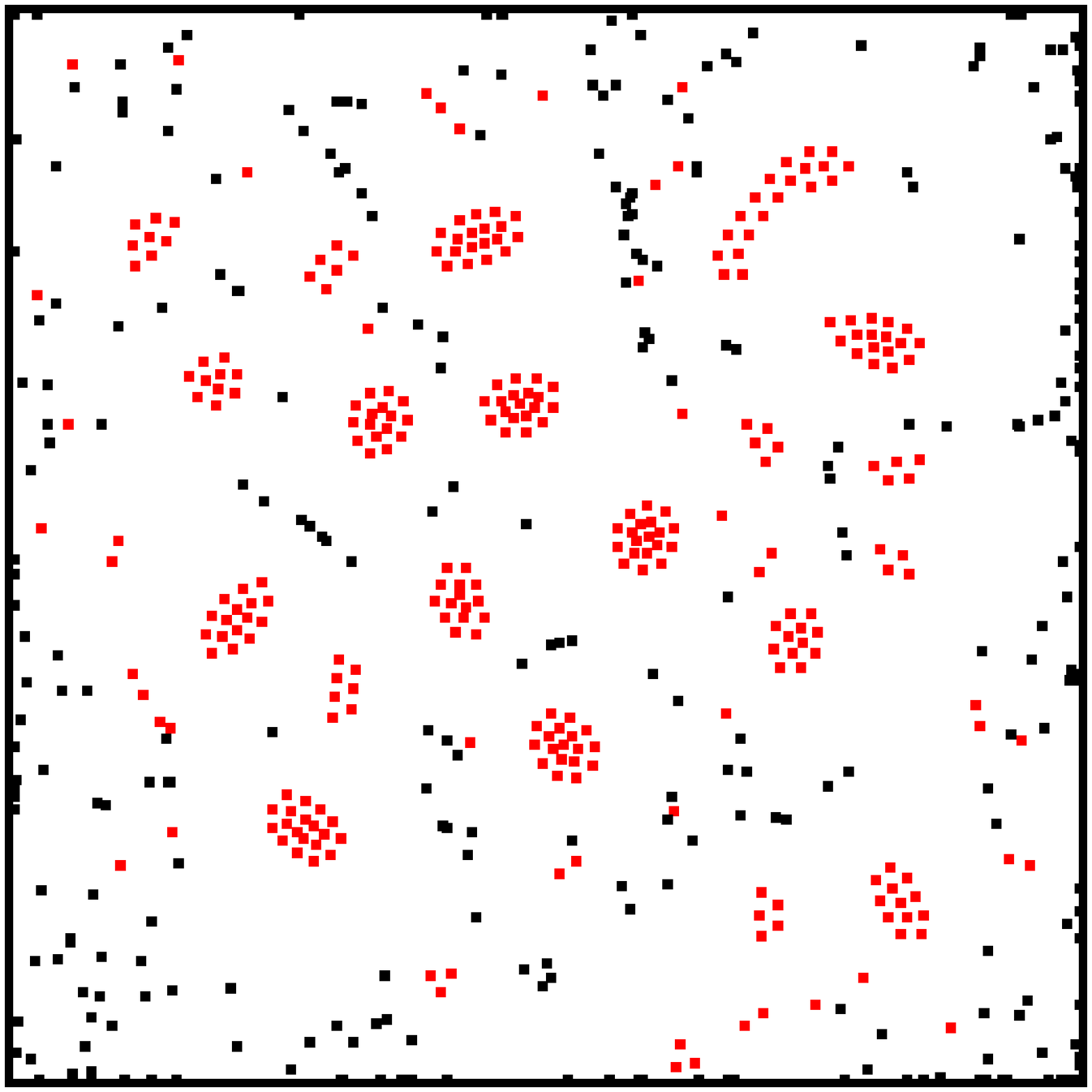}
    \includegraphics[scale=0.17]{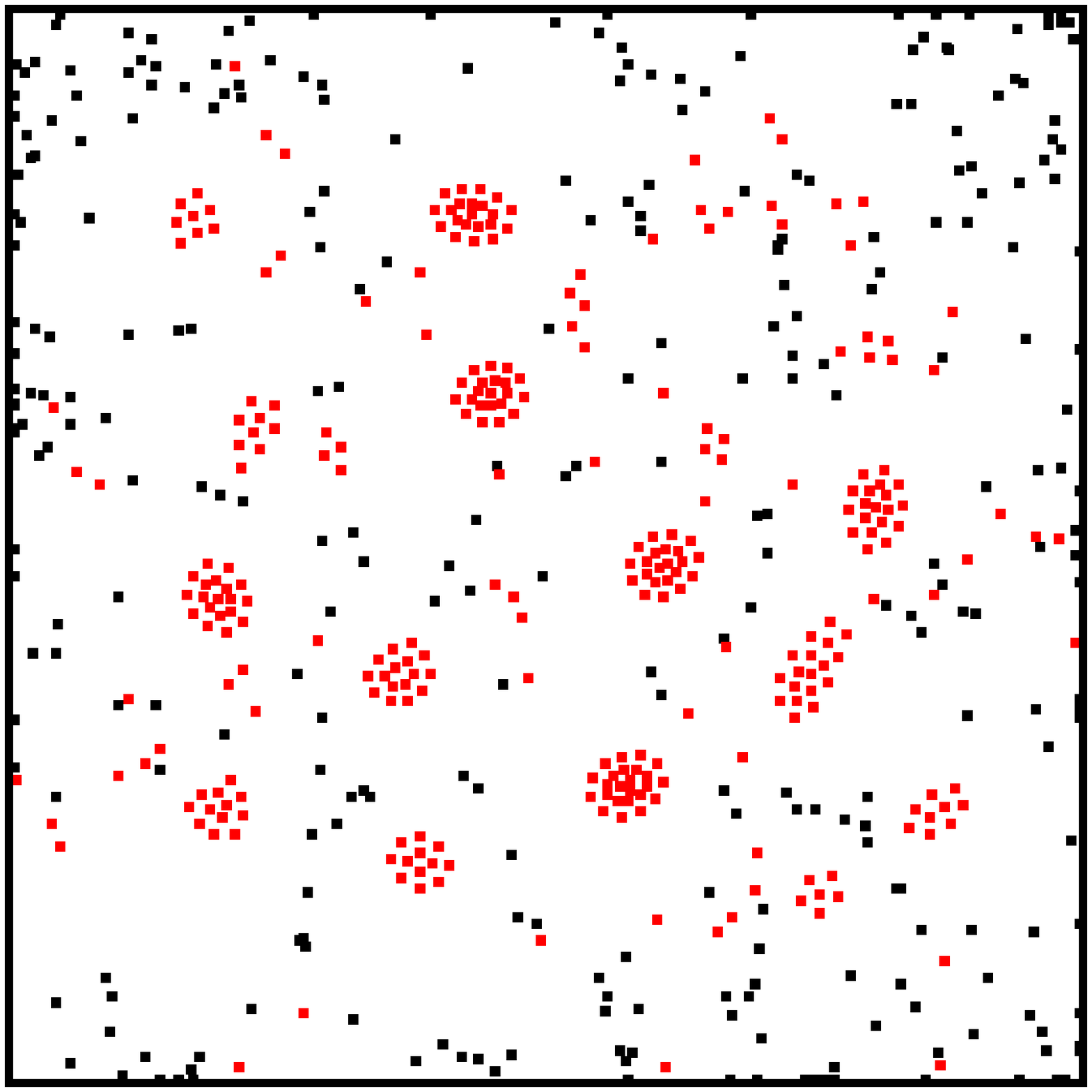}
    \includegraphics[scale=0.17]{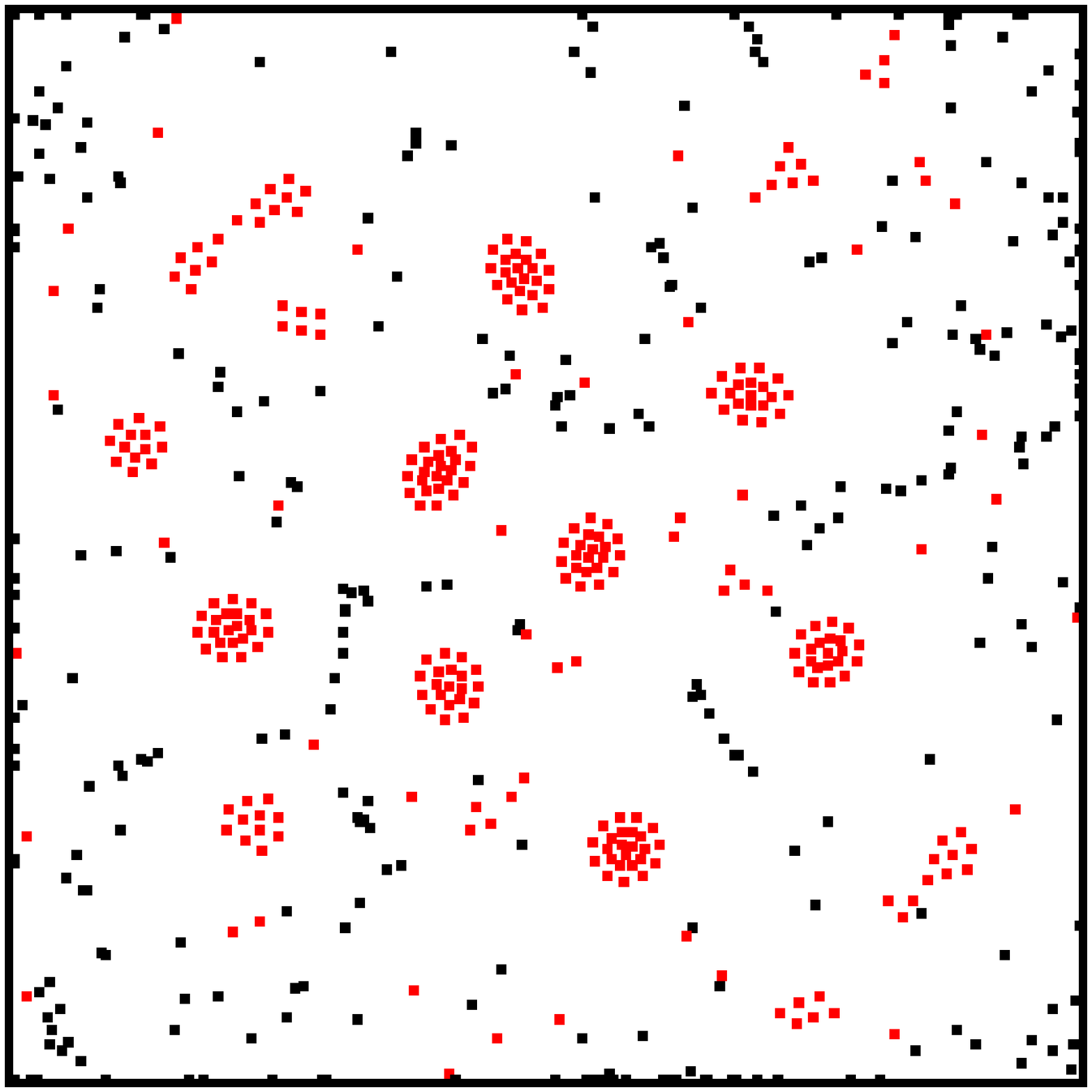}
    \includegraphics[scale=0.17]{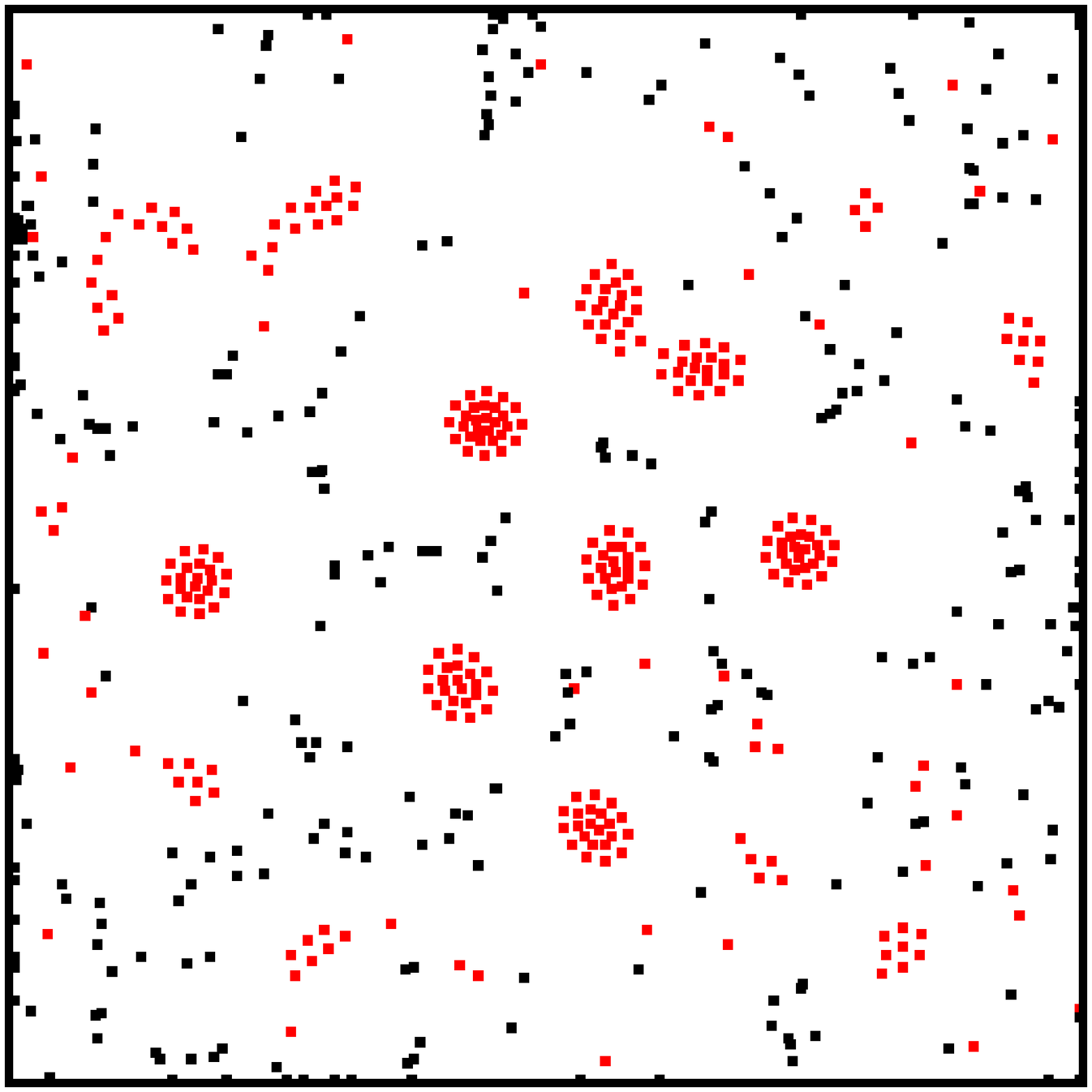}  \\
    (a) $t=50000$ \hspace{0.8cm} (b) $t=100000$ \hspace{0.8cm} (c) $t=200000$ \hspace{0.8cm} (d) $t=300000$ \hspace{0.8cm} (e) $t=500000$ \\
    (iii) \\
    \includegraphics[scale=0.17]{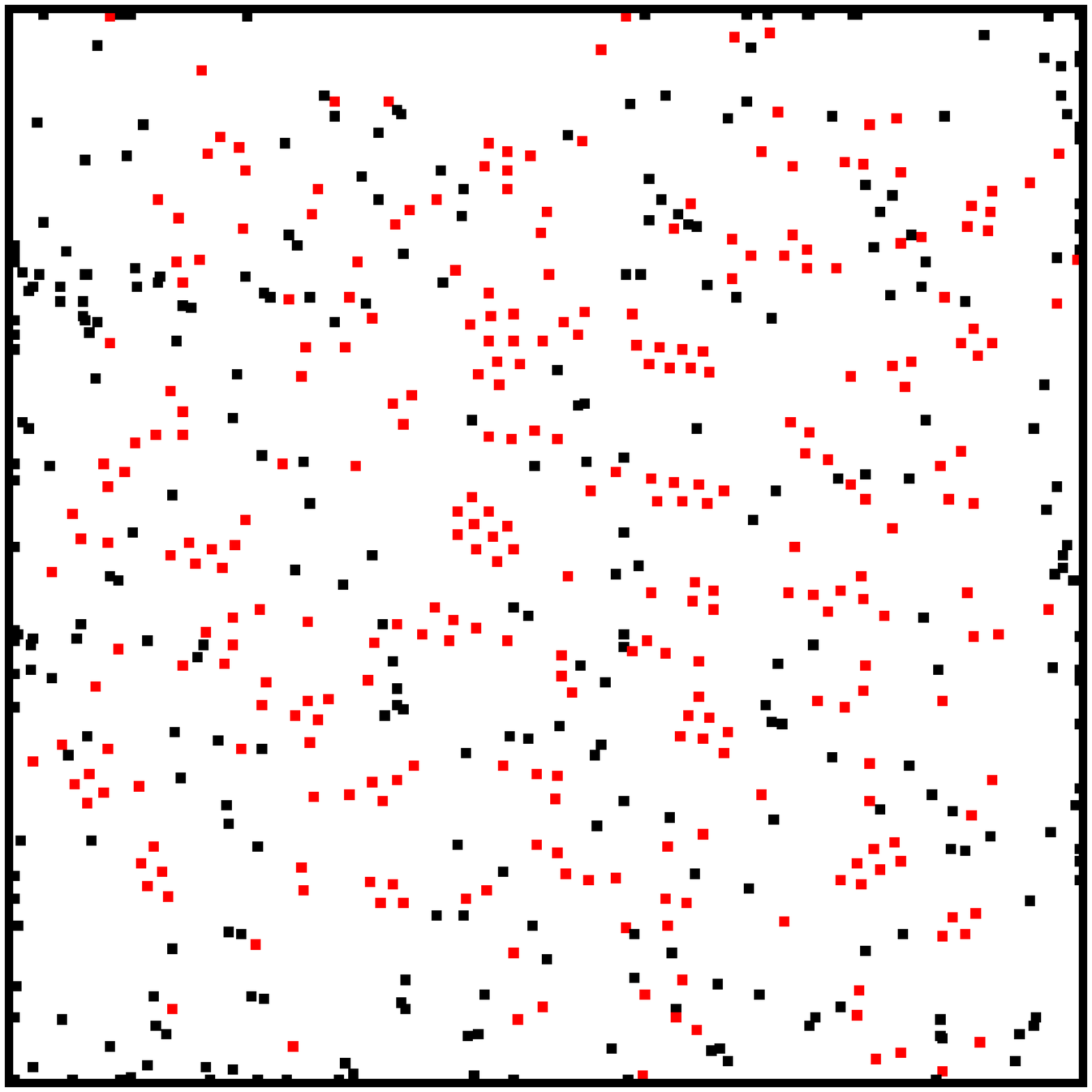}
    \includegraphics[scale=0.17]{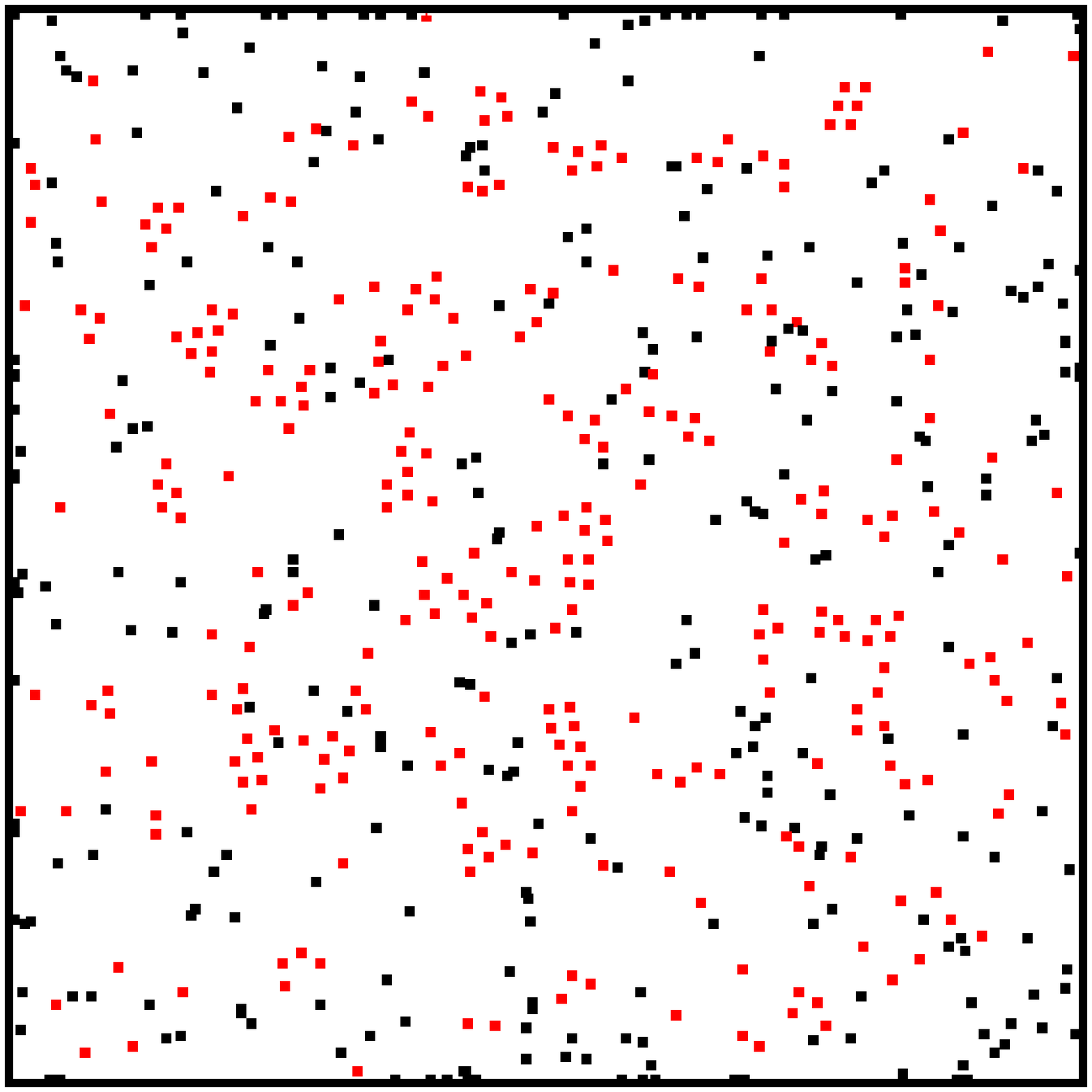}
    \includegraphics[scale=0.17]{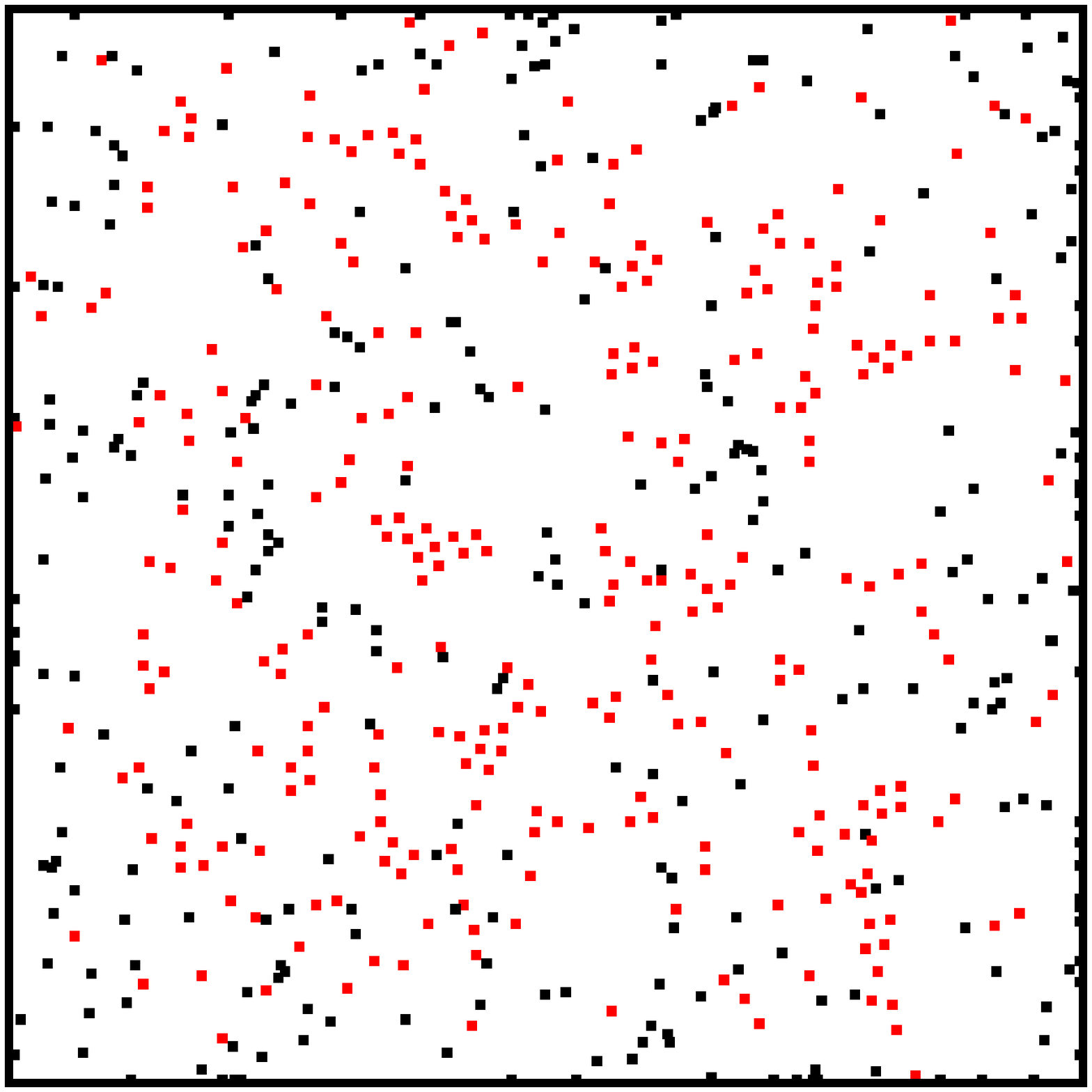}
    \includegraphics[scale=0.17]{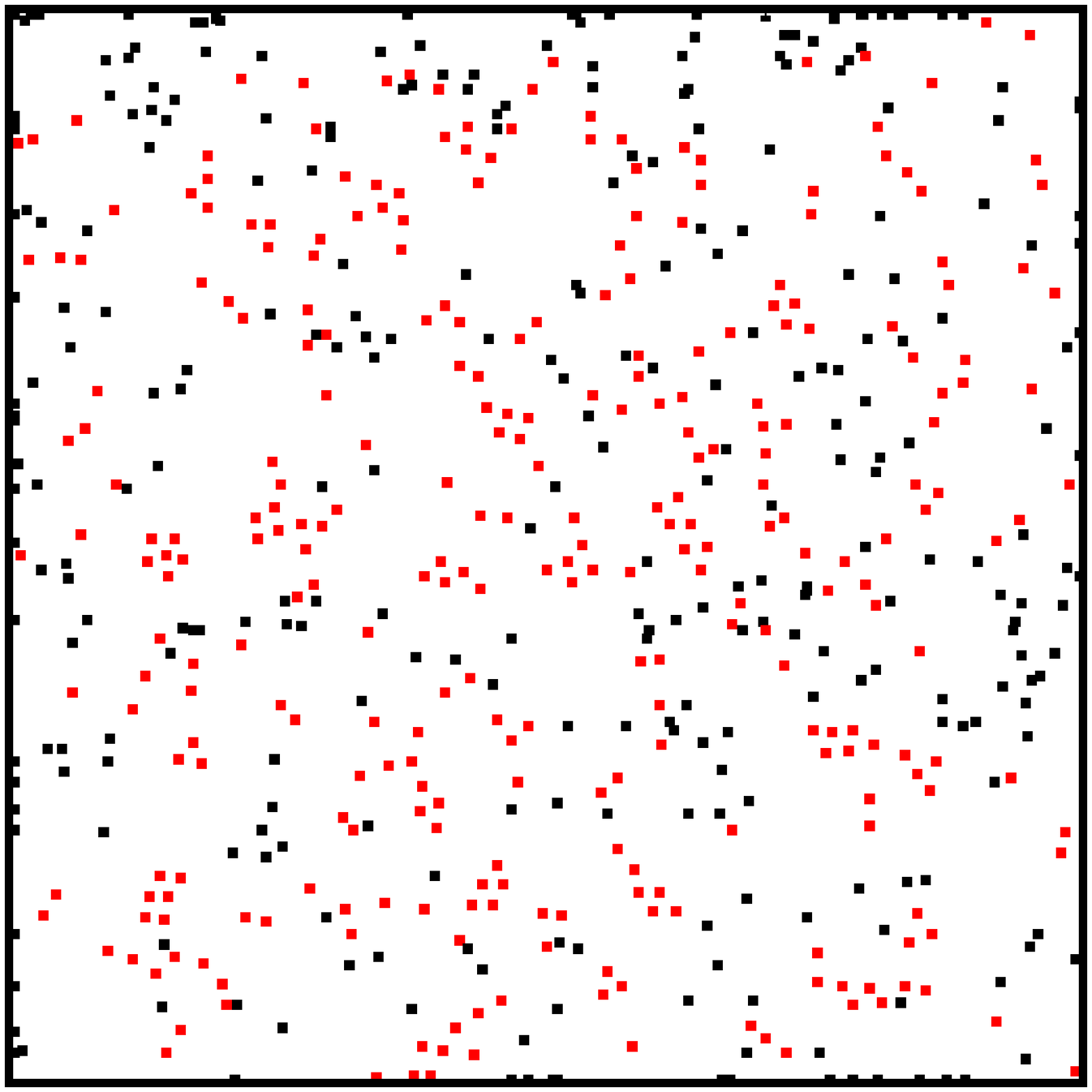}
    \includegraphics[scale=0.17]{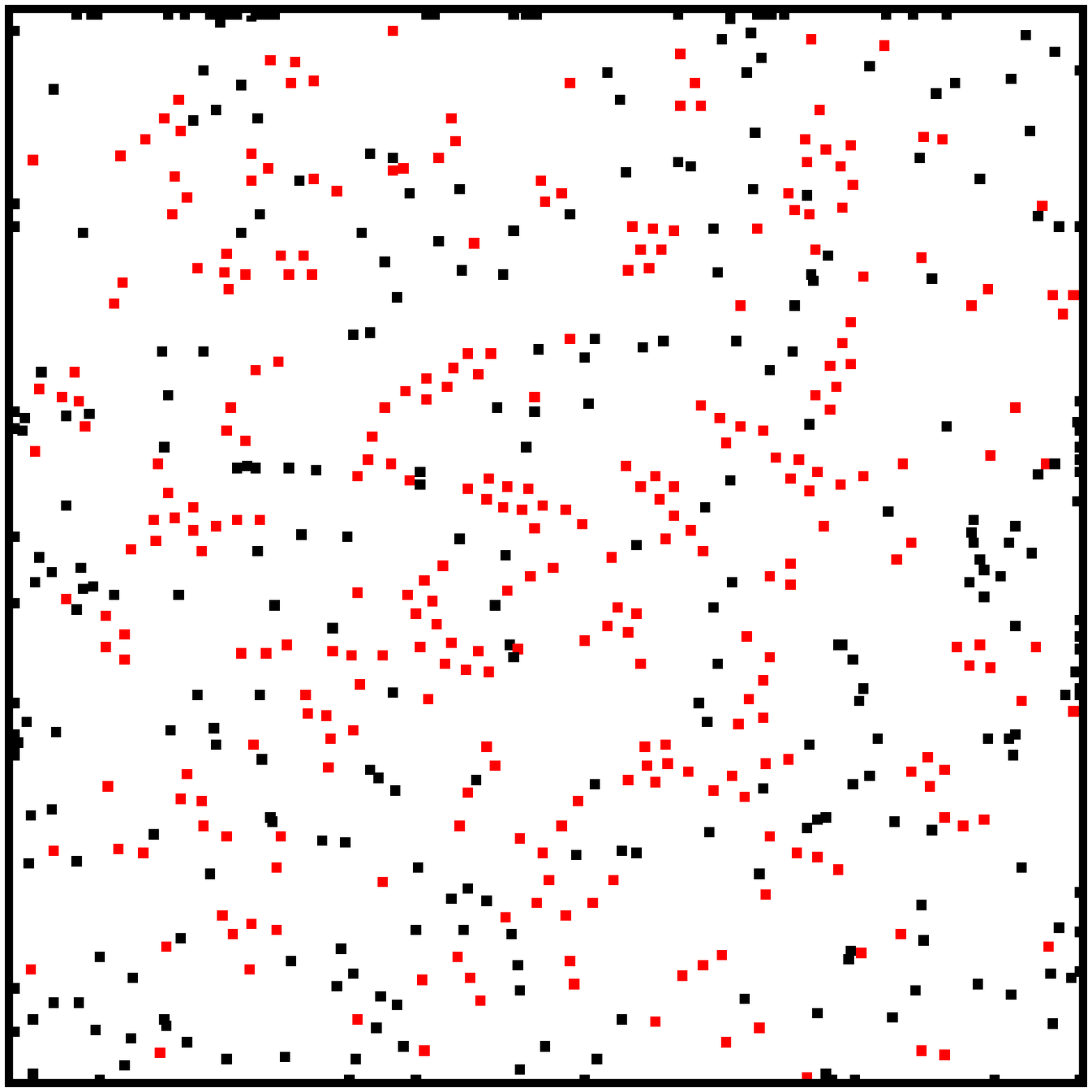}  \\
    (a) $t=100000$ \hspace{0.8cm} (b) $t=200000$ \hspace{0.8cm} (c) $t=300000$ \hspace{0.8cm} (d) $t=400000$ \hspace{0.8cm} (e) $t=500000$ \\
    \caption{Snapshots showing the pattern evolution of predators (red
    -- on-line) and preys (black -- on-line) when the attraction
    between predators is given by (i) $\beta_{predator}=0.5$, (ii)
    $\beta_{predator}=0.125$ and (iii) $\beta_{predator}=0.0625$. }
    \label{fig:04}
  \end{center}
\end{figure*}

In order to take into account the diversified behaviours observed in
nature, we now investigate the emergence of patterns considering
different values of $\beta$ for each species ($\beta = 0.5, 0.125$ and
 $0.0625$). We start by considering null interaction between preys while
the predators are able to feel their counterparts
($\beta_{predator}>0$ and $\beta_{prey}=0$). Observing the evolution
of the system when three distinct values of
$\beta_{predator}$ and $\alpha=32$ are considered (Fig.~\ref{fig:04}), we clearly
identify the emergence of predator clusters which differ according to
the $\beta_{predator}$ intensity.

When the attraction between predators is stronger
(\textit{i.e.} $\beta_{predator}=0.5$, see Figure~\ref{fig:04}-i), non-uniform
clusters emerge in a few time steps (Fig.~\ref{fig:04}-i,a). Mutual
attraction between different clusters progressively generates larger
clusters, with the smaller clusters tending to be attracted by the
larger ones (\textit{e.g.}, as is the case with the two clusters in
the middle of Figure~\ref{fig:04}-i,a). Although the clusters present
several shapes in the first steps, they evolve towards spherical
shapes because of symmetrical internal forces and increase in the
number of members (Fig.~\ref{fig:04}-i,b). After a transient, the
number of clusters seems to stabilize. The smallest clusters can move
throughout the space following the concentration of preys
(Fig.~\ref{fig:04}-i,d) while the denser clusters repel strongly the
preys. Consequently, the cluster net velocity becomes slower than that
of the preys, allowing the latter to escape. As a consequence of the
boundary conditions, such dense clusters tend to concentrate in the
central region of $O$ while the preys move toward the periphery
(Fig.~\ref{fig:04}-i,e). Therefore, we expect a resulting giant
cluster of predators to appear after a long period of time.

Emergence of clusters also occur in the second configuration with
$\beta_{predator}=0.125$ and \mbox{$\beta_{prey}=0$} (Fig.~\ref{fig:04}-ii).
However, they are visually more uniformly distributed
(Fig.~\ref{fig:04}-ii,a) when compared with the previous configuration
(Fig.~\ref{fig:04}-i,a). Since the attraction intensity is smaller,
dense clusters emerge later along the dynamics
(Fig.~\ref{fig:04}-ii,c) while the coarser clusters tend to be
maintained along time (Fig.~\ref{fig:04}-ii,d). The last obtained time
step (Fig.~\ref{fig:04}-ii,e) resembles the initial stage of the first
configuration (Fig.~\ref{fig:04}-i,a), suggesting that the system
would possibly evolve to a similar state. However, the giant cluster
is not expected to emerge because of the weaker attraction between
predators in this case.

The last snapshots sequence (Fig.~\ref{fig:04}-iii) presents the
evolution when the lowest attraction intensity between predators is
considered ($\beta_{predator}=0.0625$ and $\beta_{prey}=0$). In this
configuration, we cannot identify dense clusters up to the last time
step. However, spatio-temporal patterns of uniformly distributed small
clusters emerge and evolve in a non-stationary way with predators and
preys moving close to each other. The pattern consists of clusters
with different sizes and shapes, indicating that predators are able to
move throughout sub-space $O$, moving between clusters
(Figs.~\ref{fig:04}-iii,d and~\ref{fig:04}-iii,e). Any attempt of
escape by preys is promptly checked by predators, which can easily
enclose any group of preys since the predators are not strongly
attracted in this configuration.

When we consider null attraction between predators and different
attraction intensities between preys, the spatio-temporal pattern
tends to include uniformly distributed predators and preys (similar to
Figure~\ref{fig:03}). Actually, small clusters of preys emerge but are
readily eliminated because of the strong predators attraction,
suggesting a cyclic process. Although we can identify some larger
clusters with higher attraction intensity, the pattern is nearly
independent of $\beta_{prey}$.

\begin{figure*}
  \begin{center}
    (i) \\
    \includegraphics[scale=0.17]{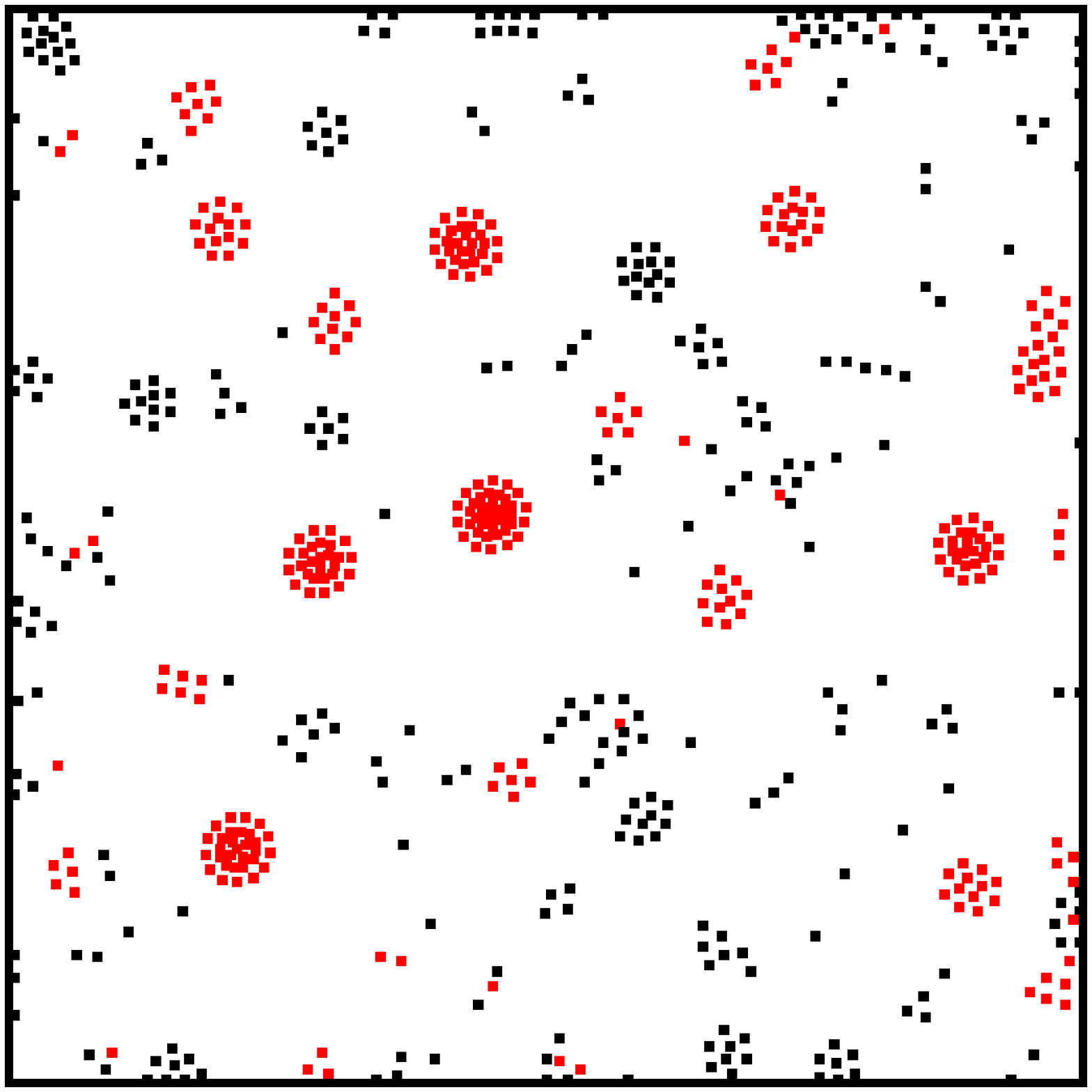}
    \includegraphics[scale=0.17]{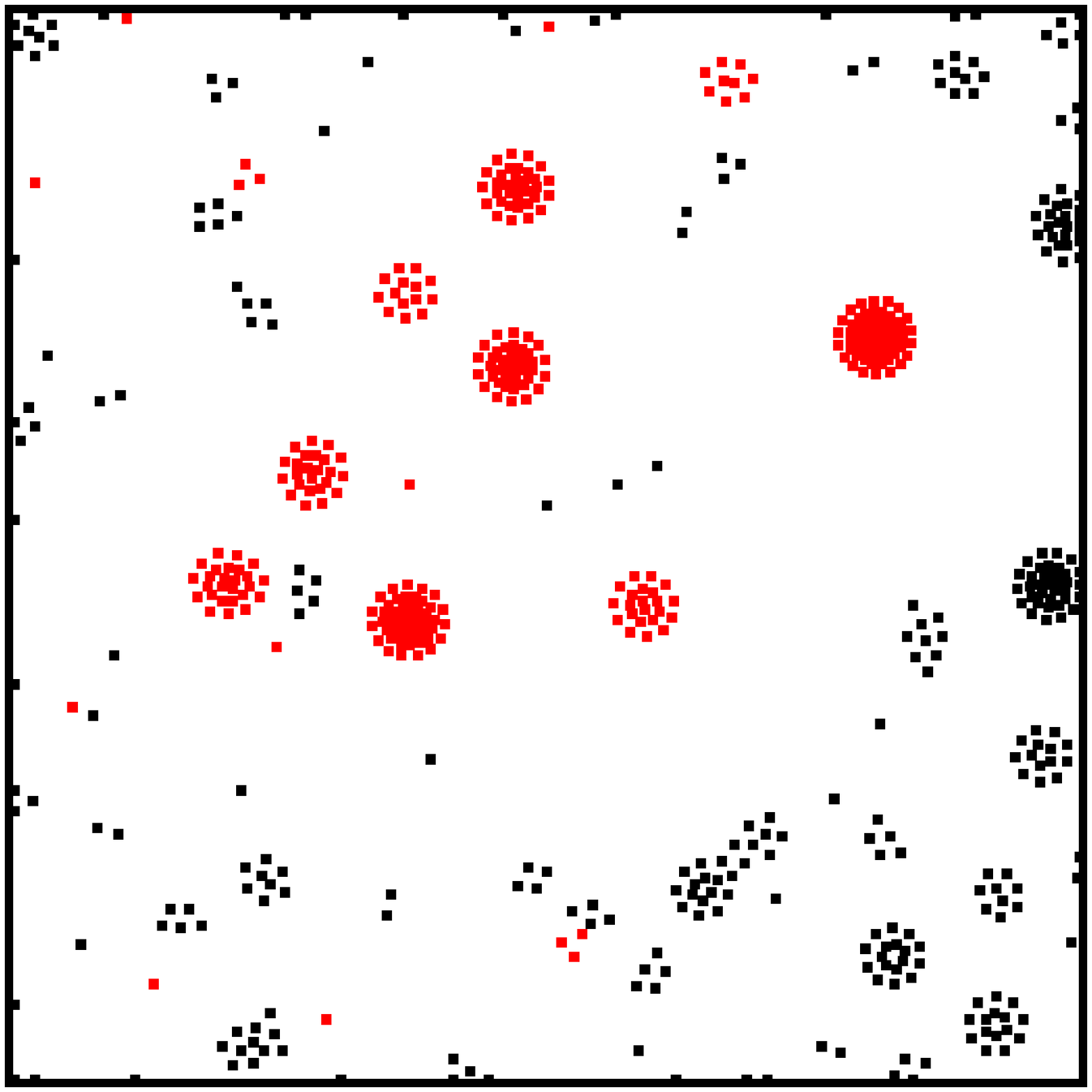}
    \includegraphics[scale=0.17]{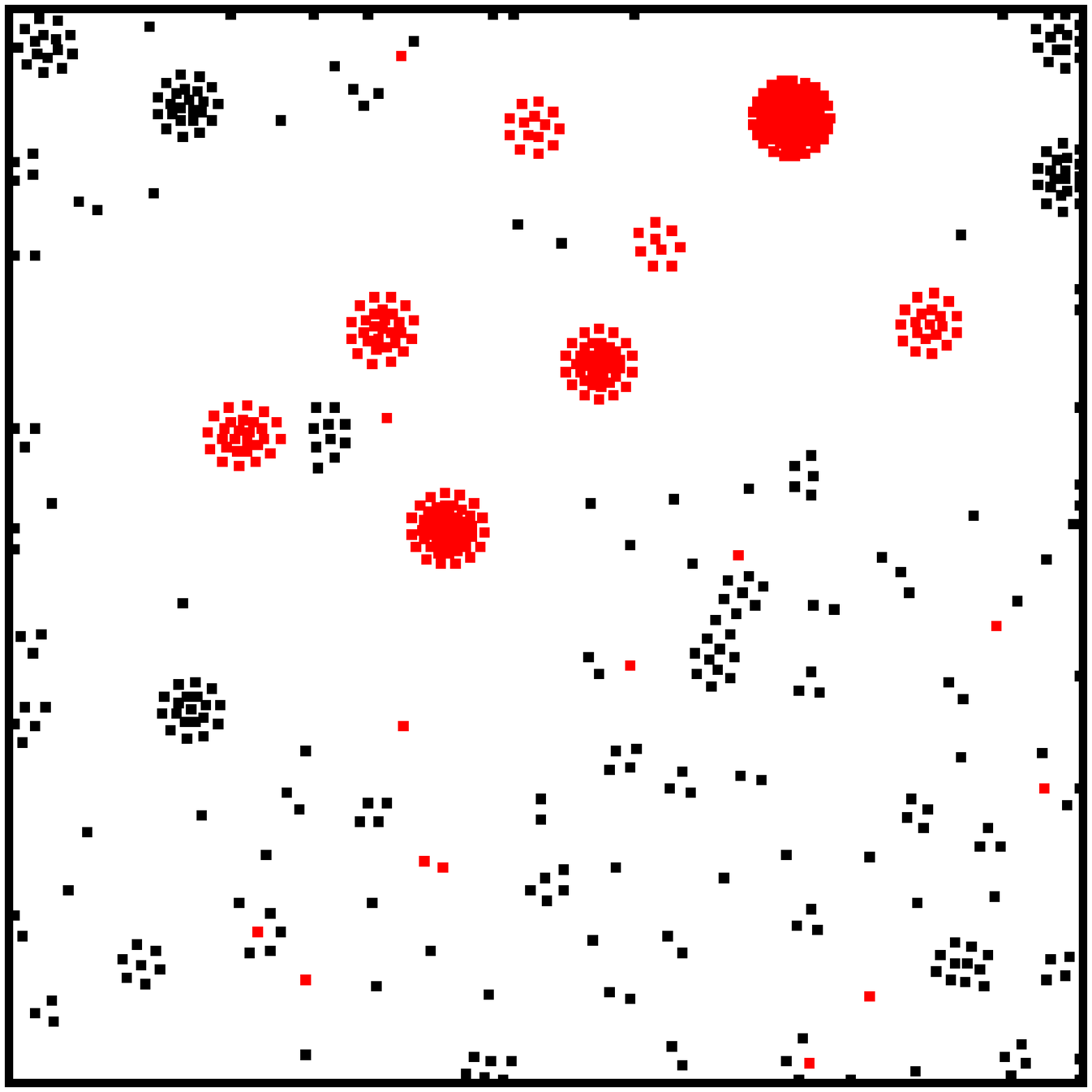}
    \includegraphics[scale=0.17]{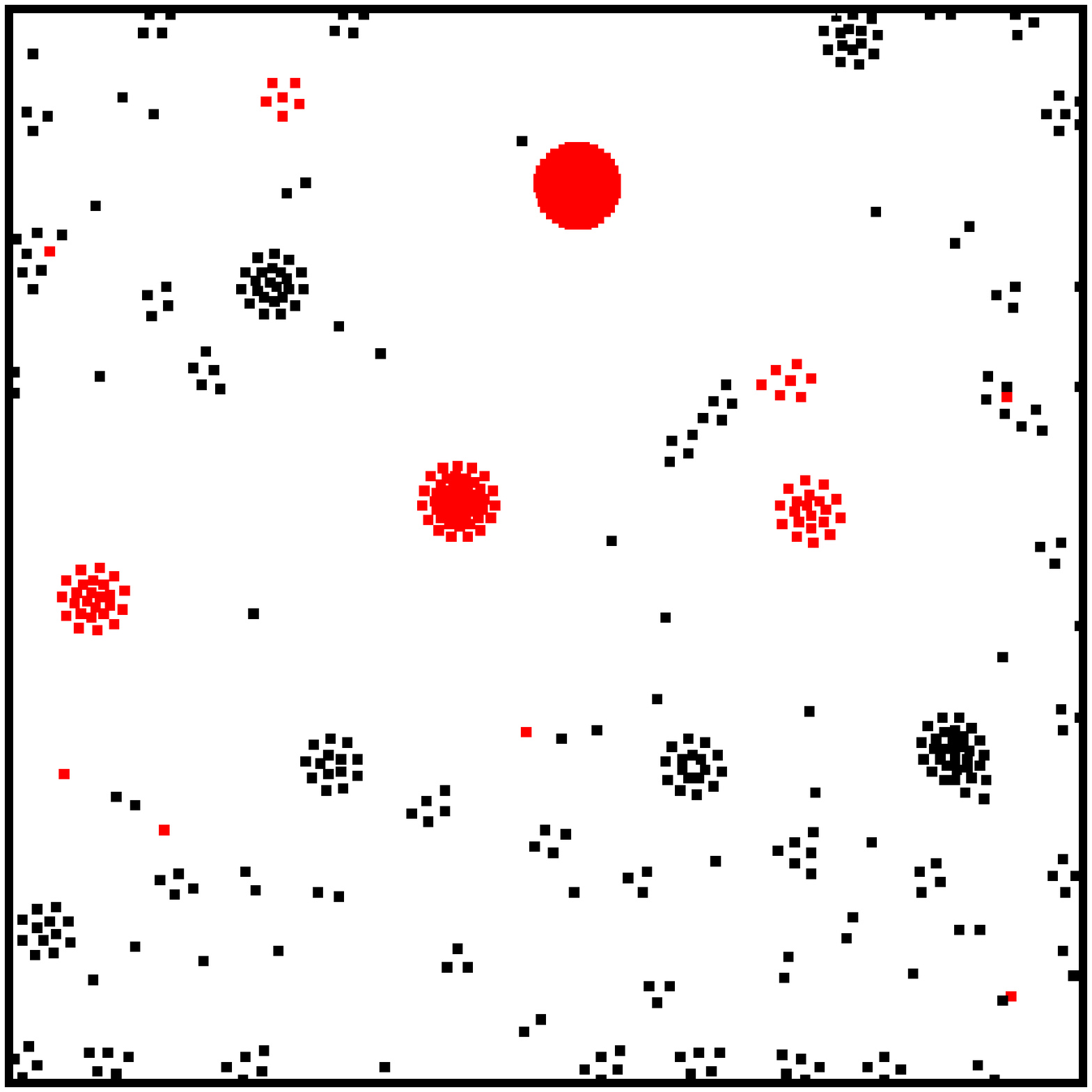}   \\
    (a) $t=10000$ \hspace{0.8cm} (b) $t=75000$ \hspace{0.8cm} (c) $t=150000$ \hspace{0.8cm} (d) $t=500000$ \\
    (ii) \\
    \includegraphics[scale=0.17]{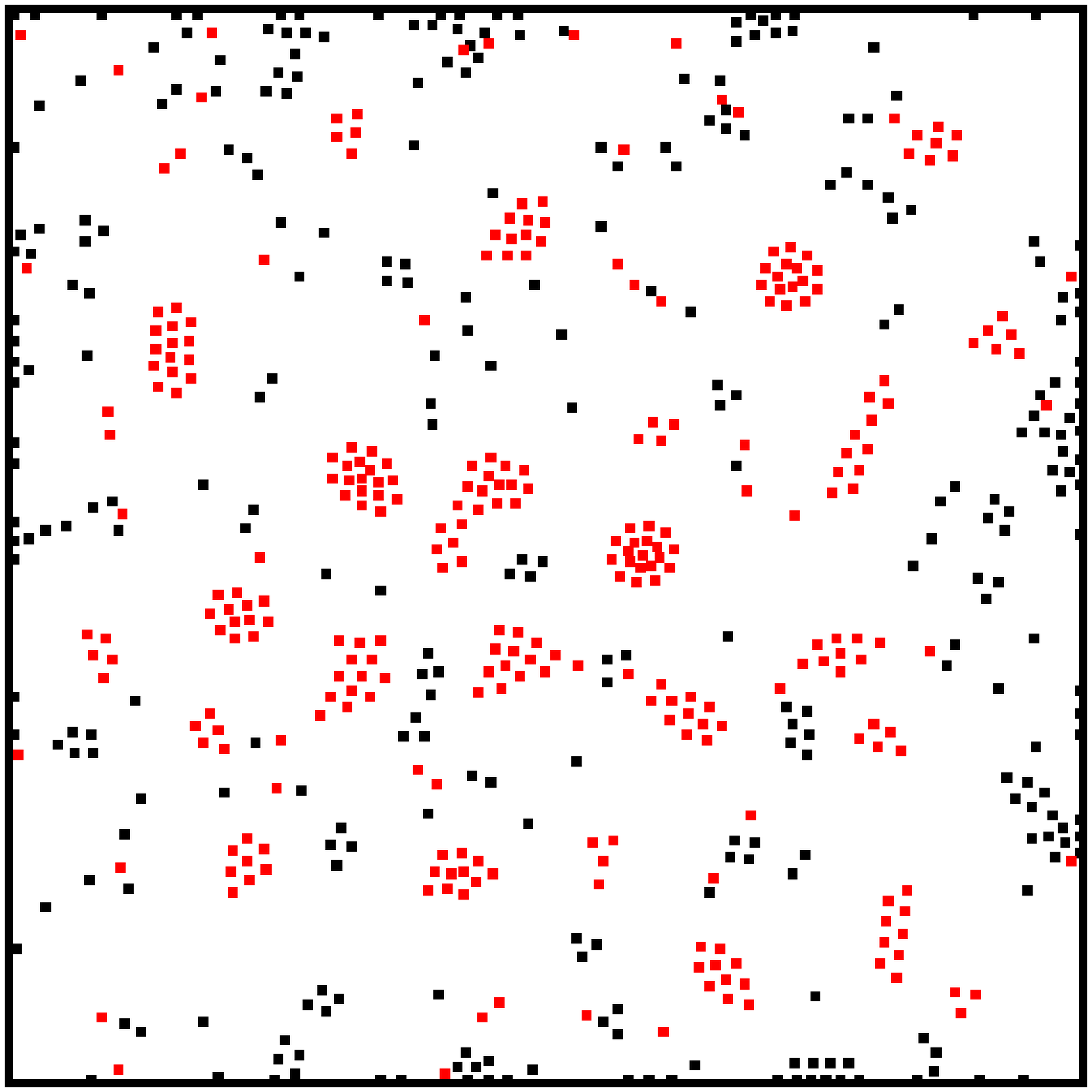}
    \includegraphics[scale=0.17]{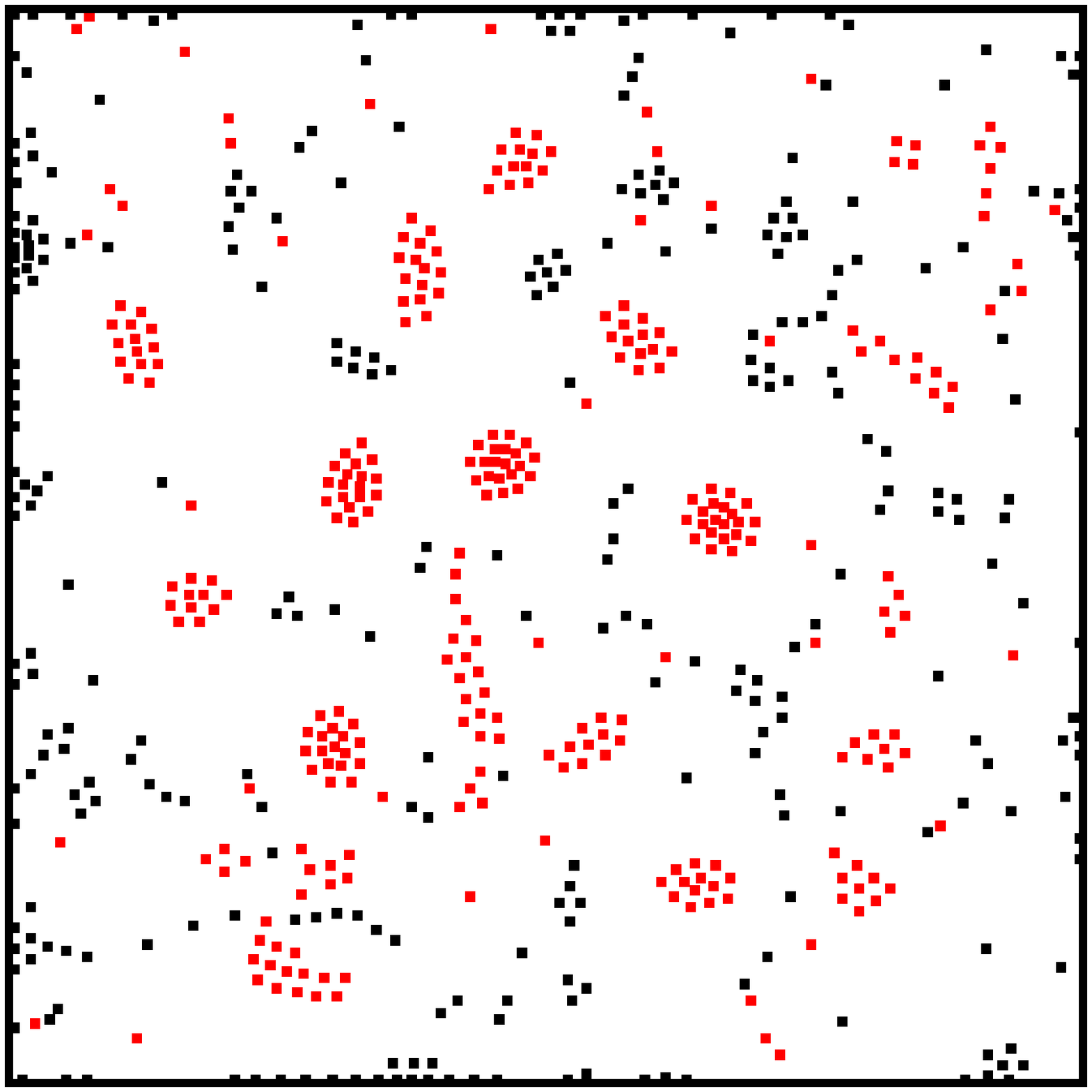}
    \includegraphics[scale=0.17]{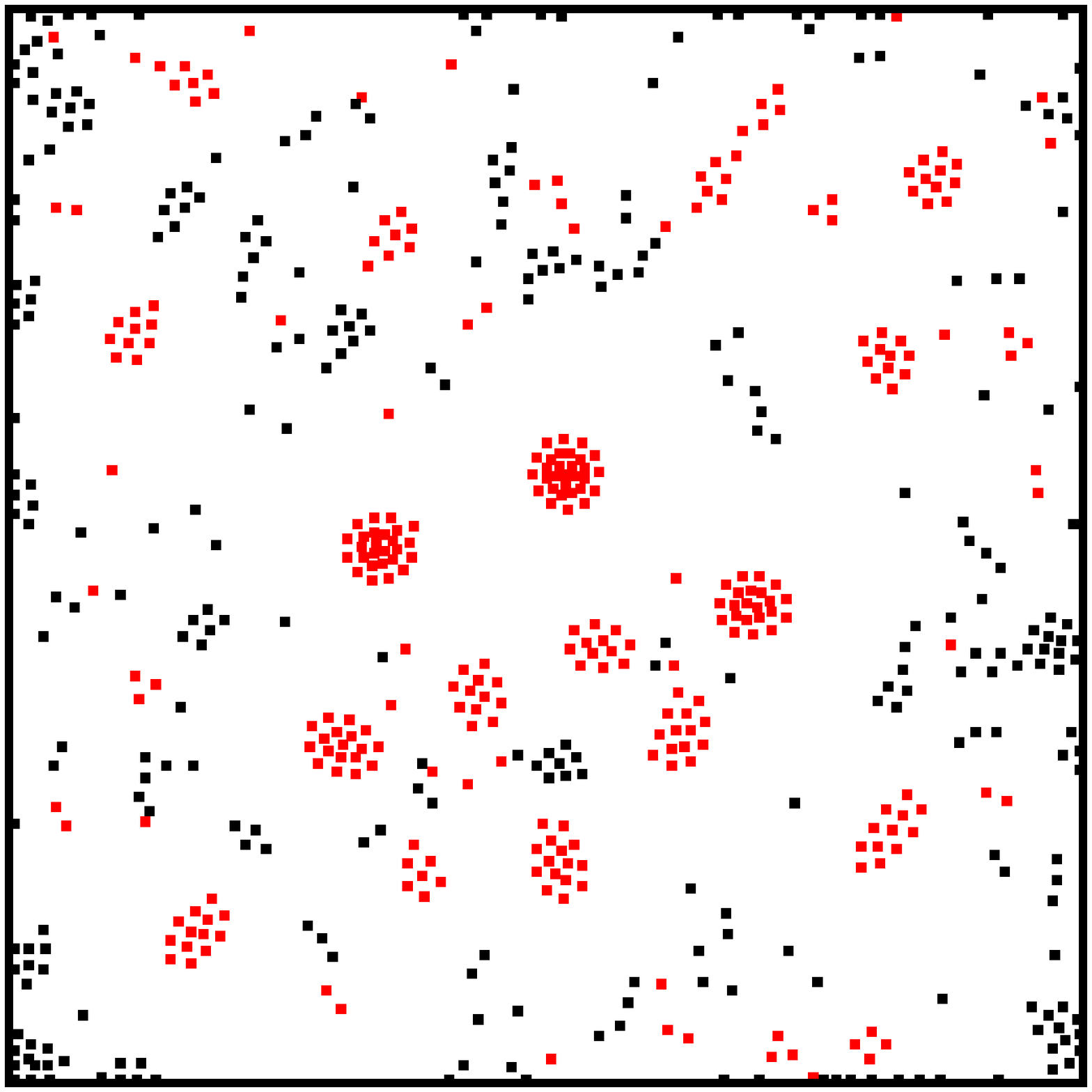}
    \includegraphics[scale=0.17]{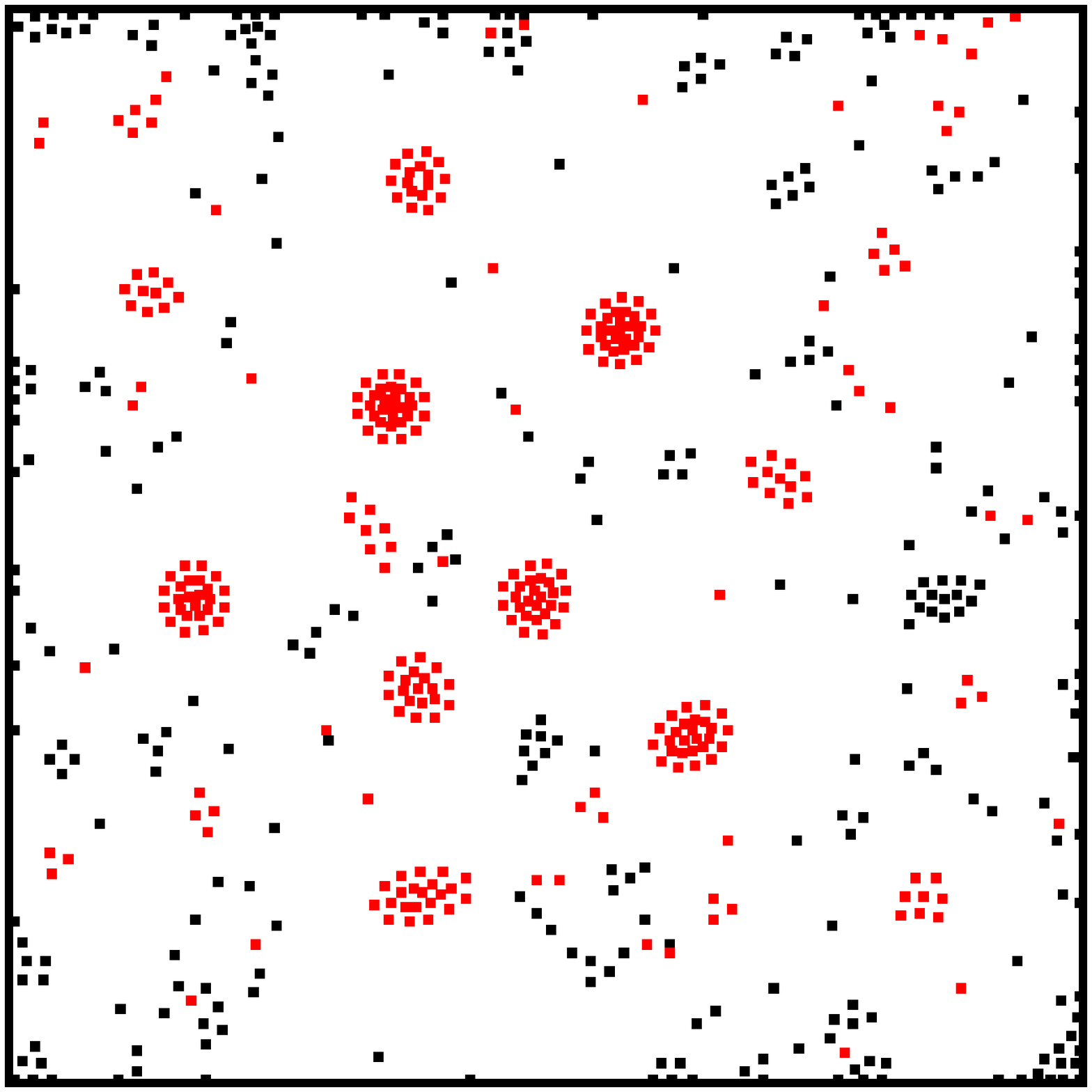}  \\
    (a) $t=50000$ \hspace{0.8cm} (b) $t=100000$ \hspace{0.8cm} (c) $t=200000$ \hspace{0.8cm} (d) $t=500000$ \\
    \caption{Snapshots showing the evolving spatio-temporal patterns of
    predators (red -- on-line) and preys (black --
    on-line). The attraction between same species individuals is given by
    (i) $\beta=0.5$ and (ii) $\beta=0.125$.}  \label{fig:05}
  \end{center}
\end{figure*}

The last experiment consists in introducing attraction between
individuals of both species, \textit{i.e.}, both predators and preys
are able to feel their counterparts
($\beta_{predator}=\beta_{prey}=\beta>0$) while they can also feel
each other (\textit{i.e.} $\alpha=32$). The high level of attraction
$\beta=0.5$ resulted a pattern of dense clusters of predators while
smaller clusters of preys have also been observed along the first time
steps (Fig.~\ref{fig:05}-i,a). The simultaneous attraction between
members of each species generated distinct regions occupied by
different species (Figs.~\ref{fig:05}-i,b to~\ref{fig:05}-i,d). The
higher concentration of preys in some clusters increased the intensity
of the sensitive field such that the clusters of predators could move
around a larger region (following the prey clusters) when compared
with the configuration with null attraction between preys
(Fig.~\ref{fig:04}). The clusters of preys often disappear and emerge
as a result of this attraction (Fig.~\ref{fig:05}-i). On the other
hand, the clusters of predators become denser over time because of
the non-uniformity in the distribution of the clusters of preys
(Fig.~\ref{fig:05}-i,d). Since the preys are non-uniformly distributed
in space, they create attractor regions.

The snapshots from Figure~\ref{fig:05}-ii show the evolution of the
system with lower attraction \mbox{$\beta=0.125$} between same species
individuals. In this case, the clusters are not so dense and they are more
uniformly spatially distributed (Fig.~\ref{fig:05}-ii,a). The preys
are also organized in clusters of smaller sizes
(Fig.~\ref{fig:05}-ii,b). The sequence of snapshots suggests a
dynamical behaviour where some clusters aggregate more predators while
the greatest part of clusters divide and recombine in new sets of
animals (Fig.~\ref{fig:05}-ii,c). The clusters of predators follow the
concentration of preys, which become higher in some regions over time.

In the last configuration where the attraction has the smallest value
($\beta=0.0625$), the density of clusters in the pattern was smaller
than in the other cases (Figs.~\ref{fig:05}-i
and~\ref{fig:05}-ii). Actually, the situation is similar to the
absence of attraction between preys and small attraction between
predators (Fig.~\ref{fig:04}-iii). However, in this case we could
identify some small clusters of preys which constantly appeared and
disappeared along time, while the clusters or predators resulted very
small.

\subsection{Prey Elimination}

\begin{figure*}
  \begin{center}
    \includegraphics[scale=0.25]{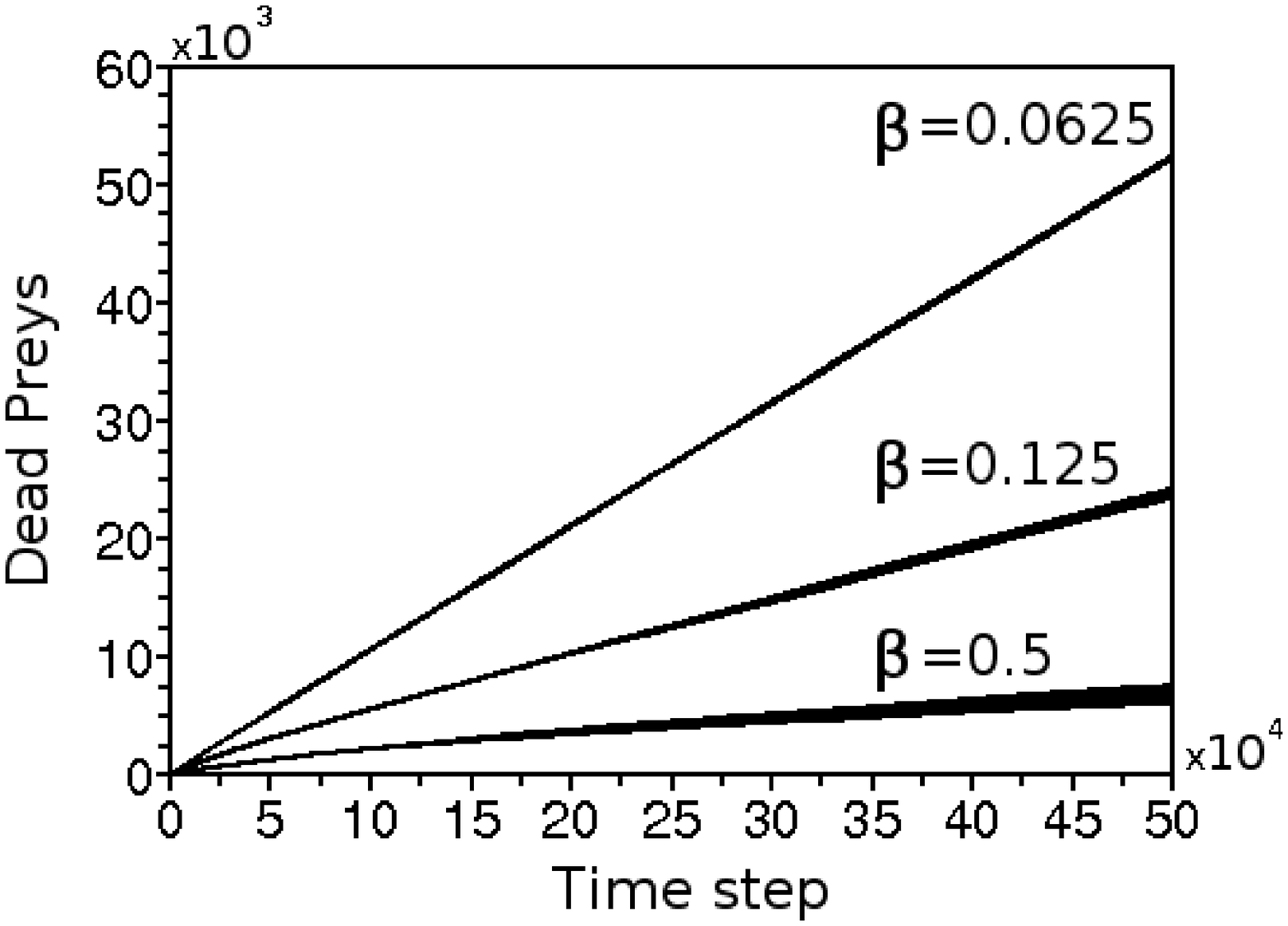}
    \includegraphics[scale=0.25]{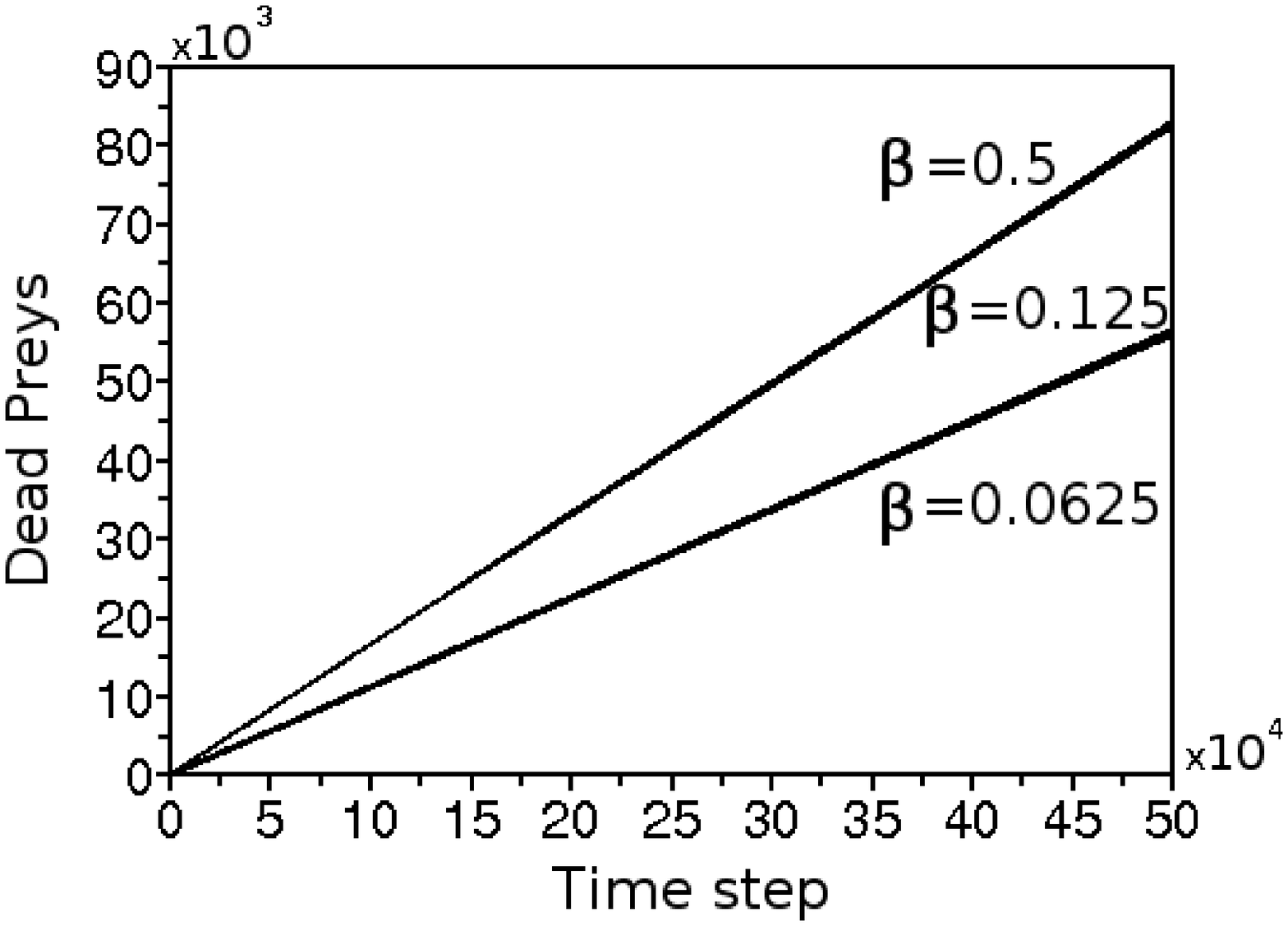}
    \includegraphics[scale=0.25]{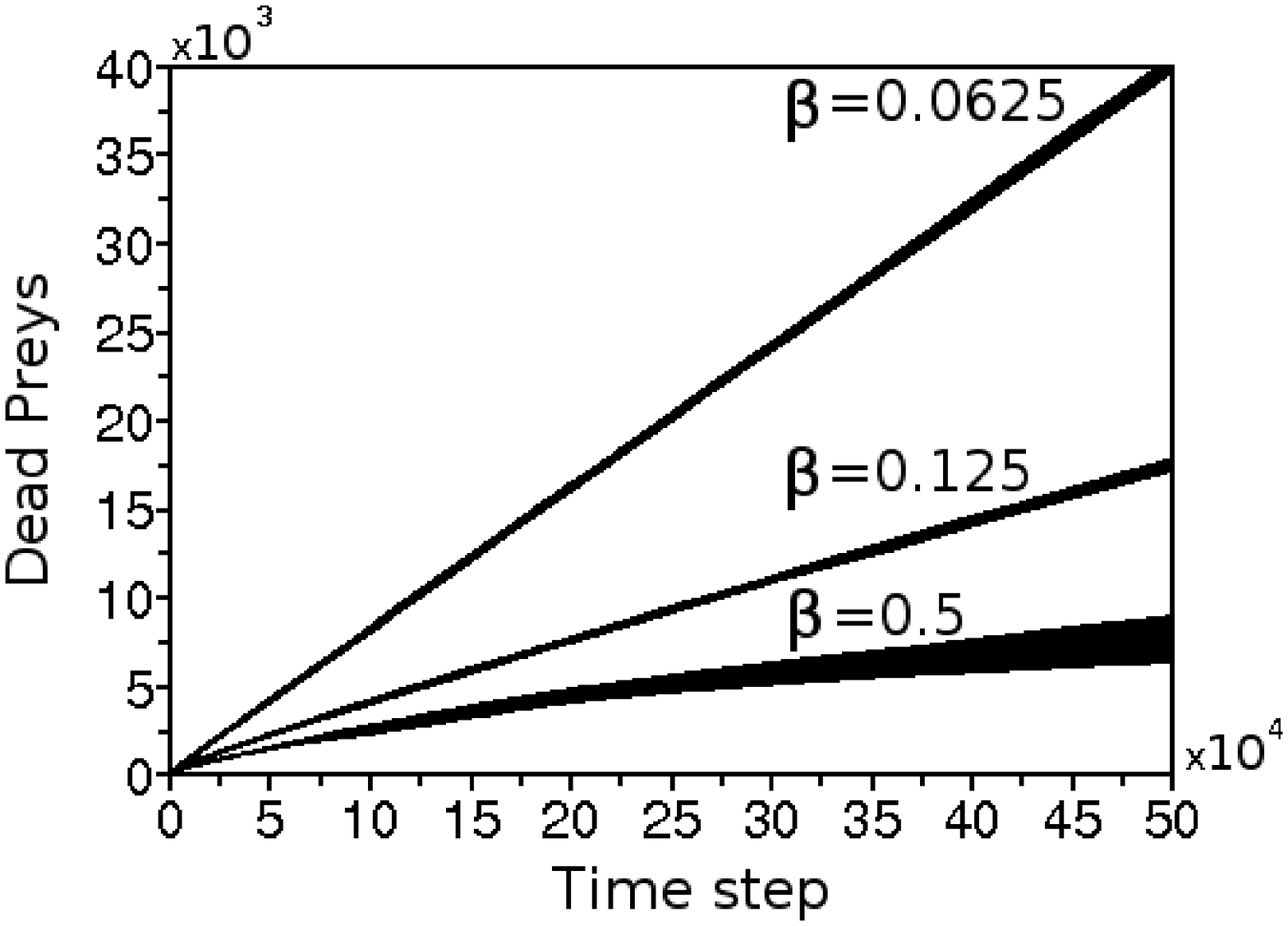} \\
    (a) $\beta_{predator}>0 \text{ and } \beta_{prey}=0$ \hspace{0.5cm} (b) $\beta_{predator}=0 \text{ and } \beta_{prey}>0$ \hspace{0.5cm} (c) $\beta_{predator}>0 \text{ and } \beta_{prey}>0$ \\
    \caption{Evolution of the amount of eliminated preys (average
    and standard deviations). We considered three intensities of
    attraction between same species animals. } \label{fig:06}
  \end{center}
\end{figure*}

Spatial cluster structure emerges in animal behaviour as a consequence
of several factors. Usually, group structure is a consequence of some
type of similarity which provides protection against external
agents. Protection not only against other individuals, but also
against ideas and other cultural behaviour. In our model, the
cluster structure depends on the attraction intensity between same
species individuals and can favor species according to the attraction
configuration. Although the preys death by itself is not necessarily
the final purpose of our model, the number of eliminated preys
indicates the importance of clusters and how they constrain the
spatial movement of predators and preys. For instance, attraction
between preys increases the sensitive field of predators because of
the larger number of preys close one another. Consequently, the
predators velocity directed to the prey clusters increases while the
cluster structure constrains the movement of single
preys. Contrariwise, the attraction between predators generates strong
fields which allow preys to escape faster than the predators can
move. This situation results in lower rates of preys deaths
(Fig.~\ref{fig:06}-a), while preys were quickly eliminated in the
previous case (Fig.~\ref{fig:06}-b).

The preys death rate is linear in almost all configurations, except in
the cases with $\beta_{predator}=0.5$, where the growth rate is faster
along the initial $200000$ time steps, slowing afterwards
(Figs.~\ref{fig:06}-a and~\ref{fig:06}-c). This two-slope behaviour is
a direct consequence of the emergence of dense predator clusters about
this time step, decreasing the preys death rate. The existence of
attraction in both species (Fig.~\ref{fig:06}-c) has similar effect as when attraction is
allowed only between predators (Fig.~\ref{fig:06}-a).
However, the cumulative number of eliminated
preys is larger in the latter case (Fig.~\ref{fig:06}-a). These
results suggest that the spatial organization of animals has crucial
importance for their survival. Therefore, according to the predator
species, there is no difference for preys to be organized in clusters
or not. Since the hunting ability of predators is similar, the preys
organization should be determined by other environmental reasons. In
this sense, the choice of preys by predators might be directly related
to the way in which the preys are spatially organized into groups.

\section{Structural Properties}

The mechanism proposed to build the complex networks generates
\textit{bi-partite} networks with two types of nodes, respectively:
predators and preys. Eliminating the connections between nodes of
different types, two sub-networks are obtained which are henceforth
named predator $\Gamma_{predator}$ and prey $\Gamma_{prey}$
networks. We shall analyse the enrollment and movement of individuals
between spatial clusters by investigating local structural properties
of both resulting networks.

\subsection{Evolution of the structural properties}

\begin{figure*}
  \begin{center}
    (i) \\
    \includegraphics[scale=0.27]{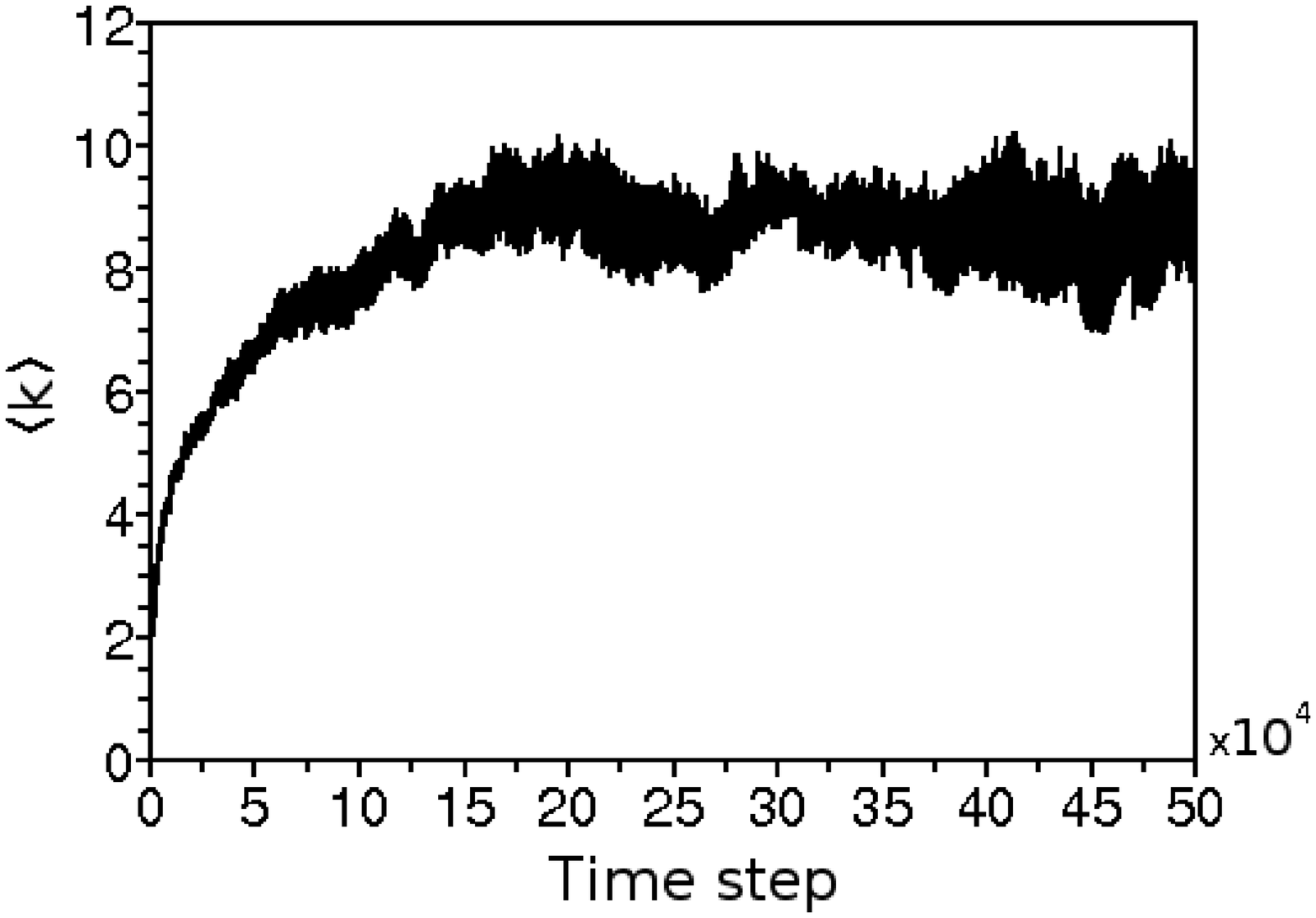}
    \includegraphics[scale=0.27]{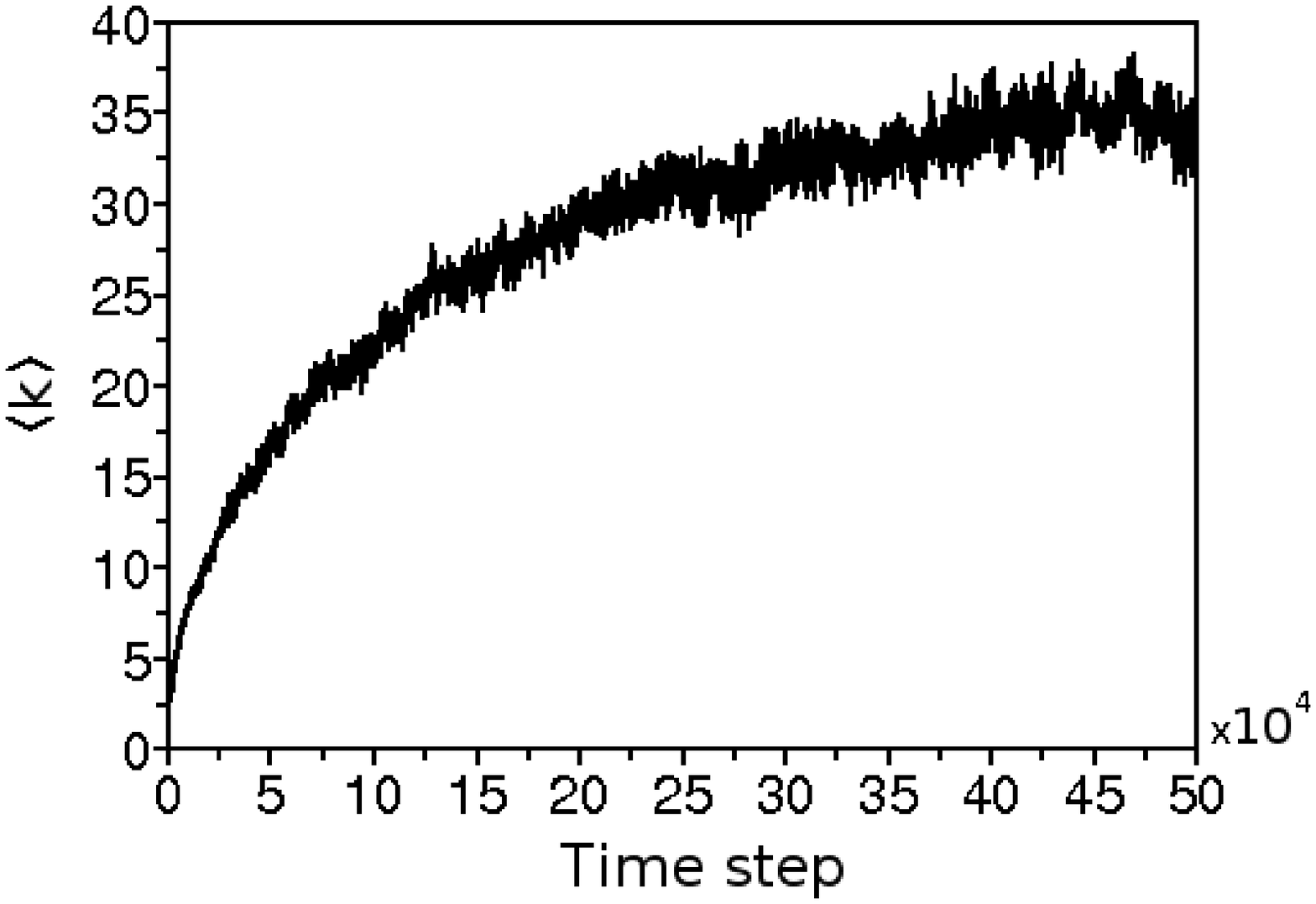}
    \includegraphics[scale=0.27]{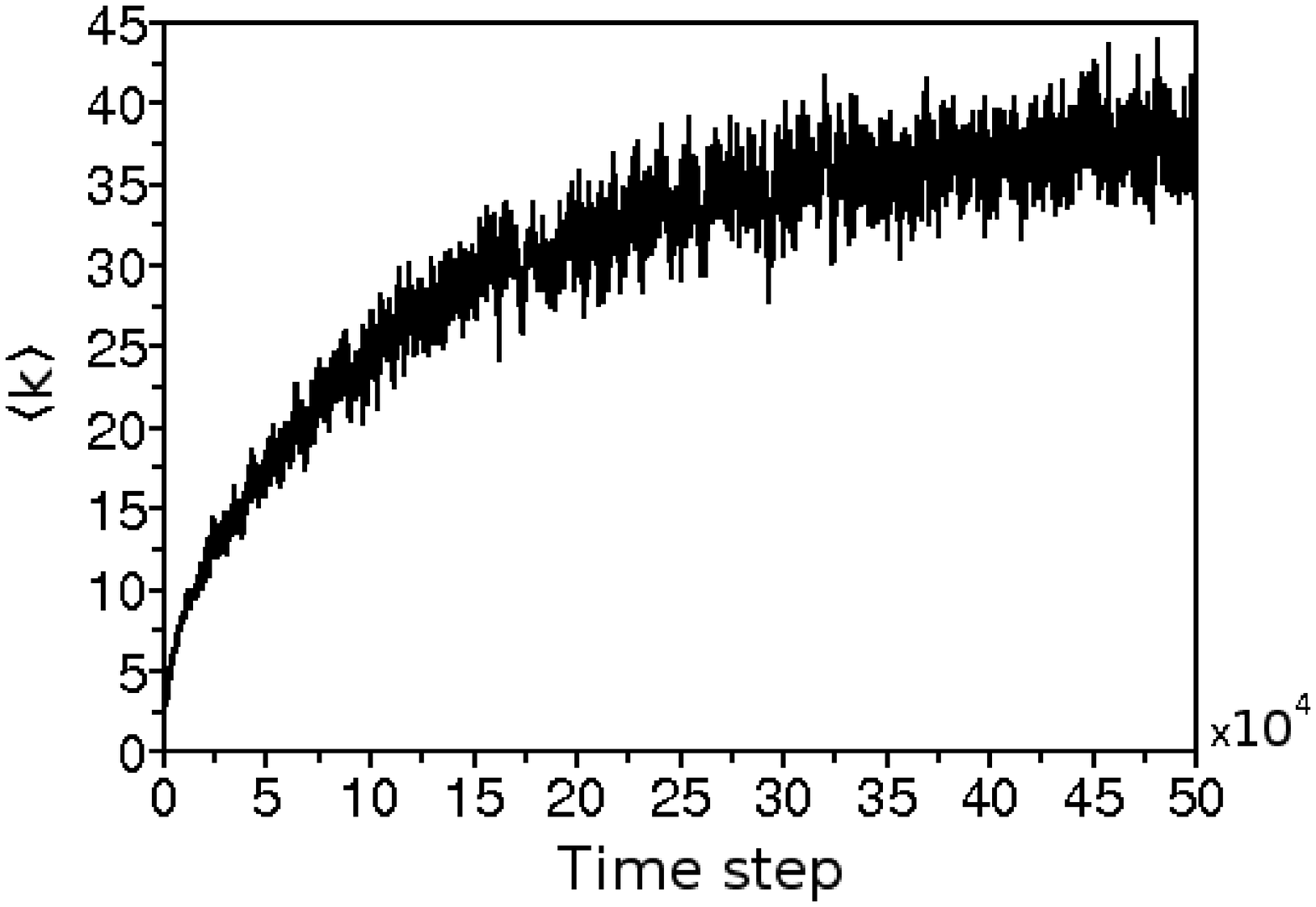}   \\
    (ii) \\
    \includegraphics[scale=0.27]{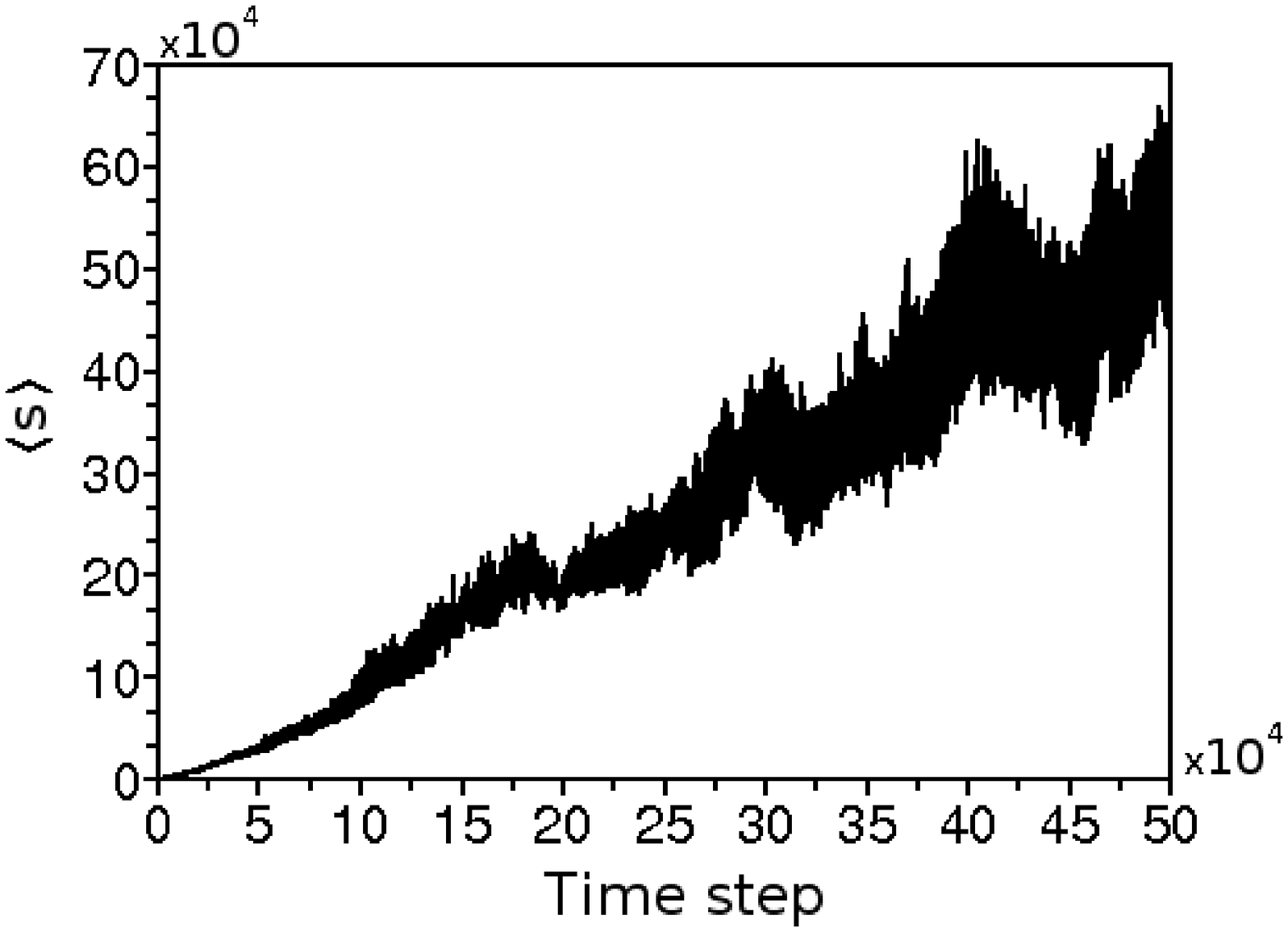}
    \includegraphics[scale=0.27]{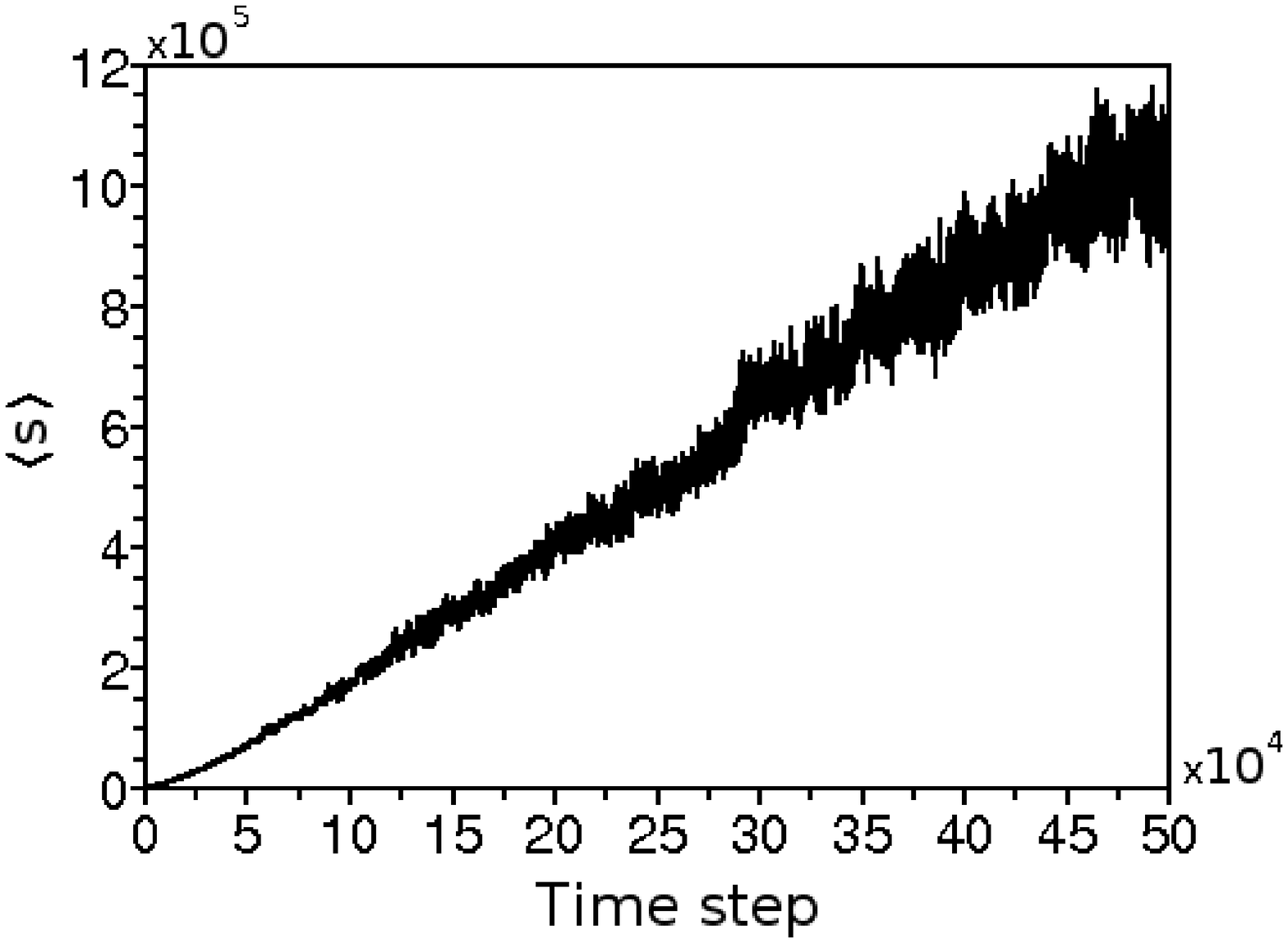}
    \includegraphics[scale=0.27]{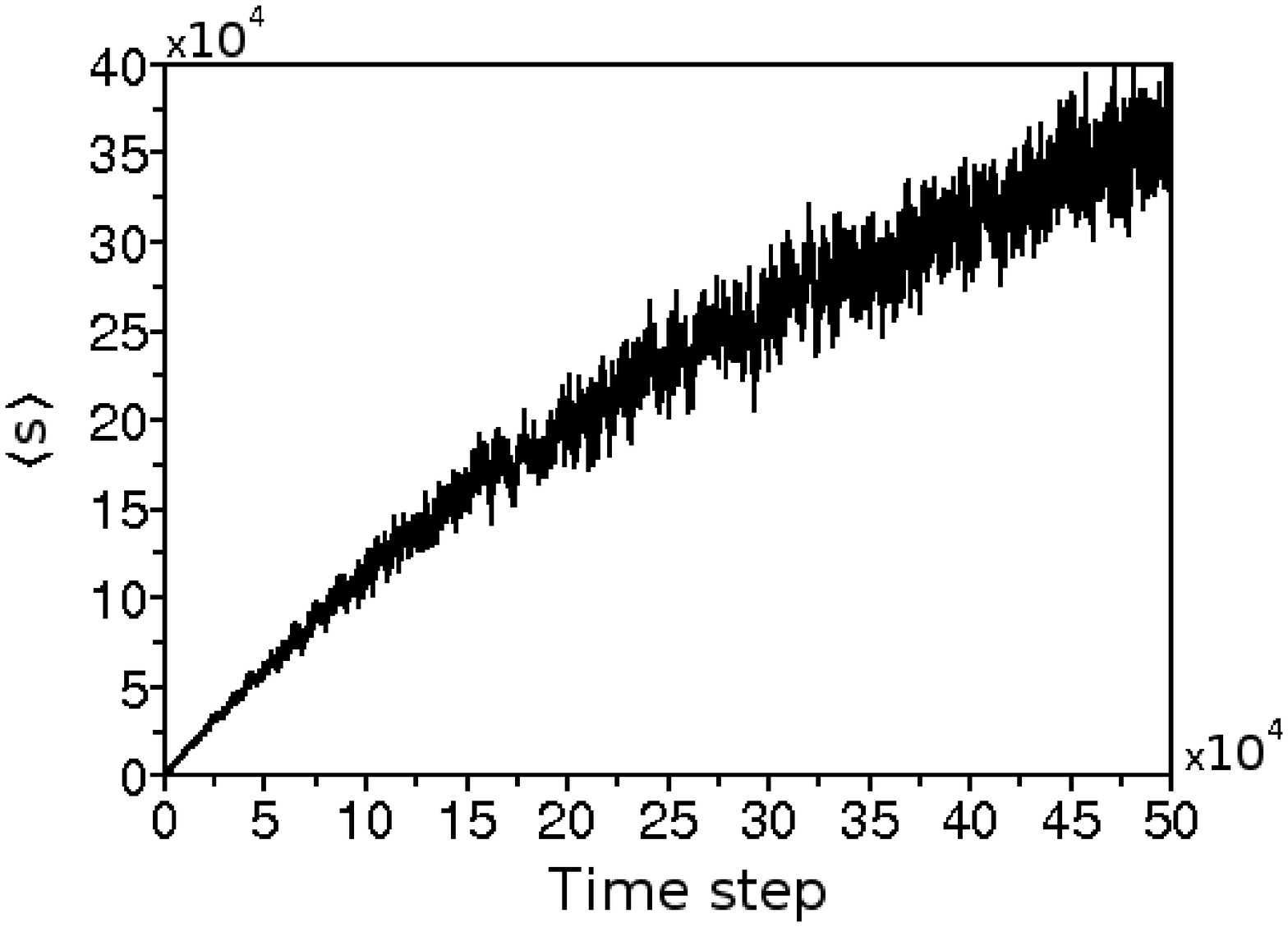}  \\
    (iii) \\
    \includegraphics[scale=0.27]{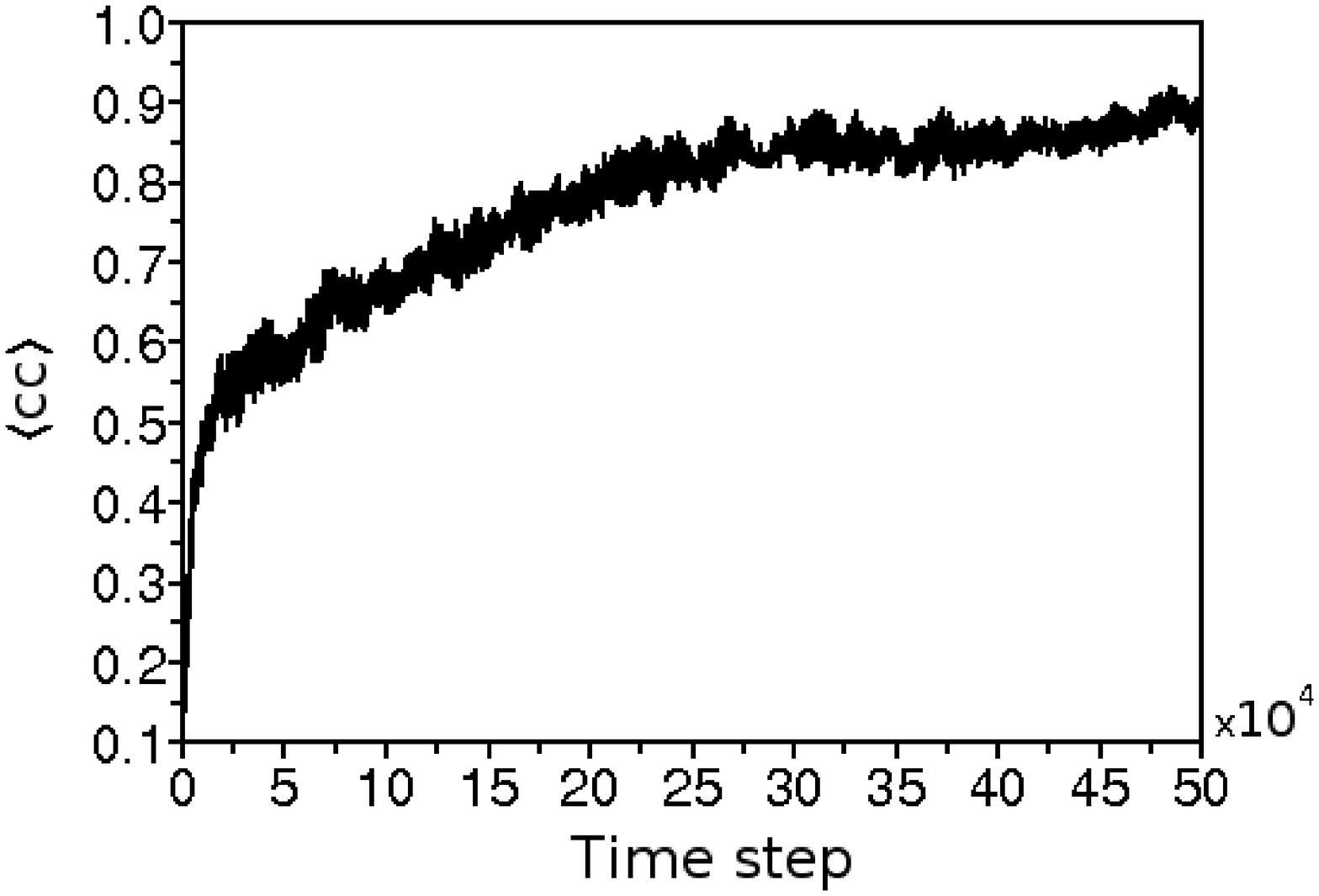}
    \includegraphics[scale=0.27]{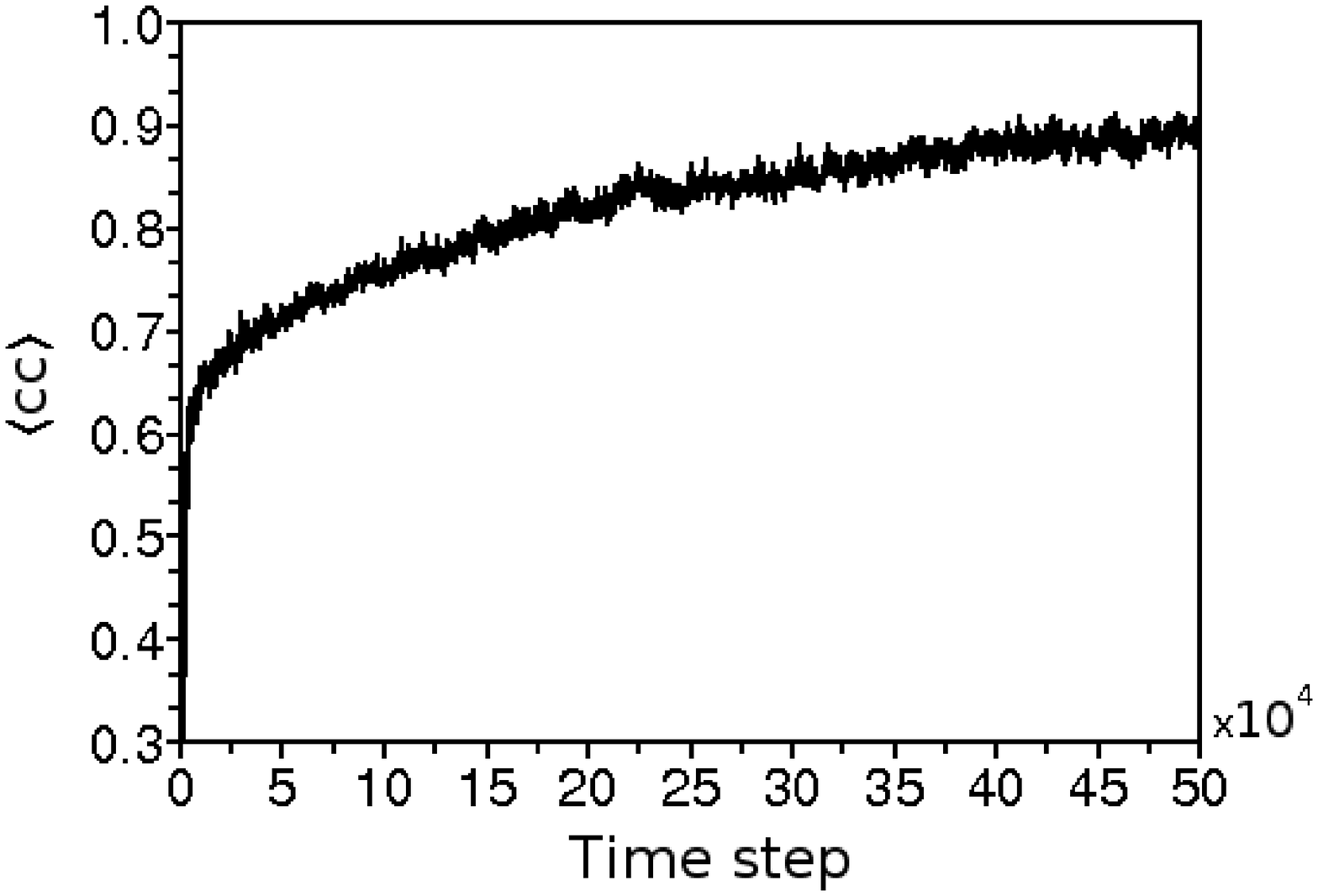}
    \includegraphics[scale=0.27]{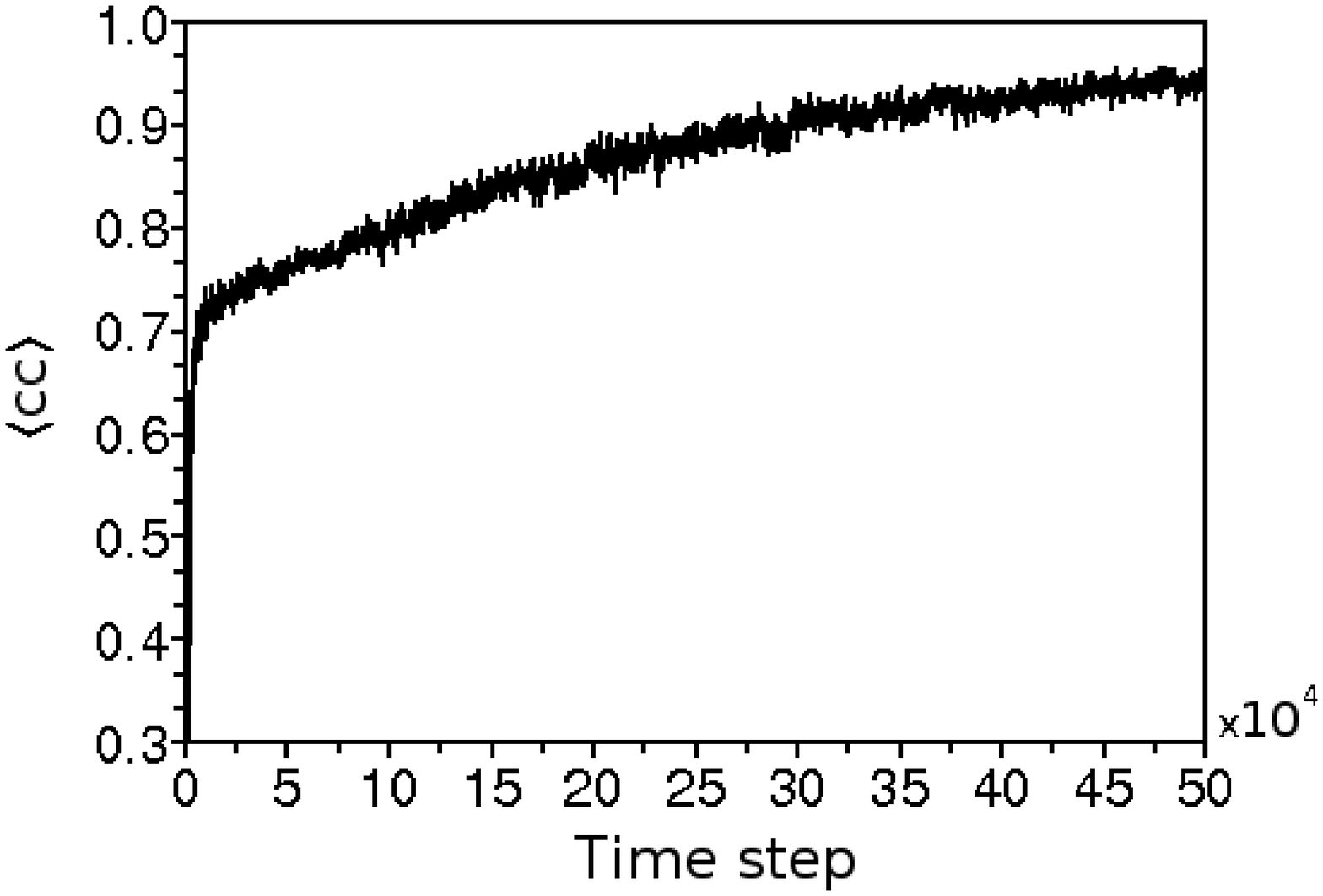}  \\
    (a) $\beta_{predator}=0.5$ \hspace{2cm} (b) $\beta_{predator}=0.125$ \hspace{2cm} (c) $\beta_{predator}=0.0625$ \\
    \caption{Evolution of the average structural properties in
    $\Gamma_{predator}$ when three intensities of attraction between
    predators and null attraction between preys are considered. (i)
    Average degree $\langle k\rangle$ ,(ii) Average strength $\langle
    s\rangle$ and (iii) Average clustering coefficient $\langle
    cc\rangle$. The standard deviation is shown to $50\%$ of the
    original values in cases (i) and (ii). } \label{fig:07}
  \end{center}
\end{figure*}

Considering the predator network $\Gamma_{predator}$. In the absence
of attraction between preys \mbox{($\beta_{prey}=0$)}, the average
degree $\langle k\rangle$ increases quickly within the $150000$ time
steps, independently of the attraction intensity between predators
(Fig.~\ref{fig:07}-i). After this stage, the average degree nearly
stabilizes about $\langle k\rangle = 8.5$ with $\beta_{predator} =
0.5$ (Fig.~\ref{fig:07}-i,a).  However, it exhibits a slower increase
in the other configurations (Figs.~\ref{fig:07}-i,b and
\ref{fig:07}-i,c), suggesting that stabilization will also be eventually
reached. Actually, because of the movements of predators between the
spatial clusters, this measurement should stabilize after the network
becomes completely connected. This will happen at $\langle k\rangle
\sim 40$ in cases of $\beta_{predator} = 0.125$
(Fig.~\ref{fig:07}-i,b) and of $\beta_{predator} = 0.0625$
(Fig.~\ref{fig:07}-i,c), which corresponds nearly to the average
number of spatial groups or number of nodes in the respective network
$\Gamma_{predator}$ (the average value is obtained after an initial
transient). Because the number of groups is smaller when
$\beta_{predator}$ is larger, the threshold becomes smaller in that
case (Fig.~\ref{fig:07}-i,a) and it possibly decrease whether the
system converges to a giant component of predators.

In the proposed growth method, the node strength relates to the
history of proximity between a node and the members of its group, and
inversely, to the rate of members exchanges between two spatial
groups. The average strength $\langle s\rangle$ evolution seems to 
take place in stages when the attraction intensity between predators
is larger ($\beta_{predator}=0.5$ --
Fig.~\ref{fig:07}-ii,a). Possibly, the stages in which the growth rate
is slower correspond to the intervals where the exchange rate between
groups is small (Fig.~\ref{fig:07}-ii,a). Since the predators spend
much time together, the weight $w_{\mu_i,\mu_j}$ of connection
$(\mu_i,\mu_j)$ in the network $\Gamma_{his}$ increases considerably
and when two or more groups merge, the average strength suddenly
rises. A lower attraction intensity implies on more circulation of
predators in space and, consequently, constant association and
disassociation to several groups. The level of proximity between
predators tends to be smaller in the configuration with
$\beta_{predator}=0.0625$ (Fig.~\ref{fig:07}-ii,c) when compared to
the other two cases (Figs.~\ref{fig:07}-ii,a and~\ref{fig:07}-ii,b),
where the predators movement is more constrained. As a consequence, in
the last two cases, the predators spend more time together with the
same partners.  The larger strength in case of $\beta_{predator}=0.125$
is a consequence of the larger number of spatial groups (nodes of
$\Gamma_{res}$).

A smaller attraction intensity between predators implies in a faster
increase in the average clustering coefficient $\langle cc\rangle$
within the first time steps (Fig.~\ref{fig:07}-iii). In the second
growing stage (from nearly $25000$ to $250000$ time steps), the growth
rate is higher for $\beta_{predator}=0.5$
(Fig.~\ref{fig:07}-iii,a).  The latter effect is possibly a result of
the decrease in the number of spatial groups in the first
configuration (Fig.~\ref{fig:04}-i). Since the clustering coefficient
measures the local connectivity (between common neighbours of a
reference node), and the growth of $\langle cc\rangle$
\mbox{(Fig.~\ref{fig:07}-iii)} is higher than the growth of the
corresponding $\langle k\rangle$ (Fig.~\ref{fig:07}-i), the
connections occur mainly between nearby spatial groups. In other words,
the movements required for following preys is considerable slow.  A
predator moves within a small Euclidean distance and then associates
to one spatial group; after a while, the same or other predator moves
again to another close spatial group in a non-stationary way.

Considering attraction between preys ($\beta_{prey}>0$) and null
attraction between predators ($\beta_{predator}=0$), the structural
properties of the evolution is similar to the situation with
$\beta_{predator}=0.0625$ and $\beta_{prey}=0$, except for the
configuration with $\beta_{prey}=0.5$. In that case, the stronger
attraction of predators by dense clusters of preys implies on more
movements of predators and, therefore, more connectivity between the
nodes in comparison to the other two configurations. The absence of
attraction between predators and the higher concentration of preys due
to mutual attraction produces an overall behaviour where the predators
have freedom to move throughout sub-space $O$. As a consequence, the average
measurements values are slightly higher in this case in comparison to
the situation with small attraction between predators
(similarly to configuration of Figure~\ref{fig:07}-c), although presenting the same stages of
evolution.

\begin{figure*}
  \begin{center}
    \includegraphics[scale=0.27]{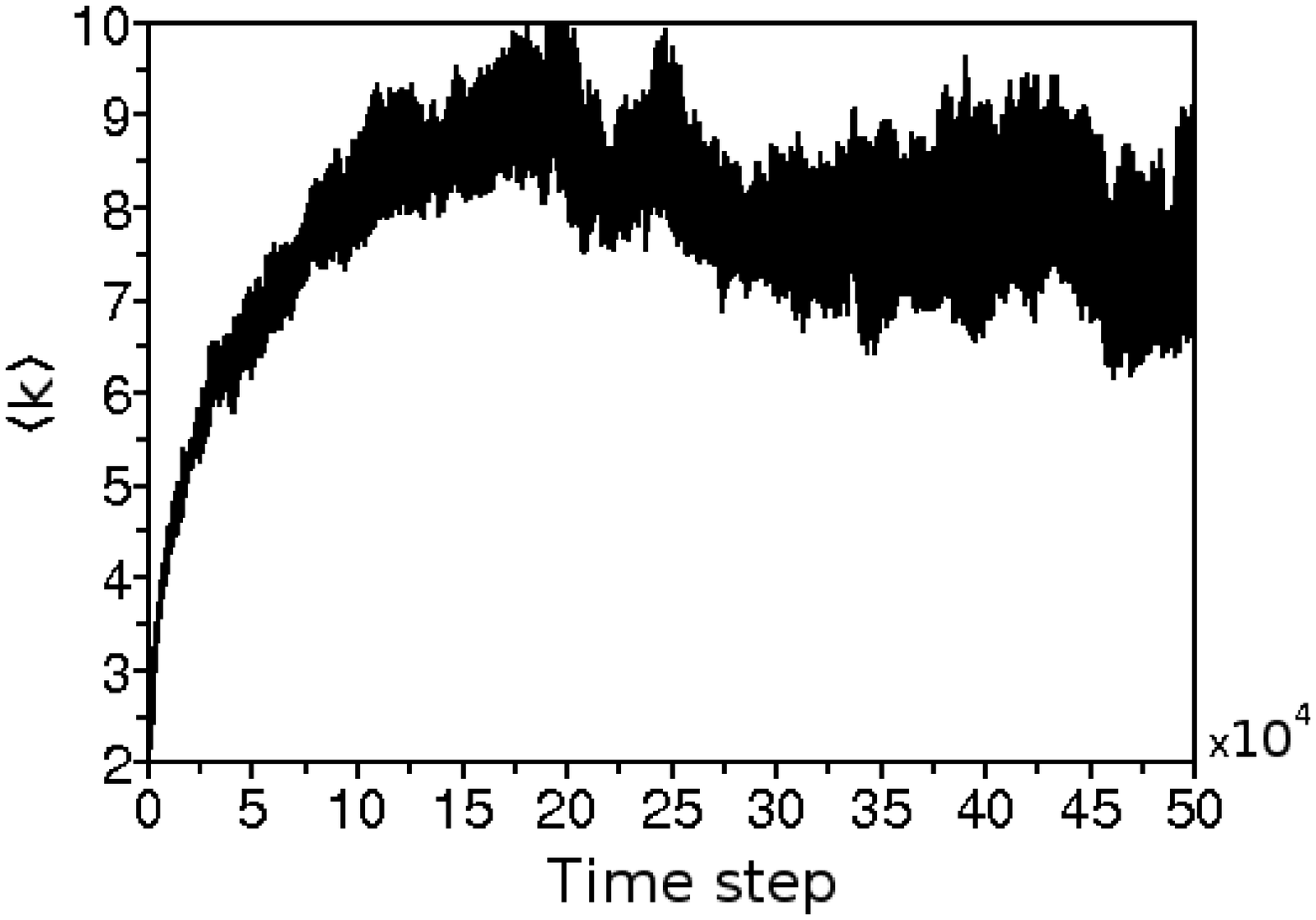}
    \includegraphics[scale=0.27]{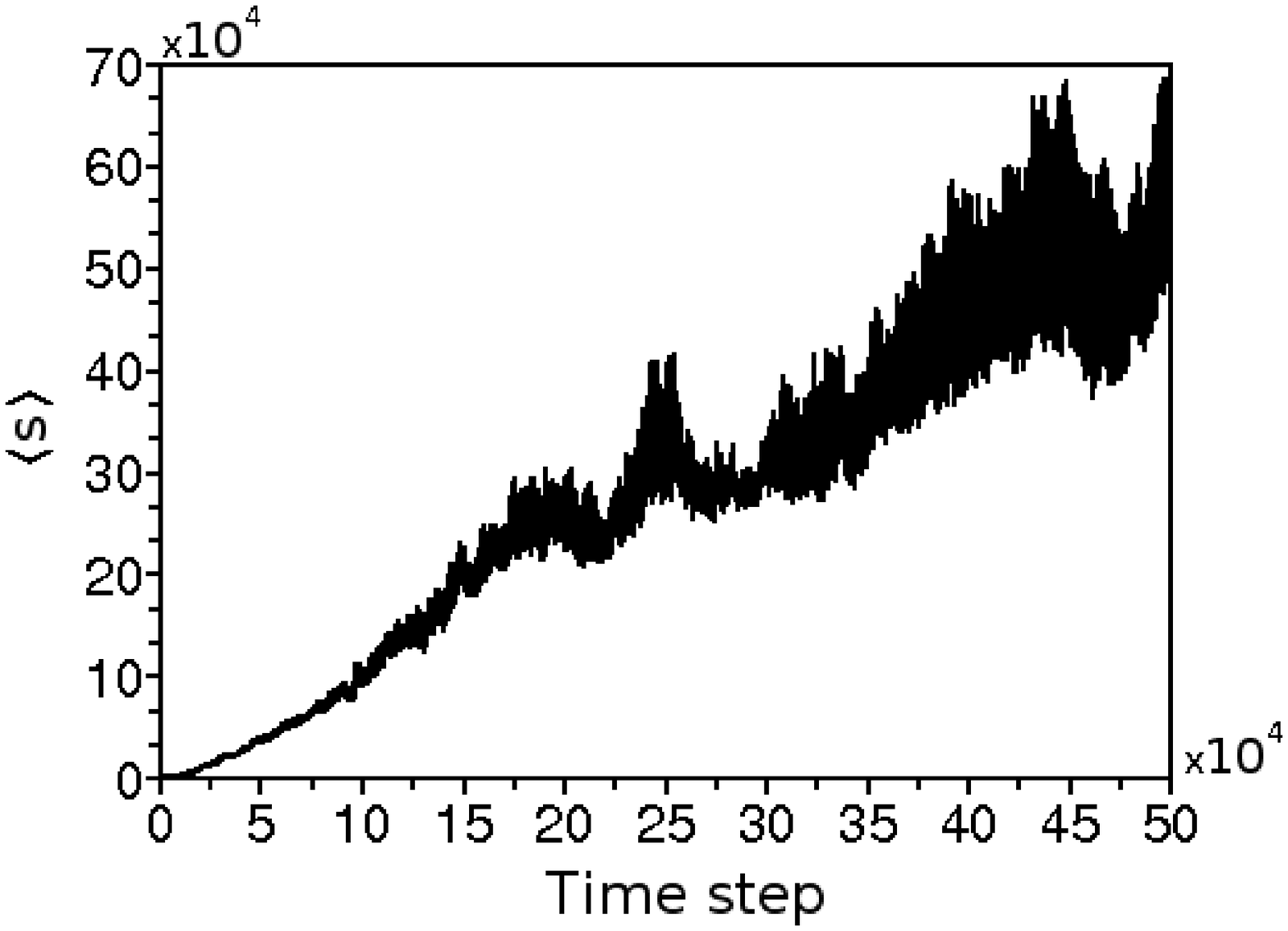}
    \includegraphics[scale=0.27]{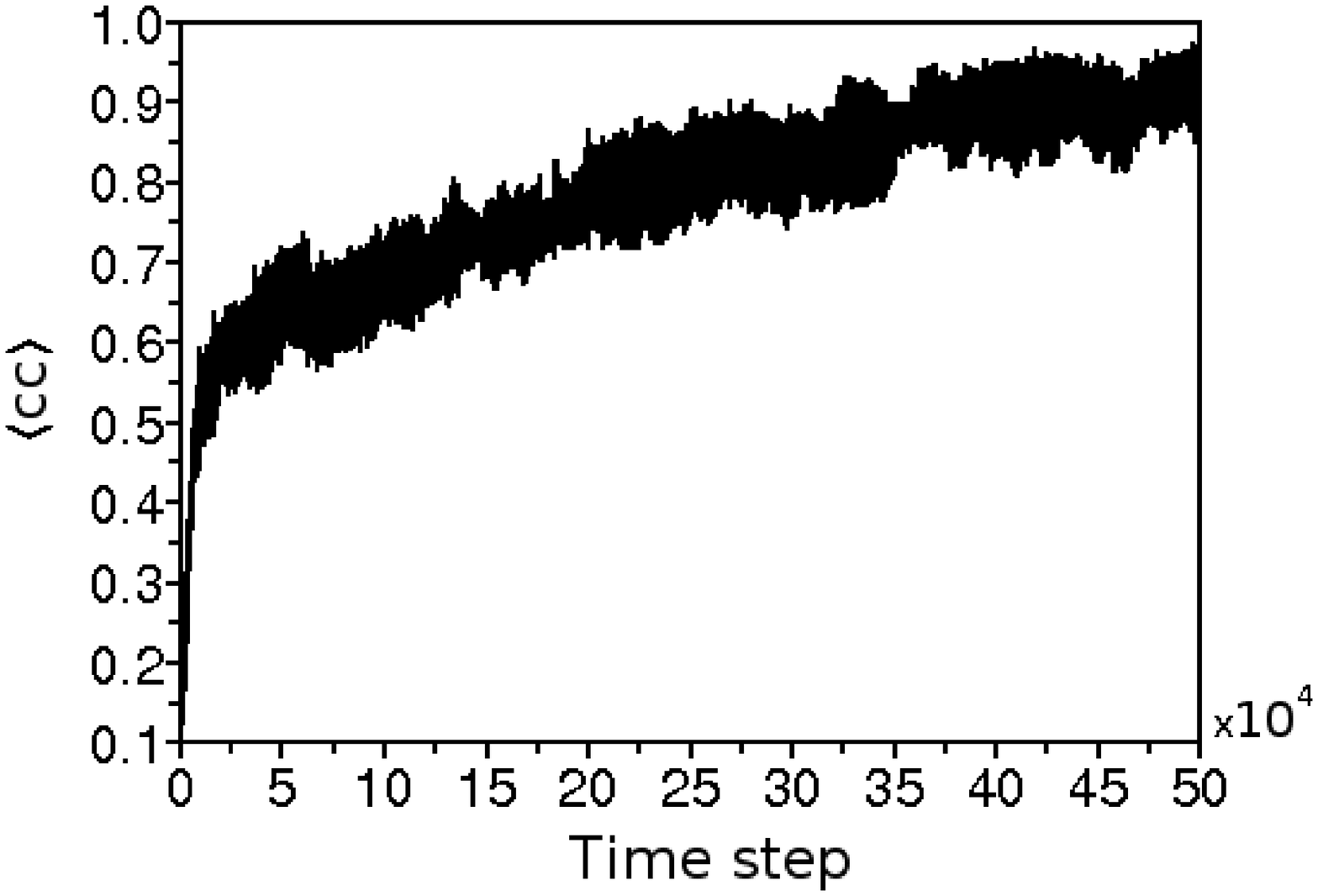}
    (a) \hspace{5cm} (b) \hspace{5cm} (c) \\
    \caption{Evolution of $\Gamma_{predator}$ average structural
    properties when the attraction between same specie individuals is
    given by $\beta=0.5$. The standard deviation is shown to $50\%$
    of the original values in (a) and (b). } \label{fig:08}
  \end{center}
\end{figure*}

In the last experiment, involving attraction between predators
($\beta_{predator}>0$) and between preys ($\beta_{prey}>0$), the
evolution of the average degree shows again an apparent stabilization
in all cases. The number of nodes in the resulting network
$\Gamma_{predator}$ (or number of connected components $\kappa$) is
about $N\sim9$ in the last time step observed (after a large
transient). Since $N$ is higher than the plateau in the graph of
Figure~\ref{fig:08}-a, the resulting network is not completely
connected yet. We verified that at the maximum value (between $150000$
and $200000$ time steps), the number of spatial groups is
approximately $1.5$ of the average degree value.
Although the node average connectivity has decreased after this
interval, the value increased if compared with the number of nodes in
the network. This effect is related to the merging of clusters, which
happens more frequently in the configuration where the attraction is
higher (Fig.~\ref{fig:05}-i). This effect is not observed in the other
two configurations, where the average number of groups is nearly
constant (about $45$) over time and the curves are similar to the case
in Figure~\ref{fig:07}-i. Comparing the number of nodes and the
stabilization threshold, we observed that the resulting network is
almost completely connected in the last time steps.  The average
connectivity among the nodes grows faster in the configuration with
the smallest value of $\beta$ as a consequence of higher mobility. The
average strength presented evolution similar to the first experiment
(Fig.~\ref{fig:07}-ii). The main difference regards the stronger
attraction between preys which generated clusters of preys. The
non-existence of uniformly distributed preys implied in less movements
between predator clusters due to absence of stimuli. Consequently, the
strength values are higher (Fig.~\ref{fig:08}-b) when compared with
the first experiment.

The decrease of the mobility between spatial groups is also observed
by comparing the evolution of the clustering coefficient in
Figure~\ref{fig:08}-c with Figure~\ref{fig:07}-iii,a. Once again, the
clustering coefficient growth rate is very high within the first time
steps (up to $10000$), maintaining a nearly constant growth
henceforth. Actually, the evolution seems to stabilize after
$t=450000$ in all configurations. Differently from the first
experiment, the system may not evolve to a state where all nodes
become connected. The effect is more probable in the configuration
with $\beta=0.5$ provided the clusters do not merge in a single giant
component.  The movement of nodes between two clusters become less
probable after a certain time step.  This is a consequence of the fact
that each cluster now has higher density and is considerably far from
other clusters. The absence of stimuli because of prey clusters
emergence and the mutual attraction inside the predator clusters also
contributes to such a decrease of movements.

\subsection{Strength Distribution}

The average measurements provide global information about the complex
network.  However, the network internal structure can be completely
different even when presenting similar average
values~\cite{Costa:review}. The internal structure is responsible to
constrain dynamical processes (\textit{e.g.}, cascade
failures~\cite{Motter:cascade,Barabasi:review} or
epidemics~\cite{Satorras:epidemic,Newman:review,Boccaletti:review}) in
the network.  Consequently, it is fundamental to characterize the
resulting network in terms of measurements related to their internal
topological properties.

\begin{figure*}
  \begin{center}
    (i) \\
    \includegraphics[scale=0.26]{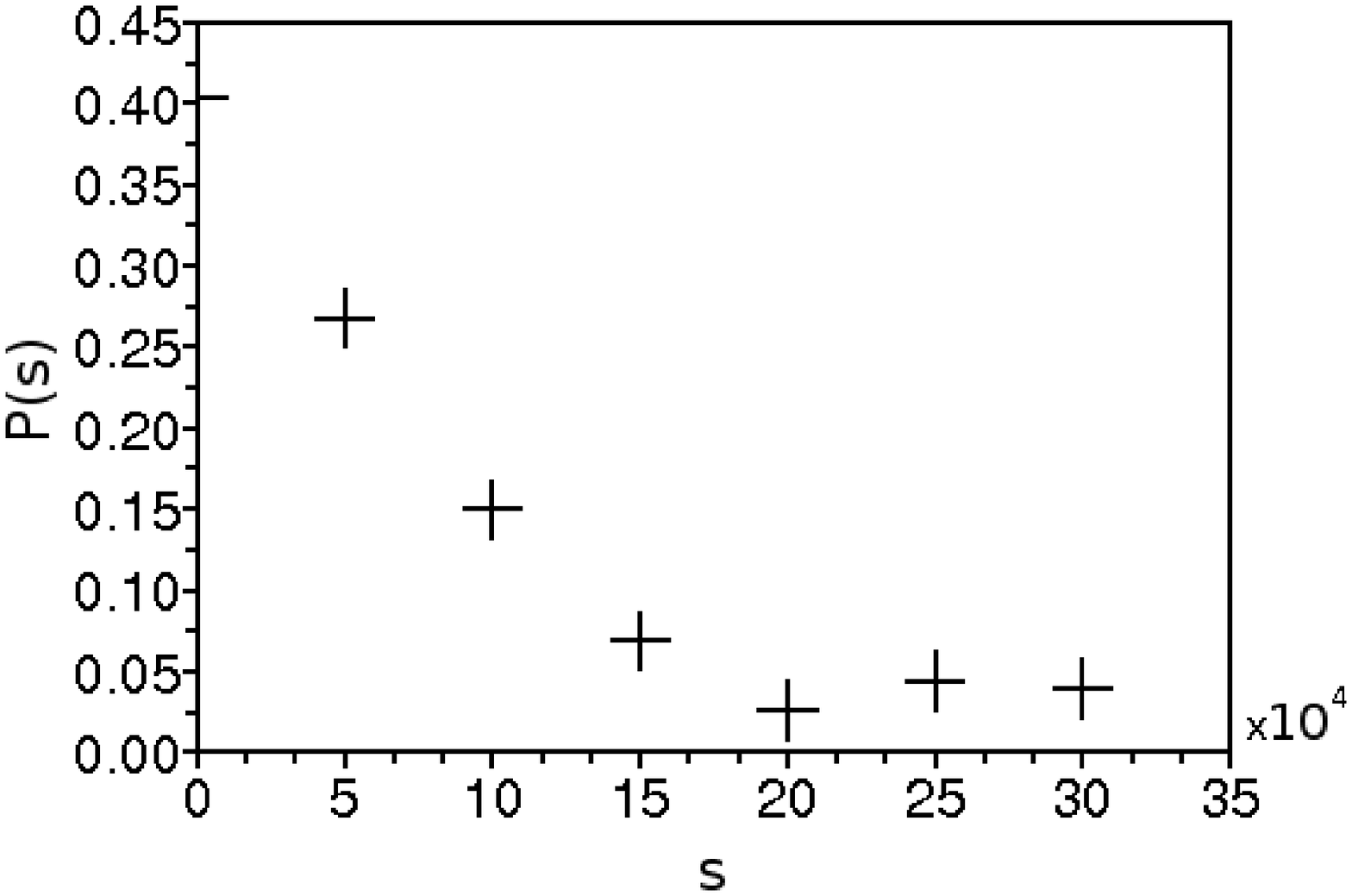}
    \includegraphics[scale=0.26]{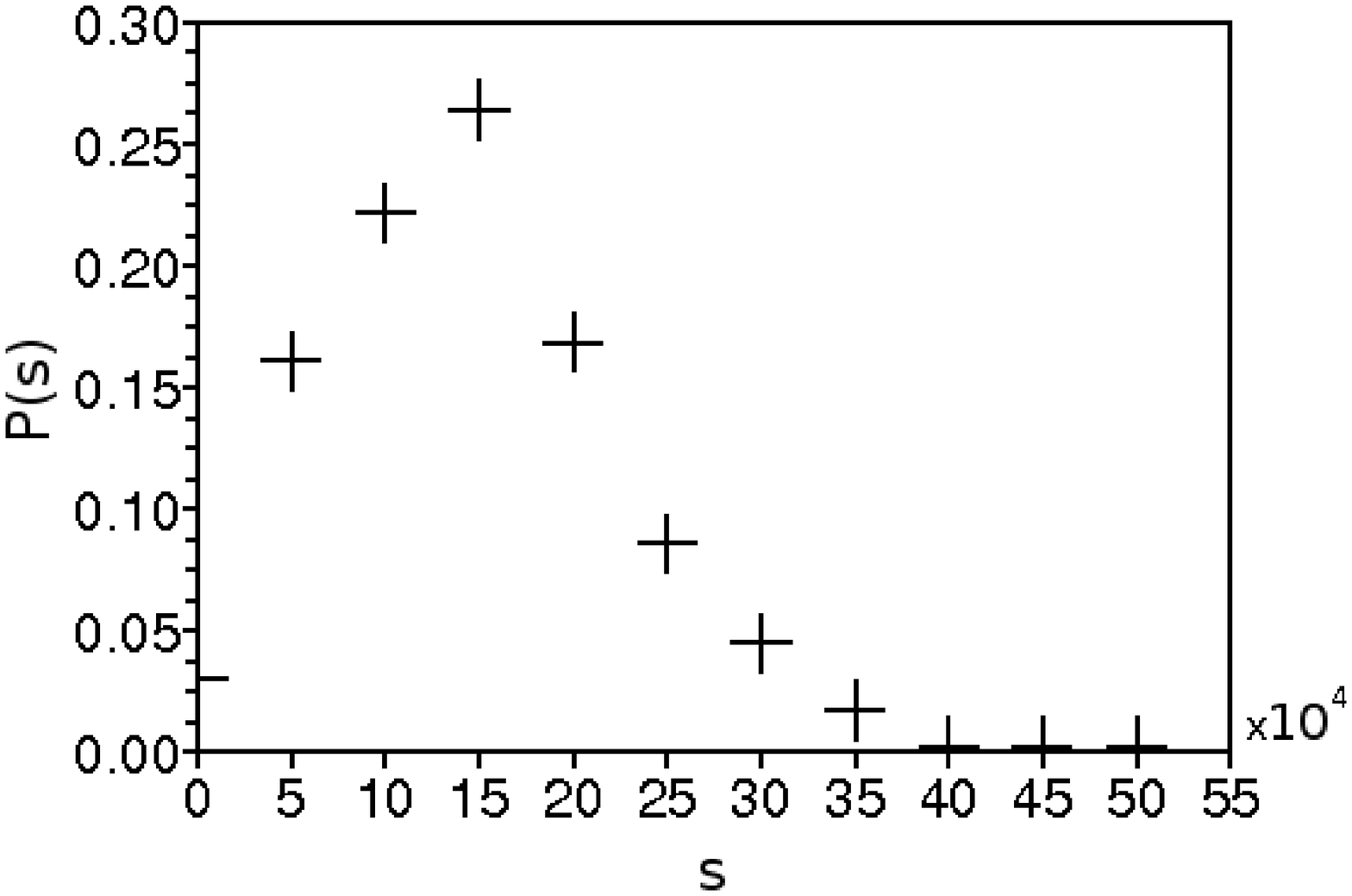}
    \includegraphics[scale=0.26]{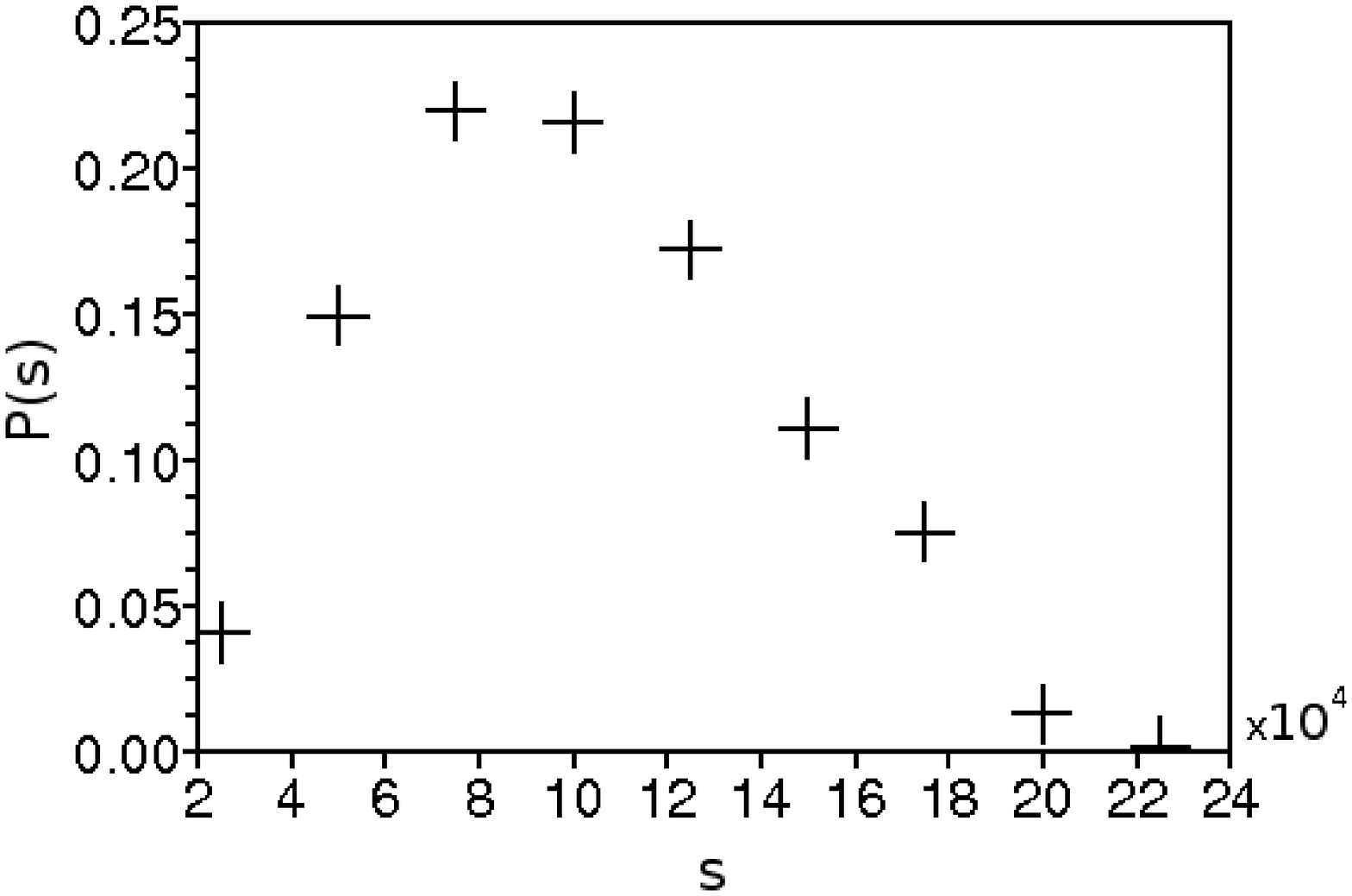}  \\
    (ii) \\
    \includegraphics[scale=0.26]{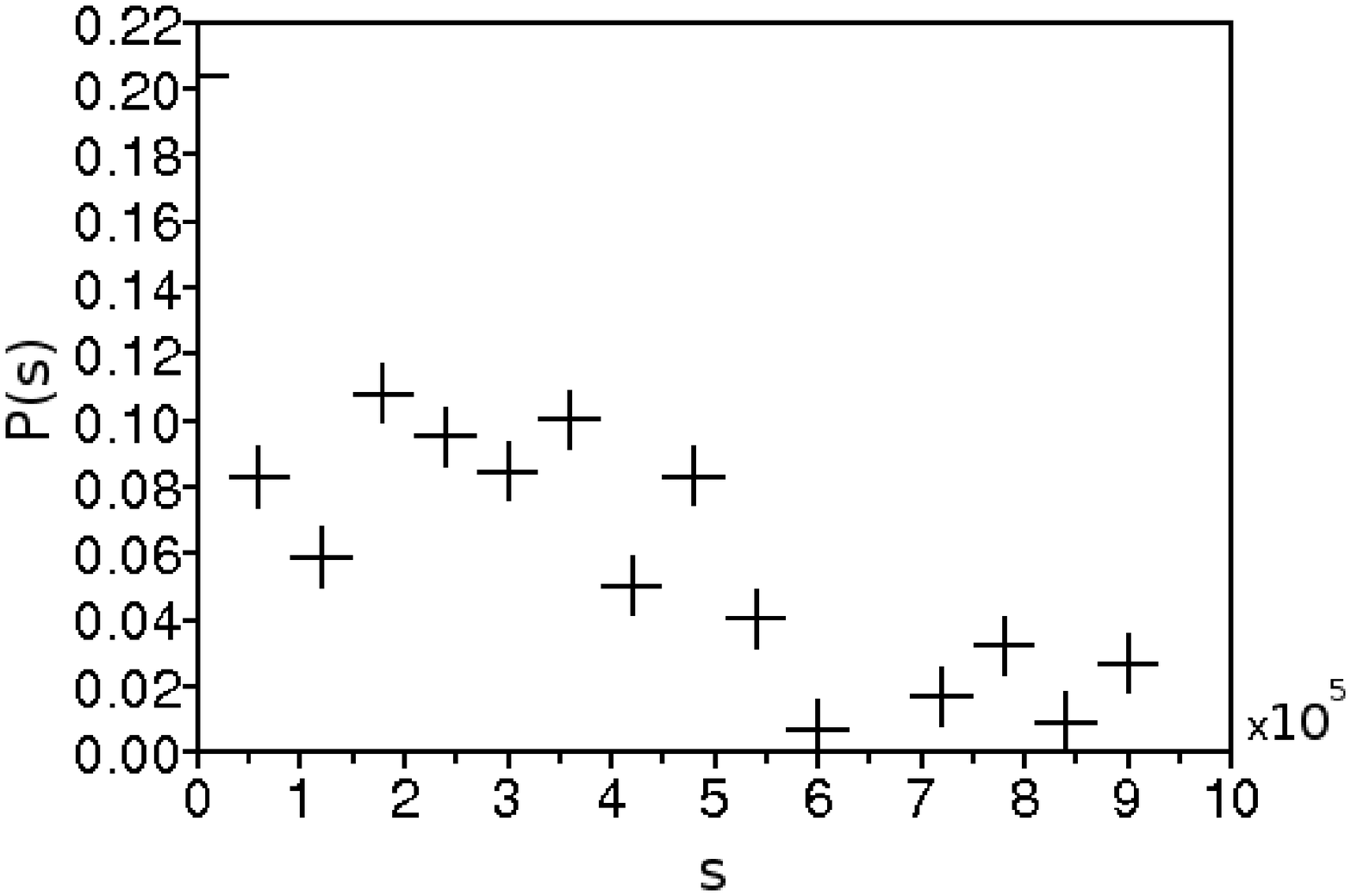}
    \includegraphics[scale=0.26]{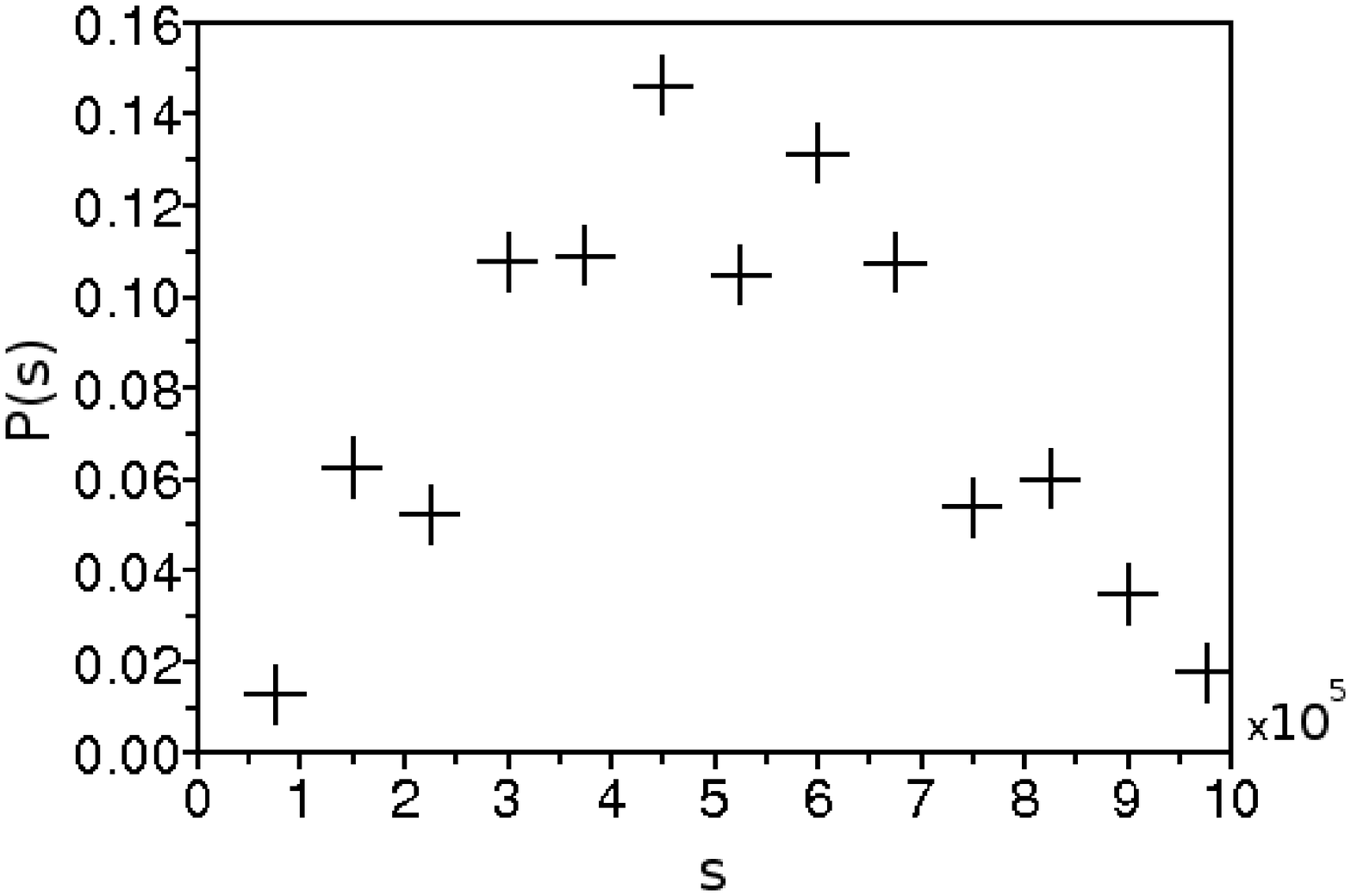}
    \includegraphics[scale=0.26]{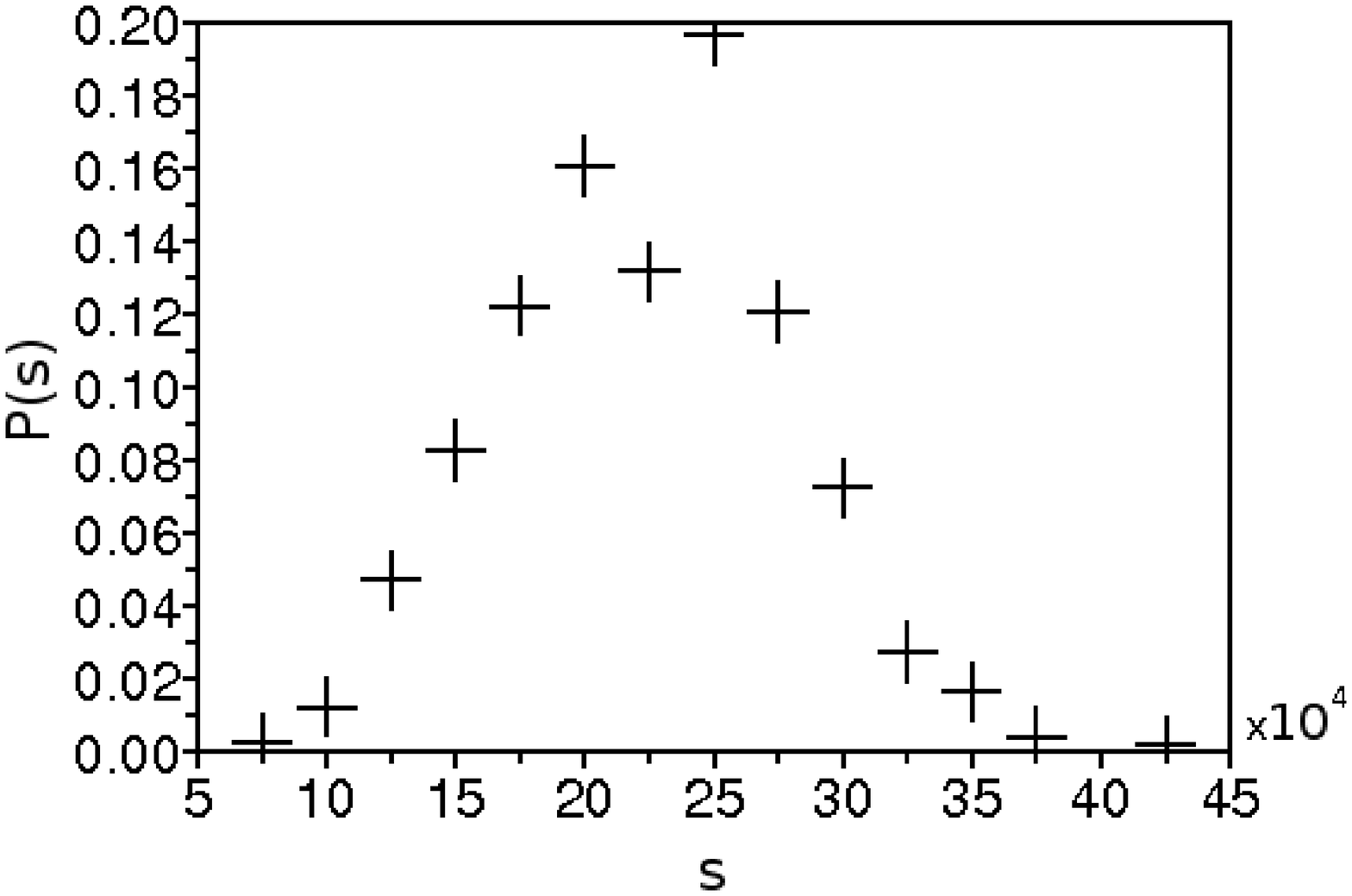}  \\
    (iii) \\
    \includegraphics[scale=0.26]{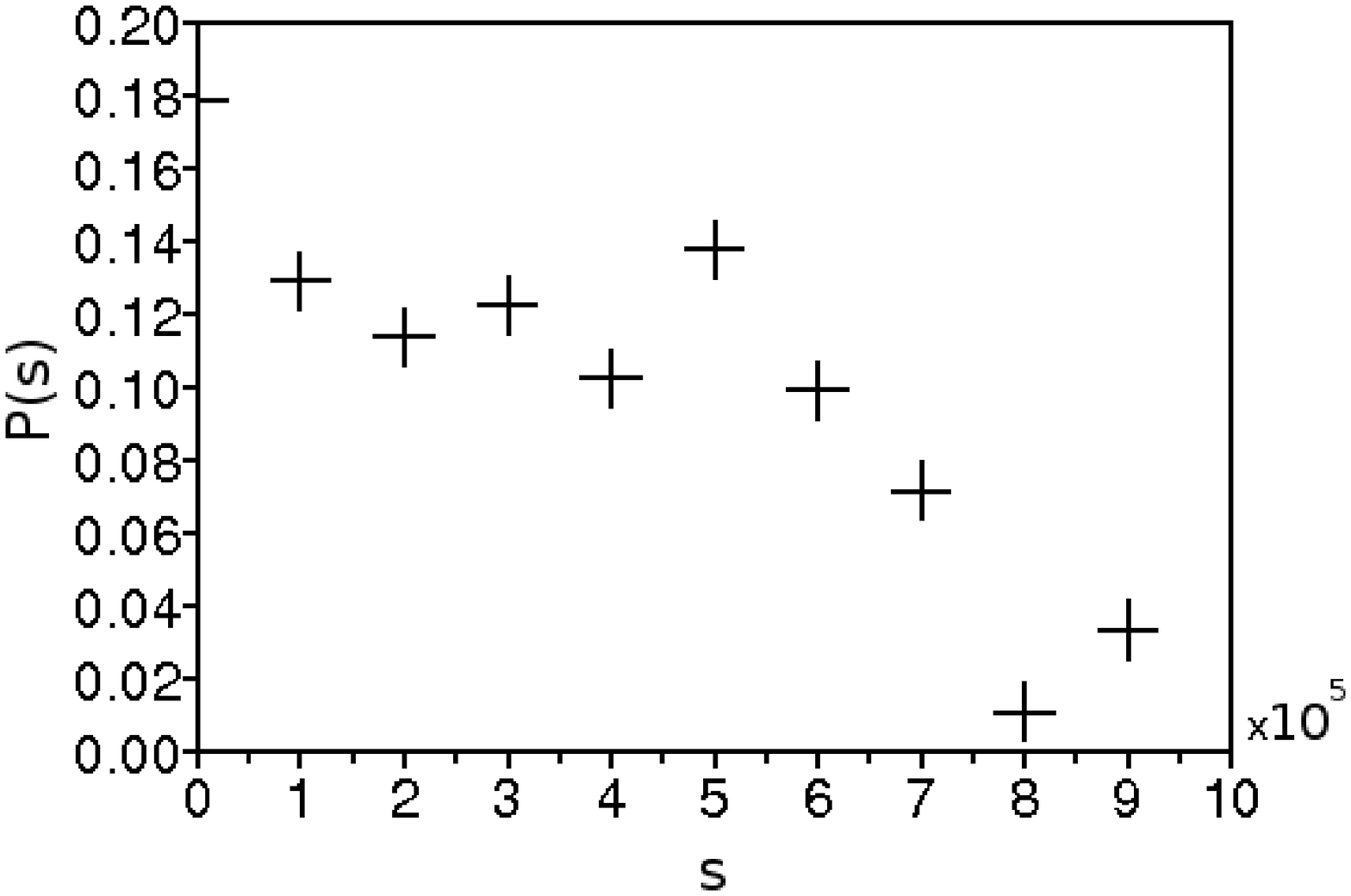}
    \includegraphics[scale=0.26]{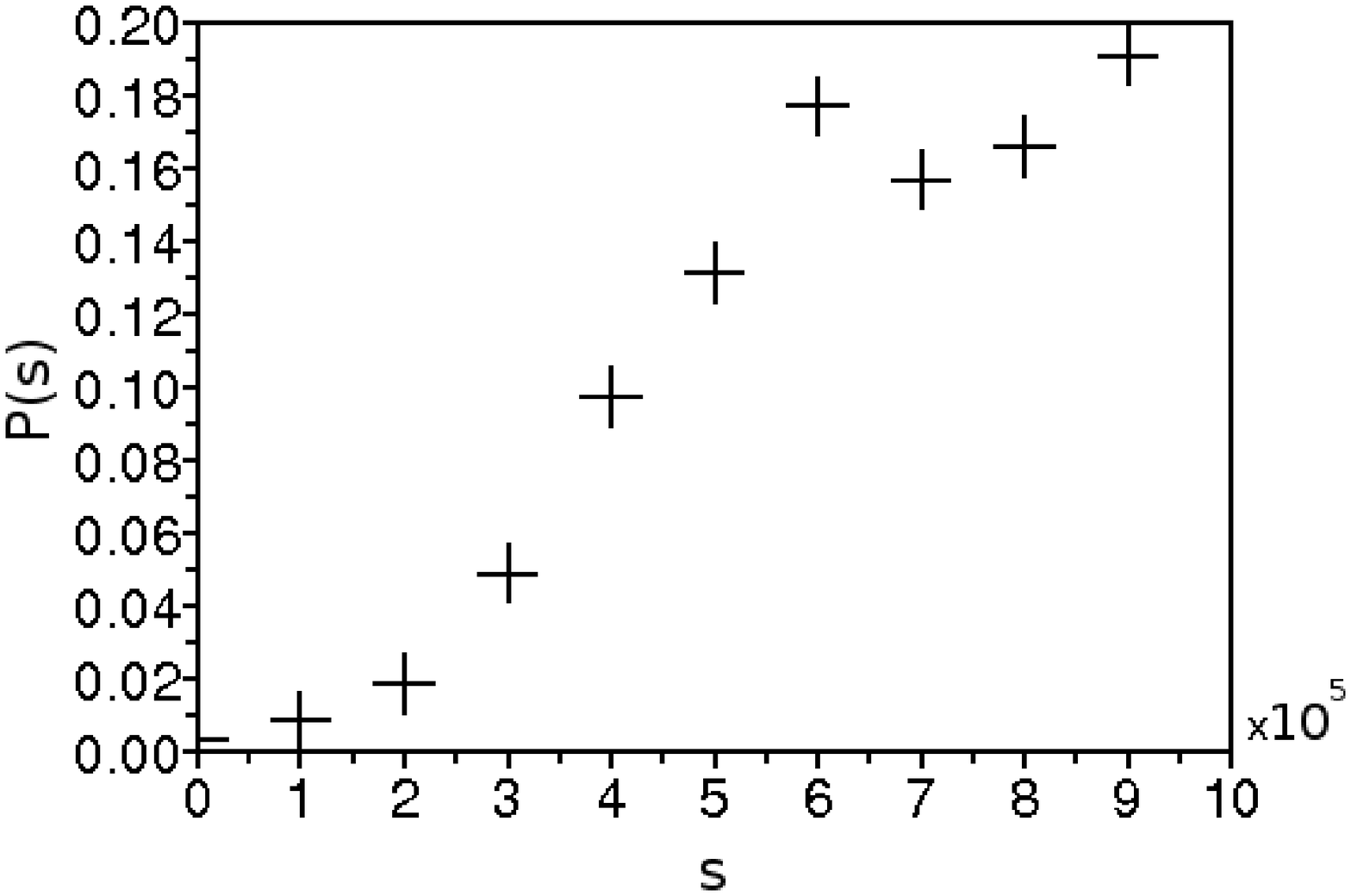}
    \includegraphics[scale=0.26]{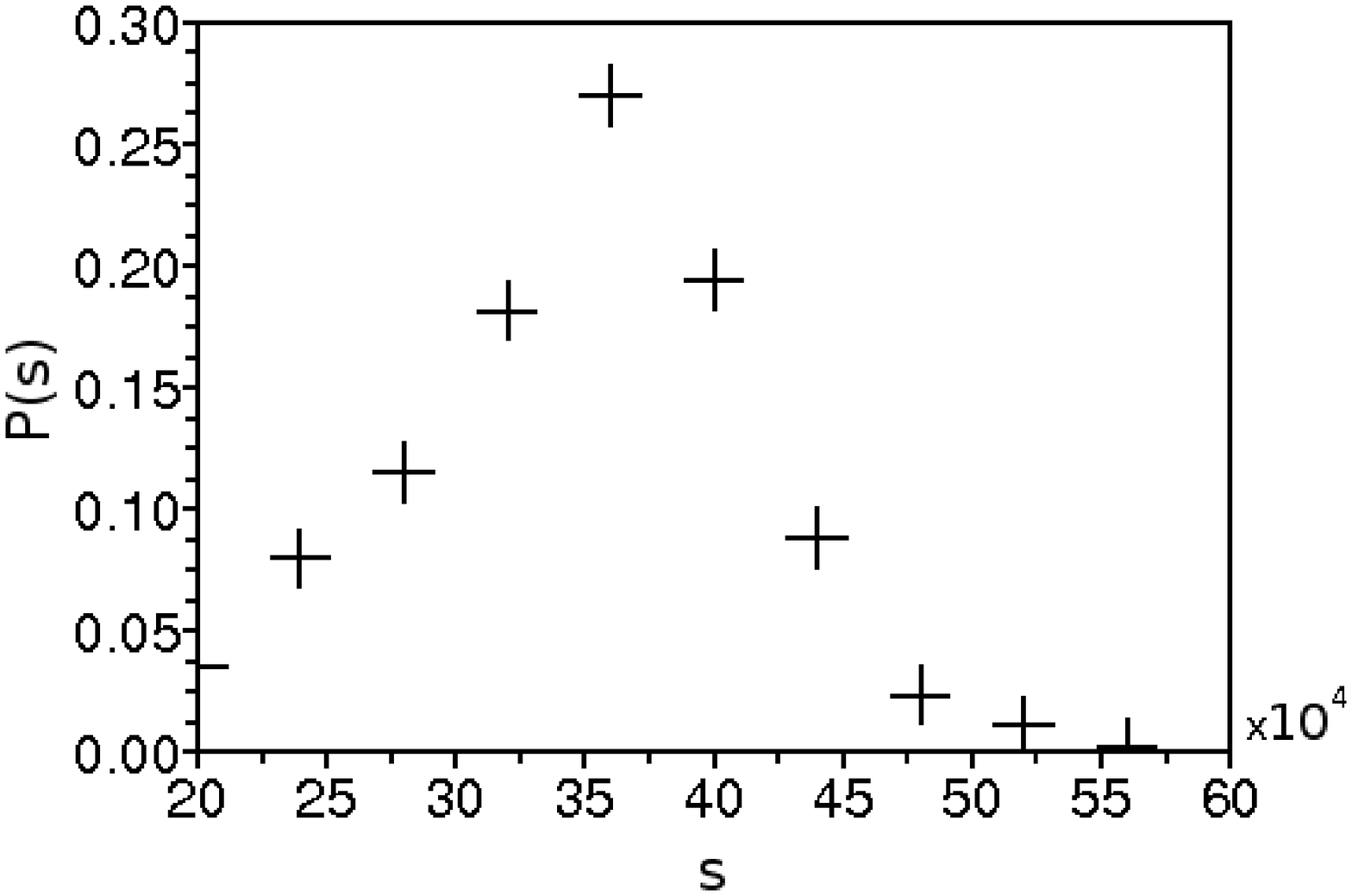}  \\
    (a) $\beta_{predator}=0.5$ \hspace{2cm} (b) $\beta_{predator}=0.125$ \hspace{2cm} (c) $\beta_{predator}=0.0625$ \\
    \caption{Strength distribution for predator networks at three time
    steps: (i) $t=100000$, (ii) $t=300000$ and (iii) $t=500000$. We
    consider three attraction intensities between same-specie
    individuals. } \label{fig:09}
  \end{center}
\end{figure*}

The strength distributions (histograms) have different shapes in the
three resulting networks (Fig.~\ref{fig:09}) when we consider
attraction between predators ($\beta_{predator}>0$) and null
attraction between preys ($\beta_{prey}=0$). Although the
configuration with $\beta_{predator}=0.5$ (Fig.~\ref{fig:09}-a)
presents undefined shape, we see that the maximum strength increases
considerable between $100000$ and $300000$ time steps. The number of
nodes with large strength values also increases at the next time steps
considered (Fig.~\ref{fig:09}-iii,a), indicating that the mobility
between groups decreases with time, which can be seen by the existence
of nearly stationary (in shape and number of members), dense spatial
clusters in Figure~\ref{fig:04}. Contrariwise, the other two
configurations (Figs.~\ref{fig:09}-b and~\ref{fig:09}-c) present a
Gaussian-like strength distribution with a characteristic scale given
by the average value (the average can be verified by comparison with
graphs of Figure~\ref{fig:07}-ii). Although presenting a
characteristic membership permanence time, some individuals constantly
change between their respective groups (decreasing the strength),
while others associate in a group for a long time (increasing the
strength) and barely move between different groups. The predators in
the last case can be seen as the core of the group, since they spent
much time together and are responsible for maintaining the group
united.

\begin{figure*}
  \begin{center}
    \includegraphics[scale=0.26]{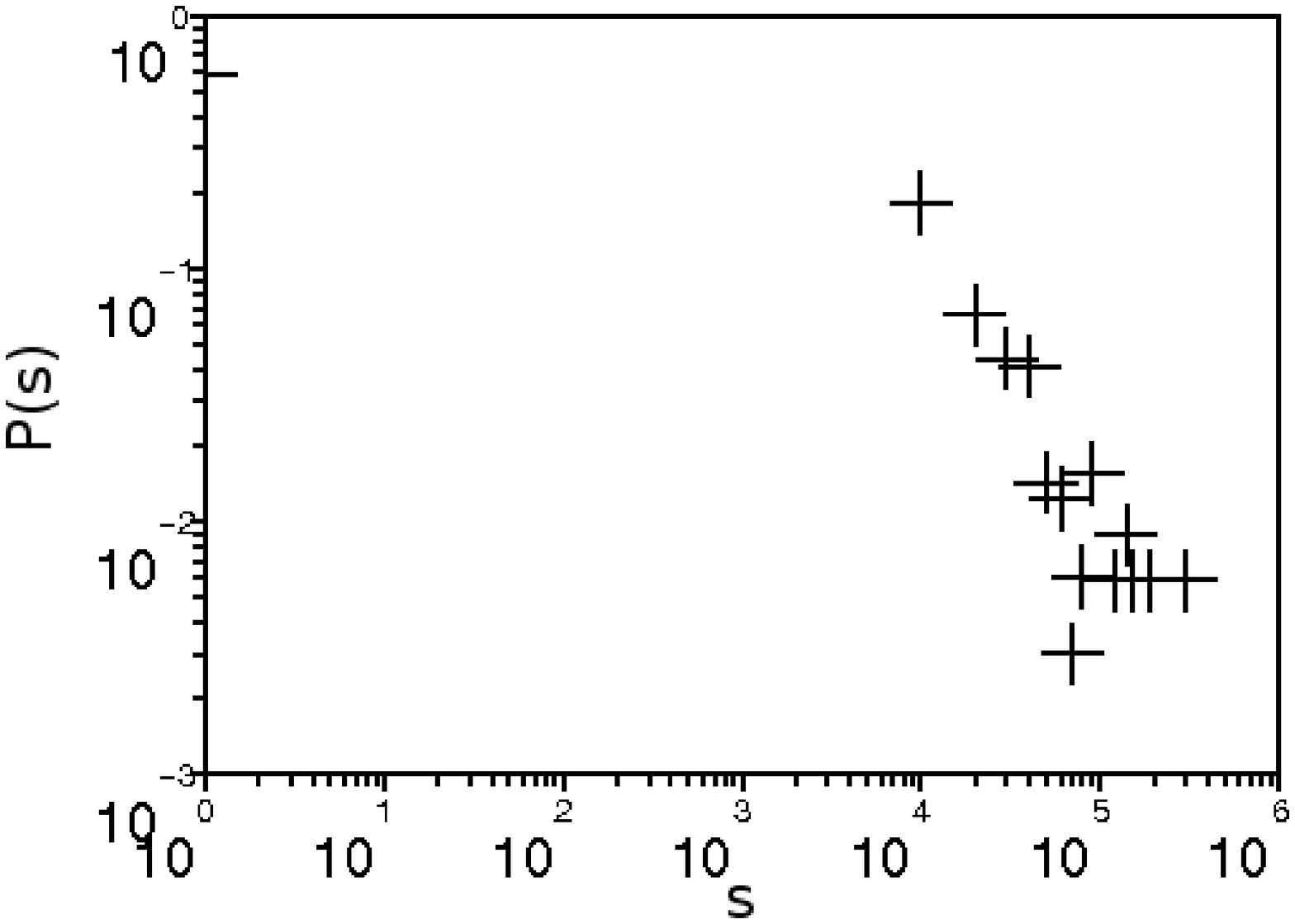}
    \includegraphics[scale=0.26]{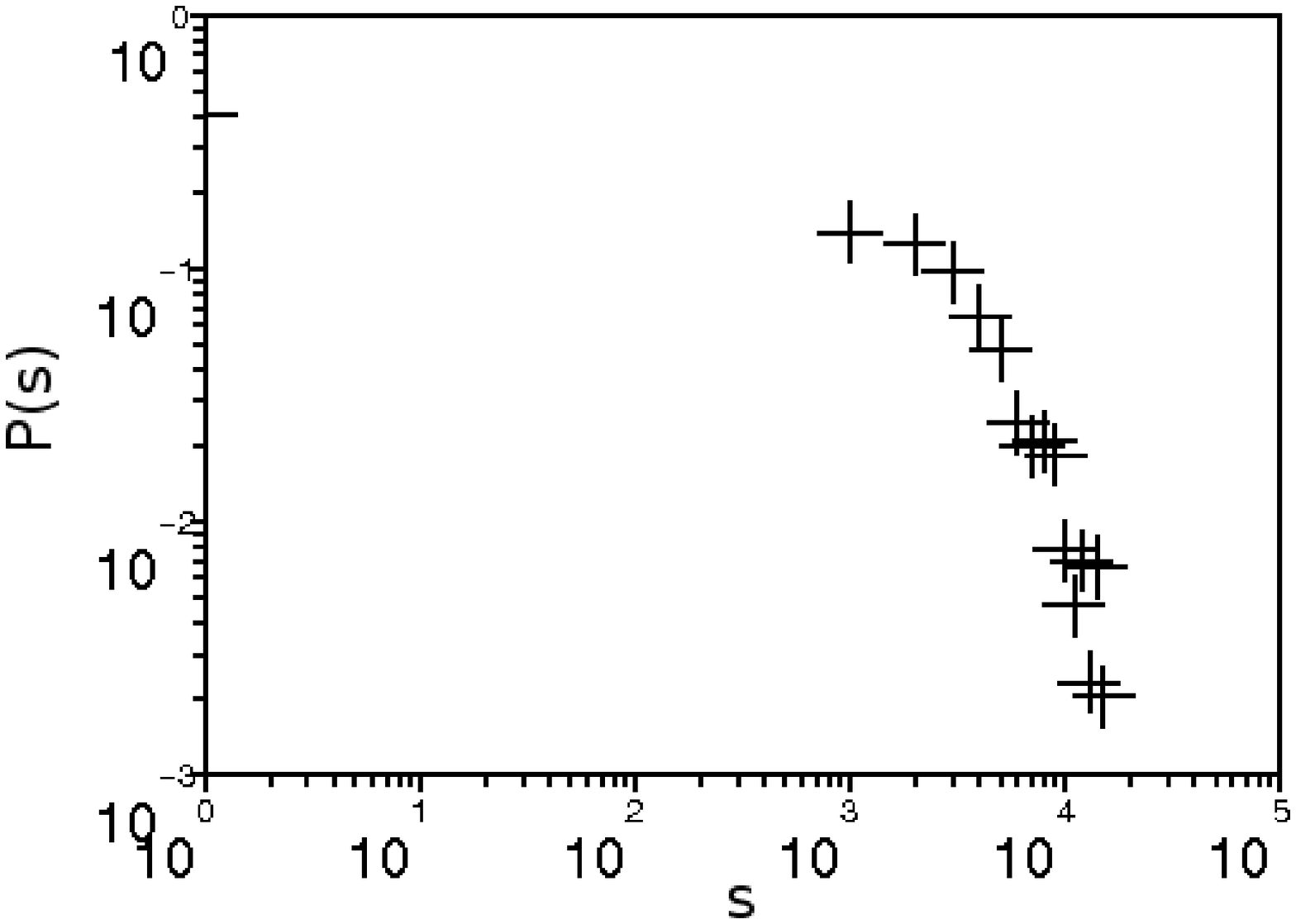}
    \includegraphics[scale=0.26]{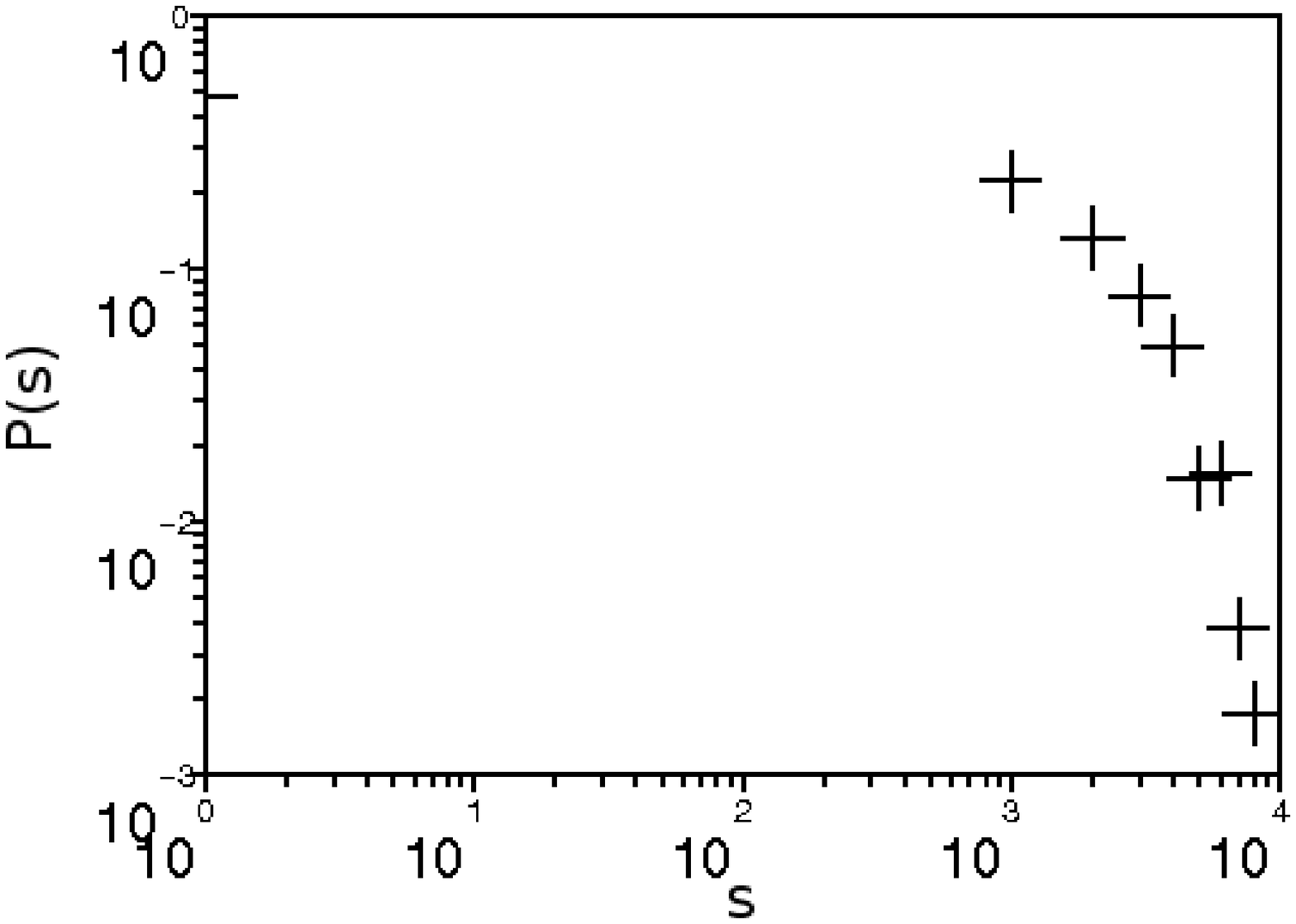}  \\
    (a) $\beta_{predator}=0.5$ \hspace{2cm} (b) $\beta_{predator}=0.125$ \hspace{2cm} (c) $\beta_{predator}=0.0625$ \\
    \caption{Strength distribution for $\Gamma_{prey}$ at $t=500000$
    when three intensities of attraction are considered between
    predators. } \label{fig:10}
  \end{center}
\end{figure*}

In the case of the prey network $\Gamma_{prey}$, the strength
distribution (Fig.~\ref{fig:10}) resulted in a completely different
function when compared to the predator network
(Fig.~\ref{fig:09}-iii). In this case the histograms remind a power
law function (especially along the largest values) with the maximum
strength value being larger in case of larger $\beta_{predator}$
(Fig.~\ref{fig:10}-a) and the other two configurations apparently
presenting two-slopes (Figs.~\ref{fig:10}-b and~\ref{fig:10}-c). The
shape of the function suggests that the system converged to an
organized state. In the last time step considered, few preys stayed
together for a long time (higher strength), while their majority moved
away or died frequently (lower strength), such that the emergent
distribution is scale-free-like (Observe the high number of null 
strength nodes). Although a low strength value could
indicate a high rate of changes between groups, the low values
observed in the preys network is mainly because of preys deaths since
the maximum strength value is much larger in the predators network and
the amount of eliminated preys decreases with larger
$\beta_{predator}$ (Fig.~\ref{fig:06}-a), while the maximum strength
value increases for larger $\beta_{predator}$ (Fig.~\ref{fig:10}).

Null attraction between predators ($\beta_{predator}=0$) and
attraction between preys ($\beta_{prey}>0$) generated a similar
pattern as in the case with $\beta_{predator}=0.125$ and
$\beta_{prey}=0$, though exhibiting a Gaussian-like strength
distribution. Comparing with the case when only attraction between
predators is considered (Fig.~\ref{fig:09}-c), we verified that the
maximum strength is considerably higher at the same time step observed
in that case. It suggests more movements between groups in the
configuration with $\beta_{prey}>0$ and $\beta_{predator}=0$ because
of the stronger attraction provided by the high concentration of
preys. Once again, the prey elimination mechanism implies in short
life-time for preys and consequently, suggesting a power law
distribution along the highest values.

\begin{figure*}
  \begin{center}
    \includegraphics[scale=0.26]{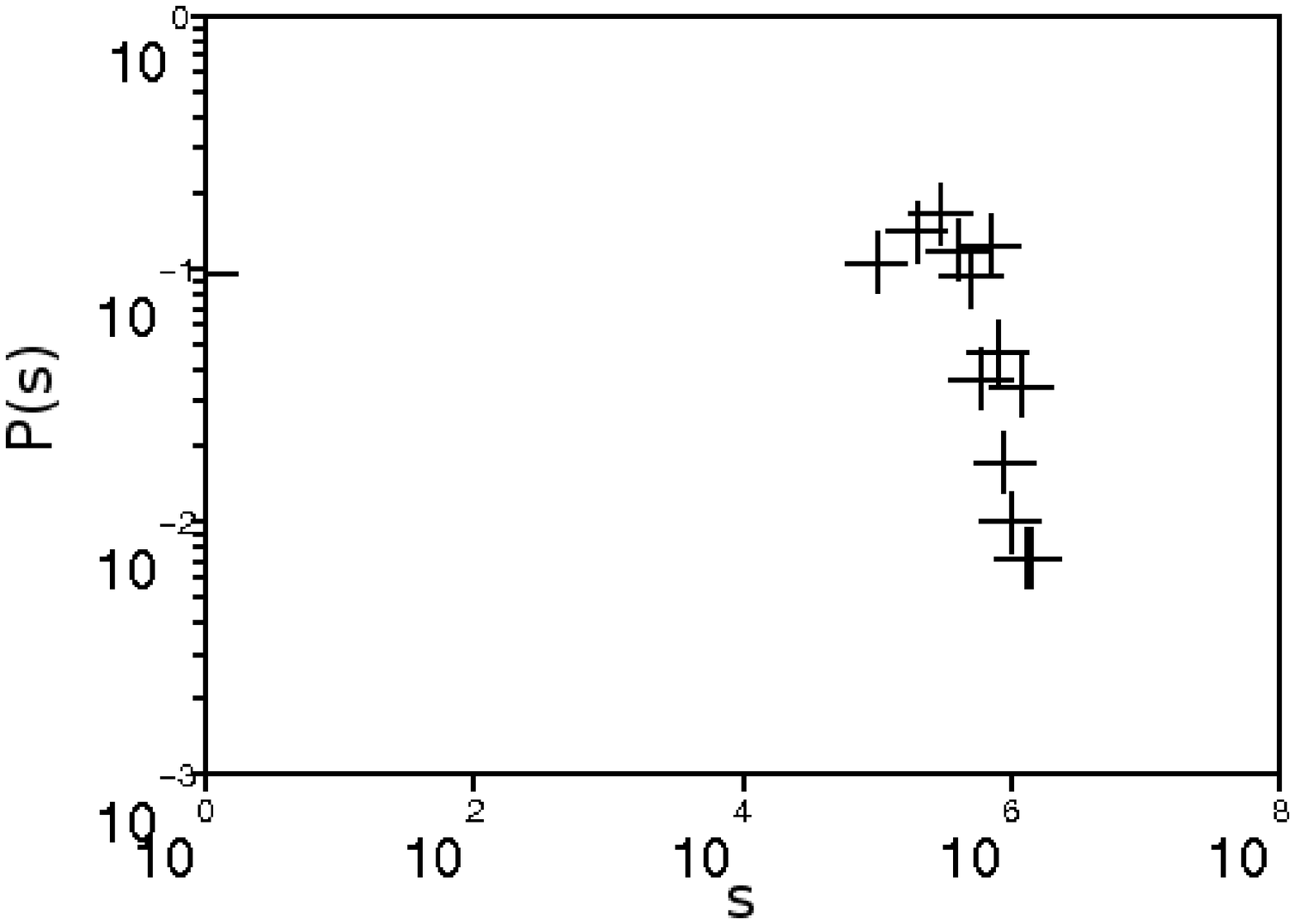}
    \includegraphics[scale=0.26]{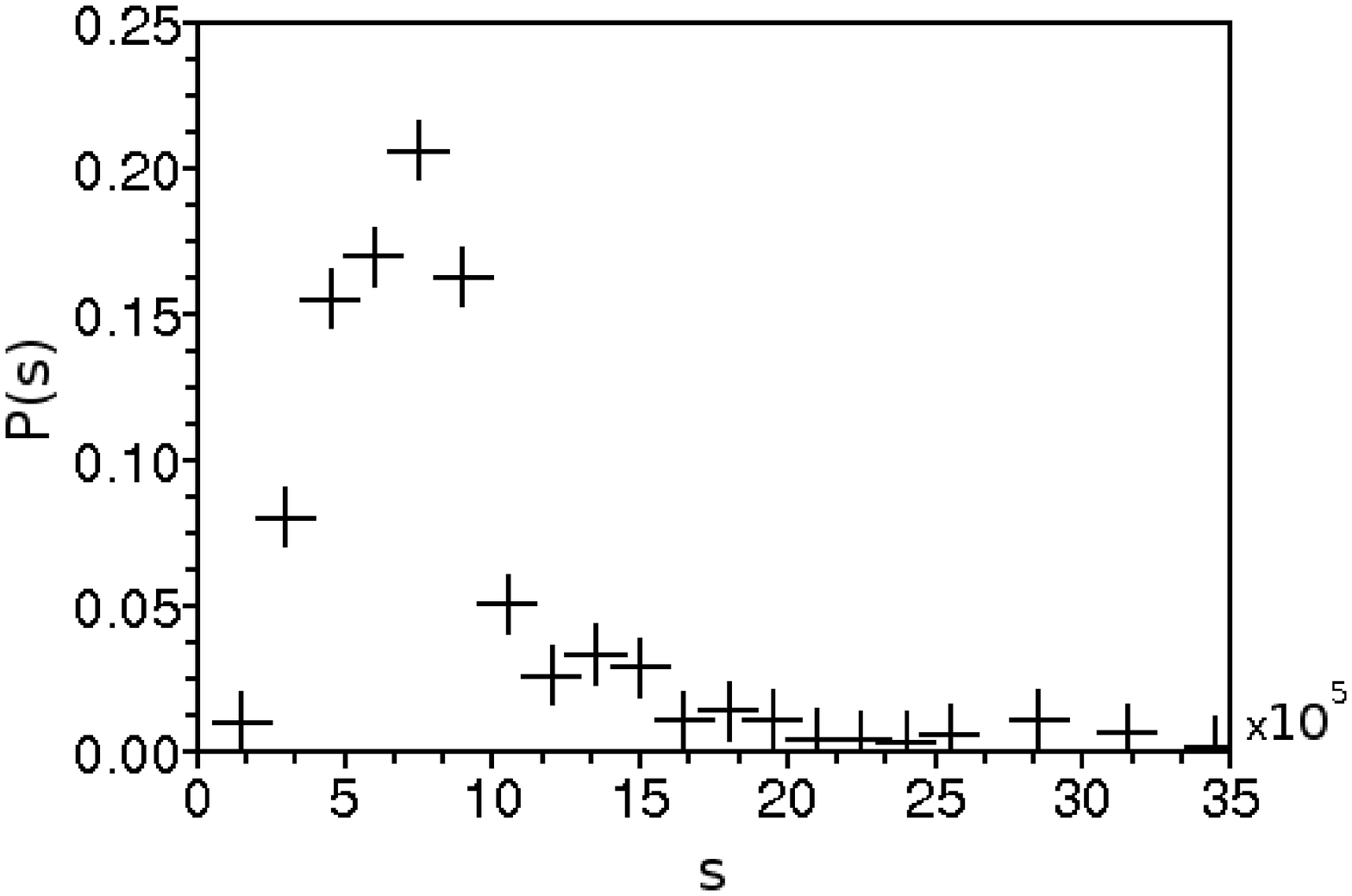}
    \includegraphics[scale=0.26]{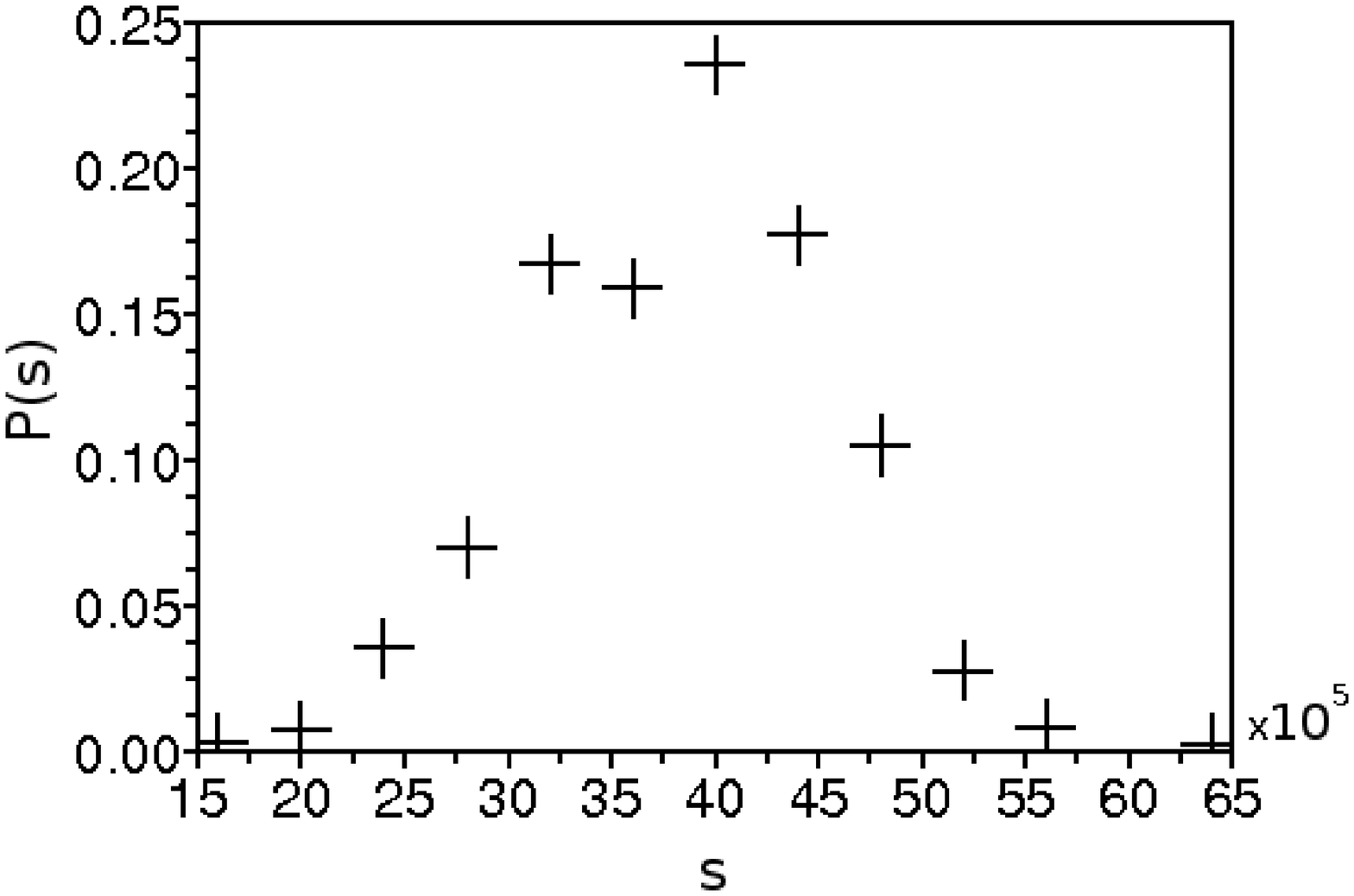}  \\
    (a) $\beta=0.5$ \hspace{3cm} (b) $\beta=0.125$ \hspace{3cm} (c) $\beta=0.0625$ \\
    \caption{Strength distribution for $\Gamma_{predator}$ at
    $t=500000$ when three attraction intensities between same specie
    individuals are considered.}  \label{fig:11}
  \end{center}
\end{figure*}

In the last experiment, considering attraction between same species
individuals, the strength distributions present some interesting
effects. The configuration with larger $\beta$ converged to a power
law-like distribution (Fig.~\ref{fig:11}-a) due to formation of dense
prey clusters which implied in faster predator clusters emergence in
some regions of the space. Dense predator clusters have strong
attraction fields and therefore, any predator close to them is readily
attracted and barely escapes from the cluster. Consequently, only
small groups change individuals frequently. In case of $\beta=0.125$,
the system evolves from a Poisson-like state to a combination of 
Gaussian-like on the left with a power-law-like on the right
(Fig.~\ref{fig:11}-b). At last, $\beta=0.0625$ (Fig.~\ref{fig:11}-c)
is similar to the cases with small (Fig.~\ref{fig:09}-c) or null
attraction between predators, suggesting that a reasonably weak
attraction intensity will result in the same non-stationary state. In
other words, weak attraction between predators and/or between preys
does not produce different collective behaviour in the system.

\begin{figure*}
  \begin{center}
    \includegraphics[scale=0.26]{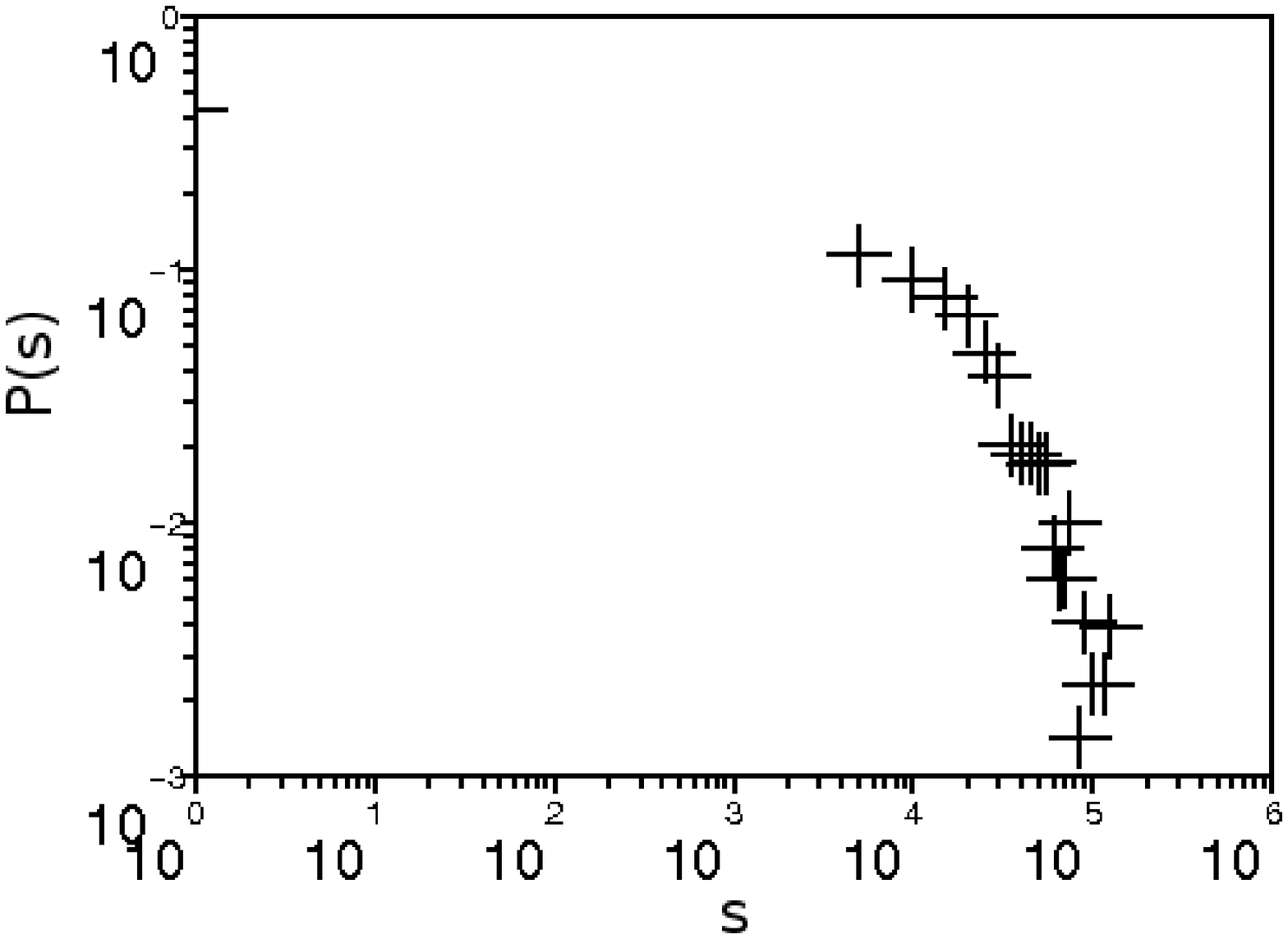}
    \includegraphics[scale=0.26]{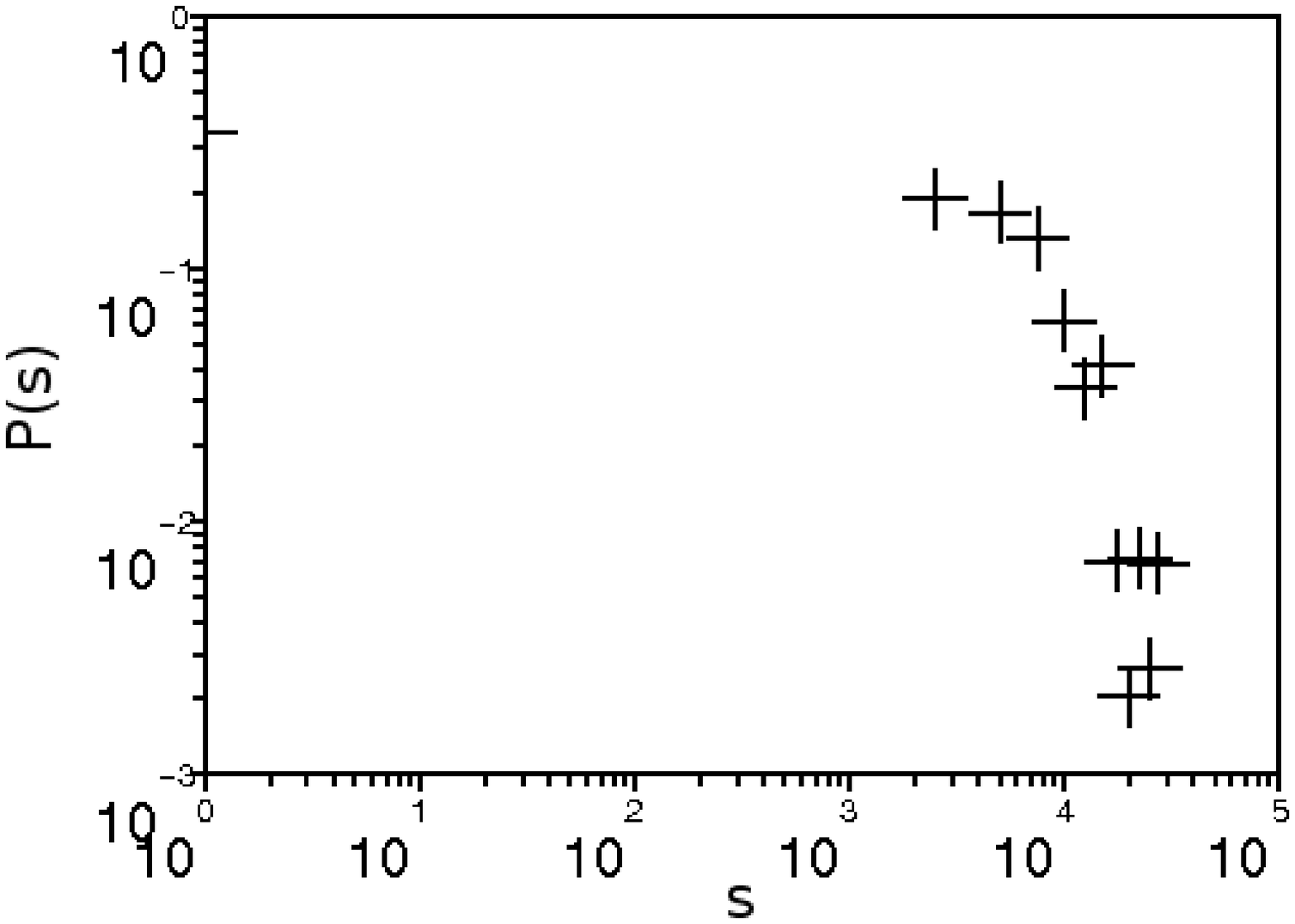}
    \includegraphics[scale=0.26]{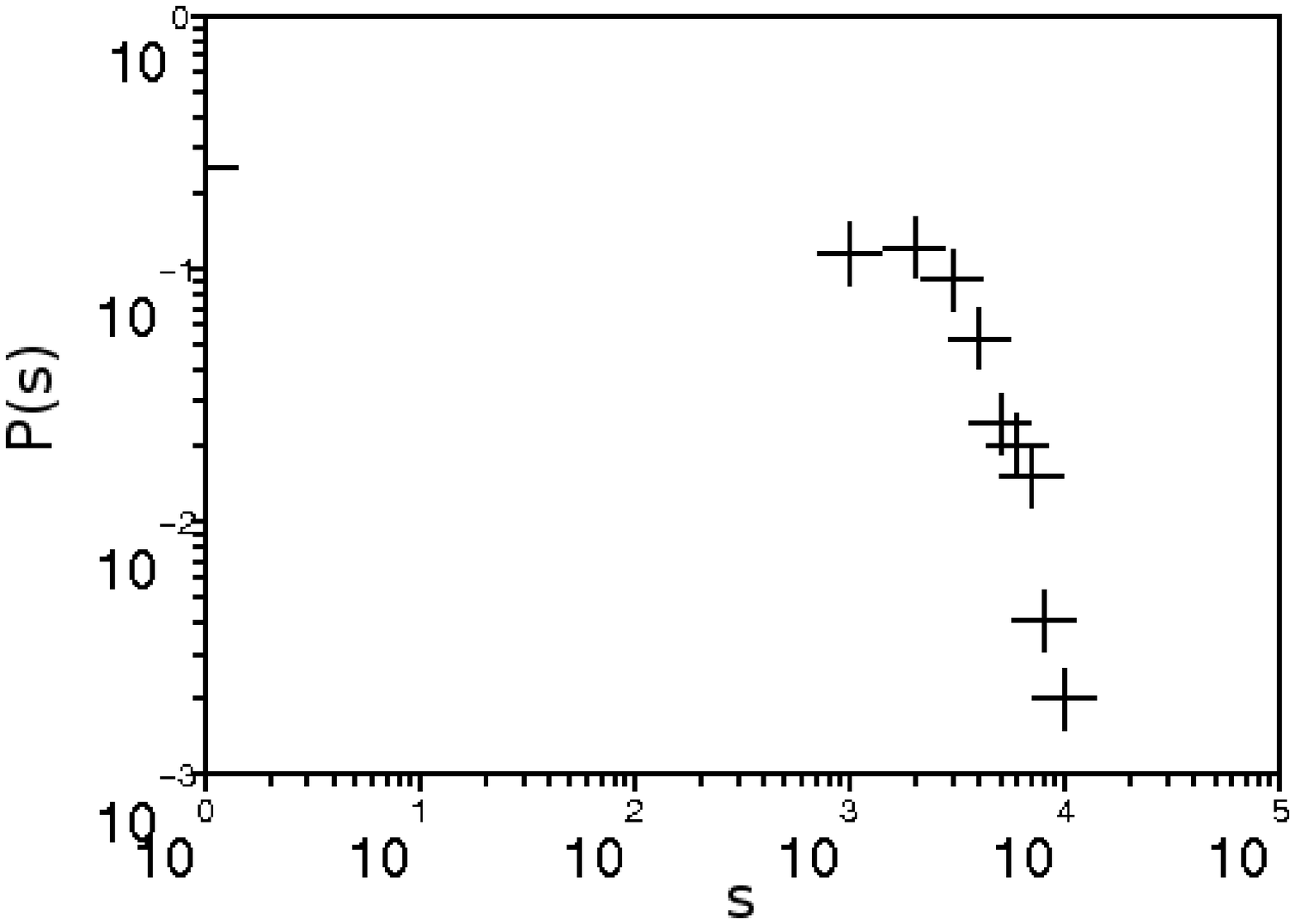}  \\
    (a) $\beta=0.5$ \hspace{3cm} (b) $\beta=0.125$ \hspace{3cm} (c) $\beta=0.0625$ \\
    \caption{Strength distribution for $\Gamma_{prey}$ at $t=500000$
    with three attraction intensities between same species
    individuals. } \label{fig:12}
  \end{center}
\end{figure*}

The power law-like distribution is observed in the prey network
$\Gamma_{prey}$ for large values as in the other configurations
(Fig.~\ref{fig:12}). The effect confirmed the hypothesis that such a
scale-free structure emerged because of the elimination of preys and
not because of the attraction mechanism between two individuals. The
longer life-time of preys is identified by the larger maximum strength
observed in this case in comparison to the other configurations
(Fig.~\ref{fig:10}).

\begin{table*}
\label{tab:01}
\begin{center}
    \begin{tabular}{ | p{1.4cm} | p{2.3cm} | p{3.6cm} | p{4.2cm} | p{4.2cm} | }
    \hline
    $\beta_{Predator}$\linebreak $\beta_{Presa}$ & \centering 0  & \centering 0.0625  & \centering 0.125 &  \centerline{0.5}     \\ \hline
    \centering 0
    & No clusters & Clusters of predators\linebreak \textbf{Predator:}\linebreak ``Gaussian''\linebreak \textbf{Prey:}\linebreak Two-regime ``Power-law''
    & Dense clusters of predators\linebreak \textbf{Predator:}\linebreak ``Gaussian''\linebreak \textbf{Prey:}\linebreak ``Two-regime Power-law''
    & Dense clusters of predators\linebreak \textbf{Predator:}\linebreak Unidentified shape\linebreak \textbf{Prey:}\linebreak ``Power-law'' \\ \hline
    \centering 0.0625
    & Some small clusters\linebreak \textbf{Predator:}\linebreak ``Gaussian''\linebreak \textbf{Prey:}\linebreak ``Power-law''
    & Clusters of predators and small clusters of preys\linebreak \textbf{Predator:}\linebreak ``Gaussian''\linebreak \textbf{Prey:}\linebreak ``Power-law'' &  & \\ \hline
    \centering 0.125
    & Small clusters\linebreak \textbf{Predator:} ``Gaussian''\linebreak \textbf{Prey:}\linebreak ``Power-law''
    &
    & Some dense clusters of predators and clusters of preys\linebreak \textbf{Predator:}\linebreak ``Gaussian''/``Power-law''\linebreak \textbf{Prey:}\linebreak ``Power-law''    &              \\ \hline
    \centering 0.5     & Small clusters of preys\linebreak \textbf{Predator:}\linebreak ``Gaussian''\linebreak \textbf{Prey:}\linebreak ``Power-law''
    &  &
    & Dense clusters of predators and some dense dense clusters of preys\linebreak \textbf{Predator:}\linebreak ``Power-law''\linebreak \textbf{Prey:}\linebreak ``Power-law'' \\ \hline
\end{tabular}
\end{center}
\caption{Summary of the main properties (spatial and topological) according to different attraction intensities between same species individuals. }
\end{table*}

\subsection{Clustering Coefficient Distribution}

\begin{figure*}
  \begin{center}
    (i) \\
    \includegraphics[scale=0.26]{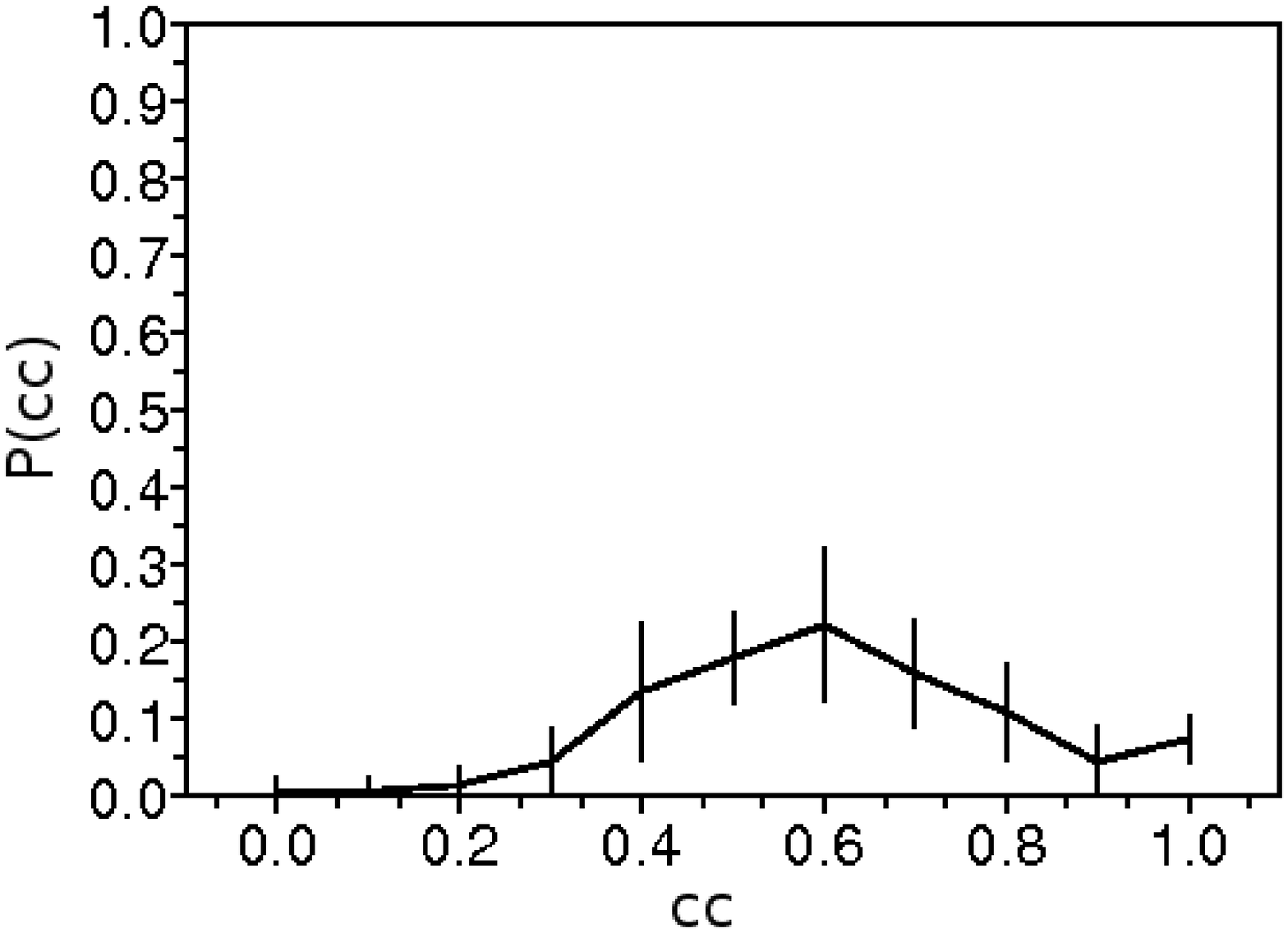}
    \includegraphics[scale=0.26]{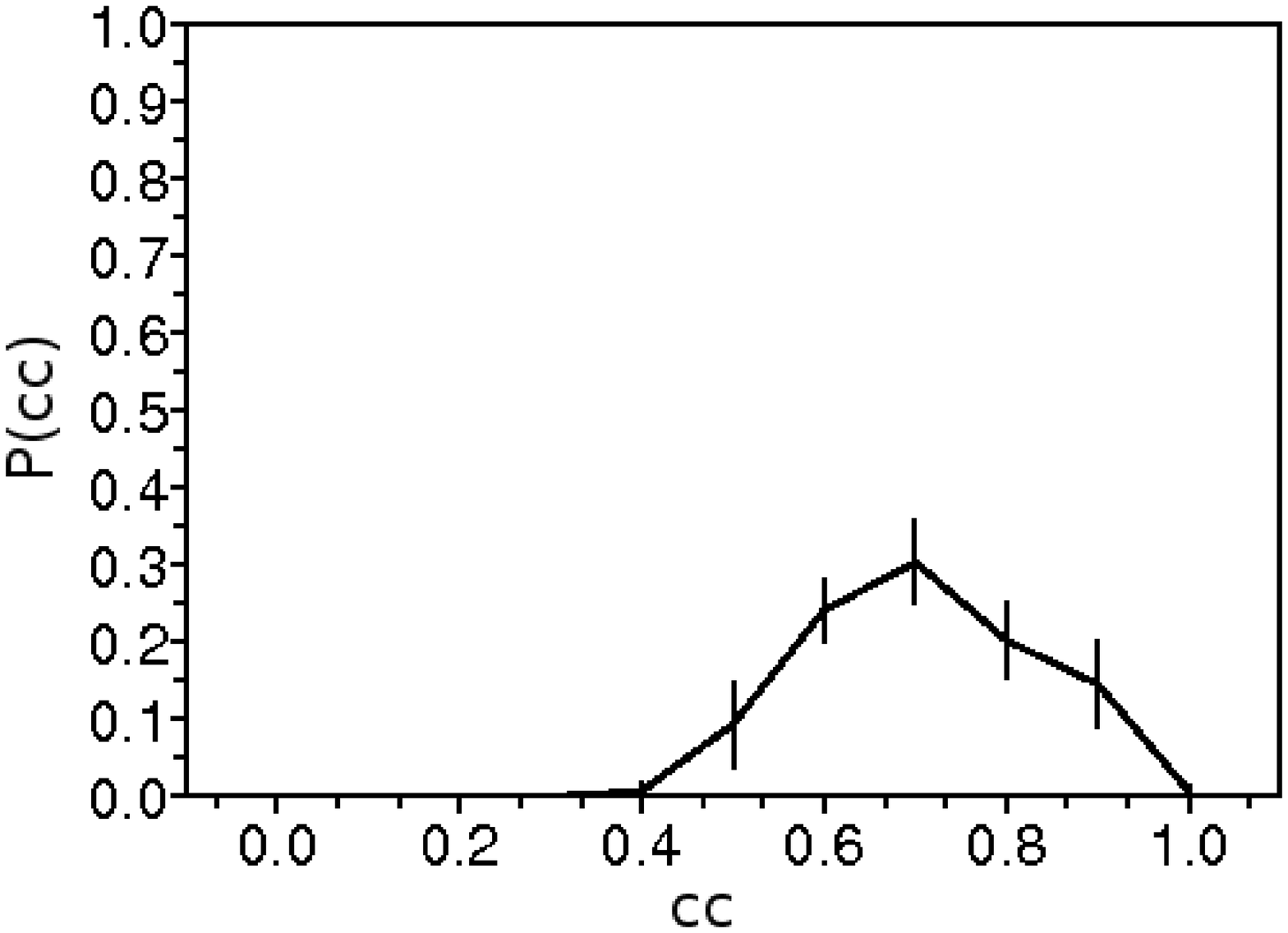}
    \includegraphics[scale=0.26]{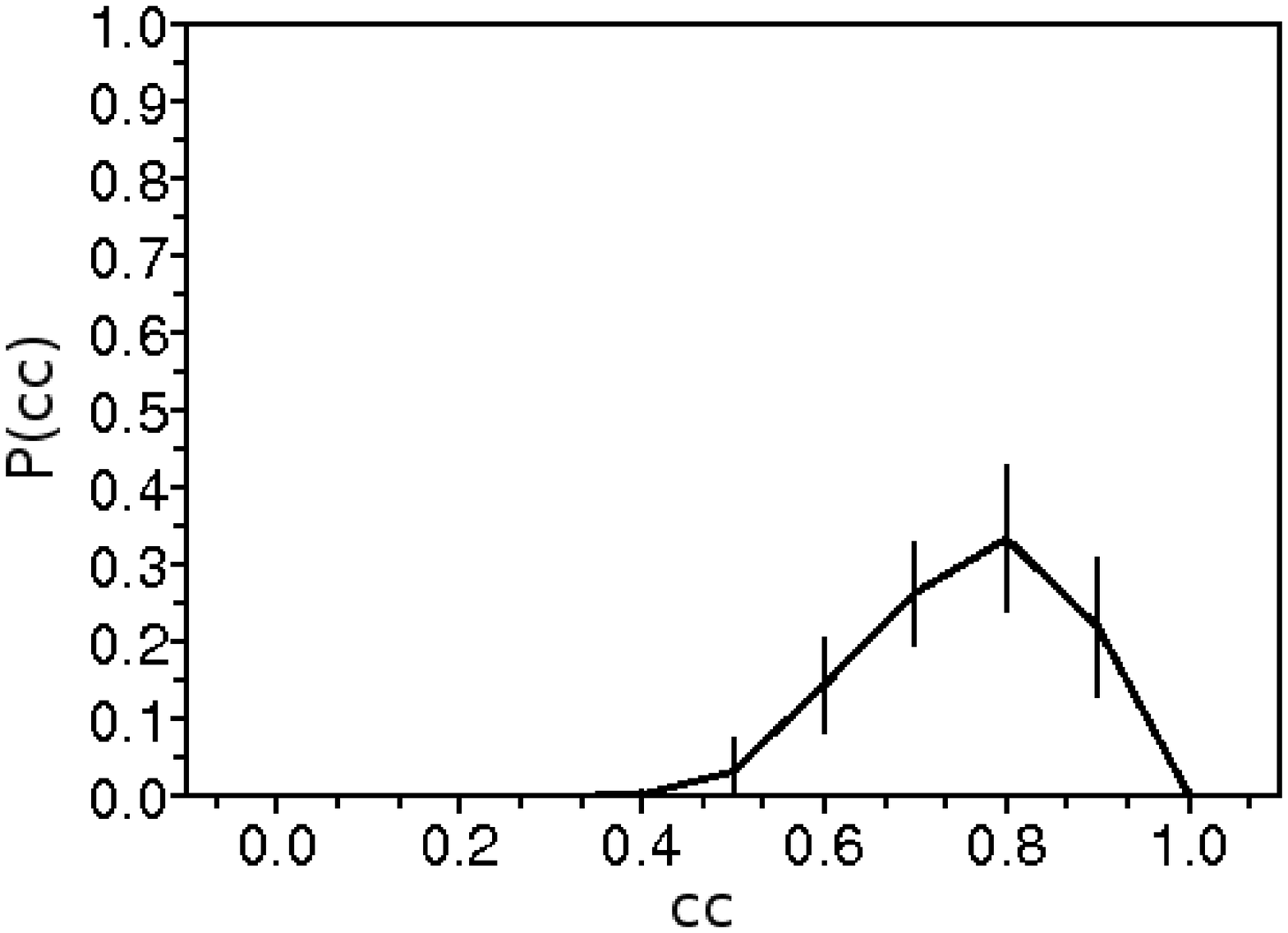}   \\
    (ii) \\
    \includegraphics[scale=0.26]{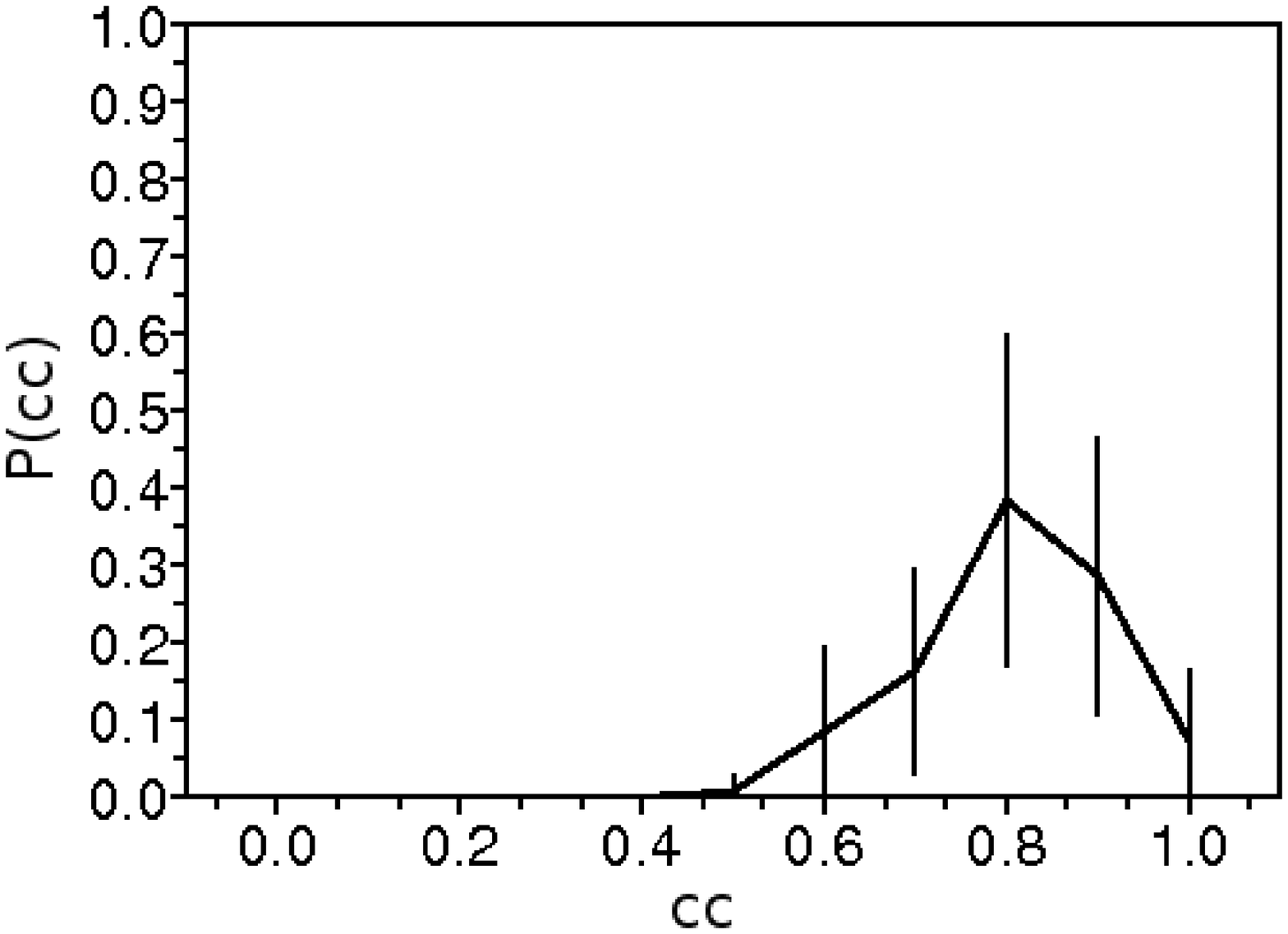}
    \includegraphics[scale=0.26]{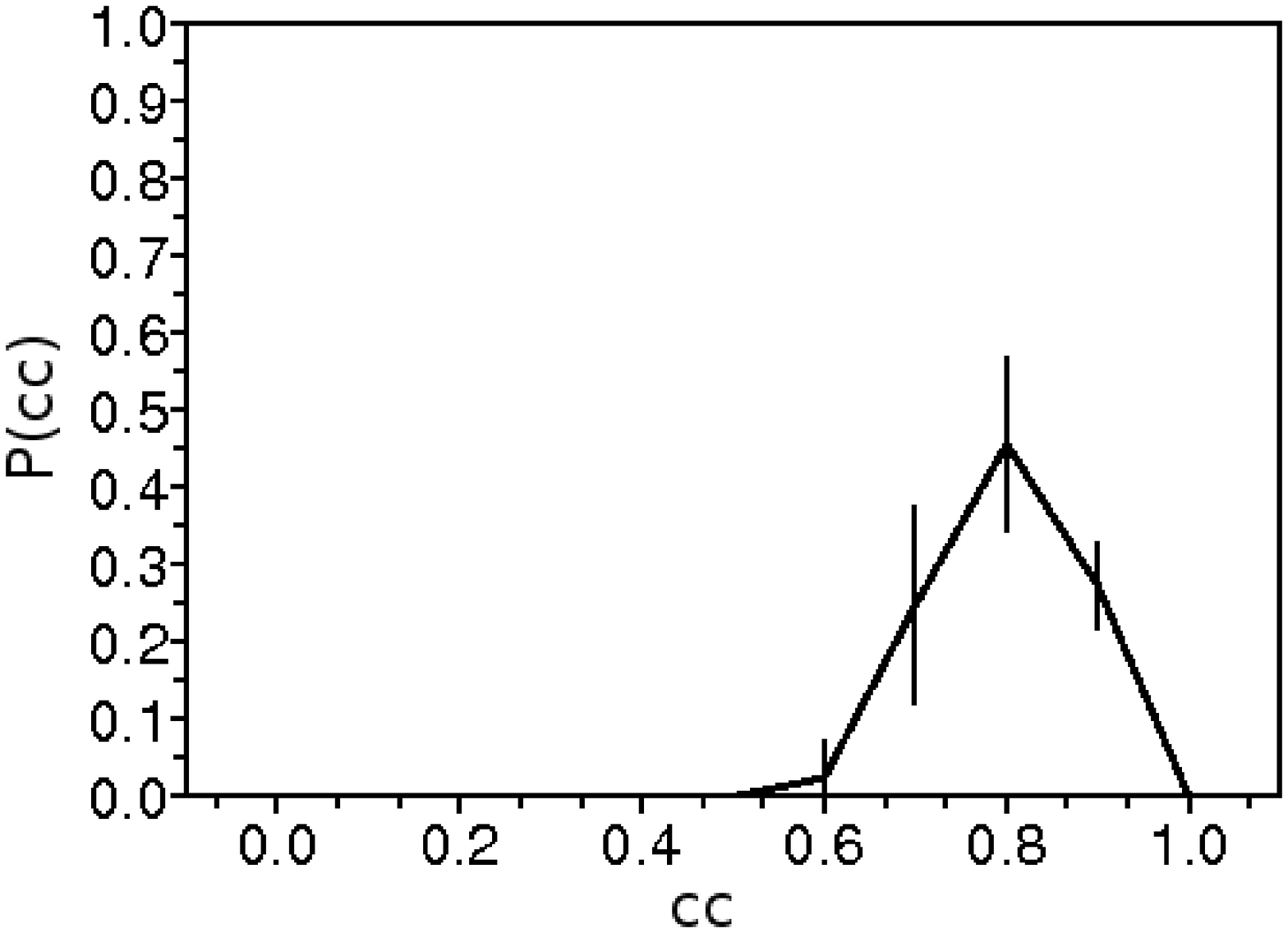}
    \includegraphics[scale=0.26]{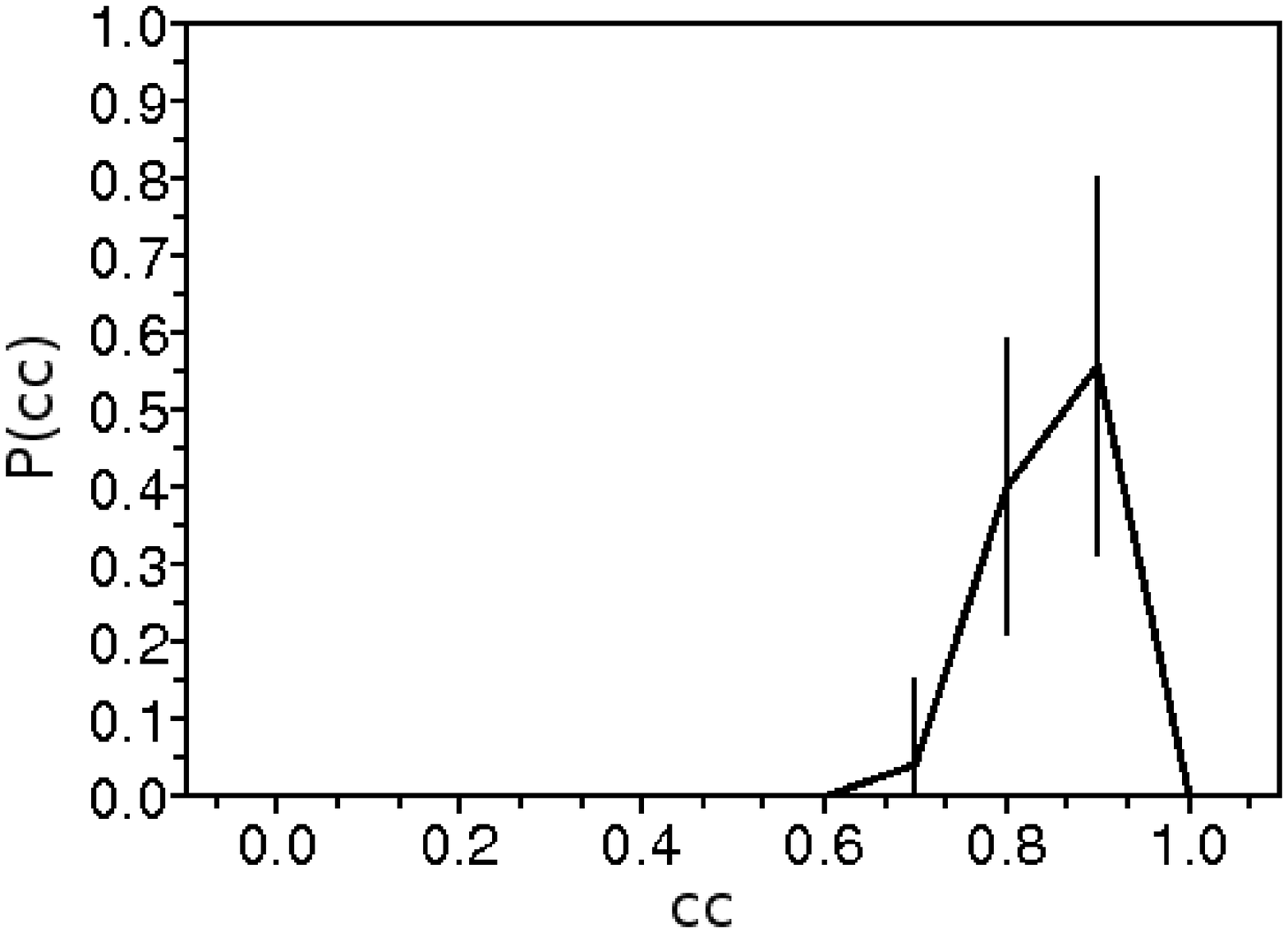}  \\
    (iii) \\
    \includegraphics[scale=0.26]{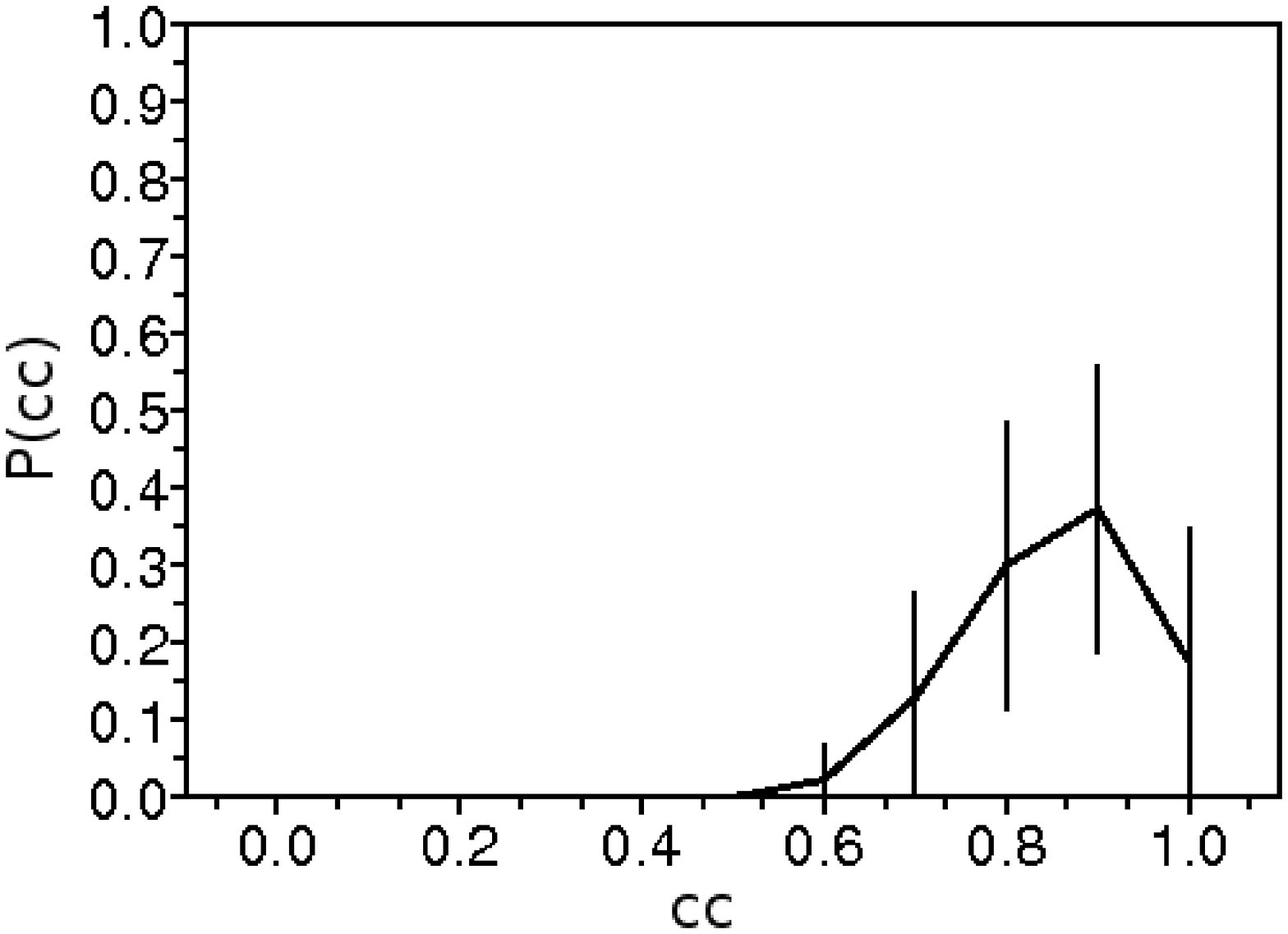}
    \includegraphics[scale=0.26]{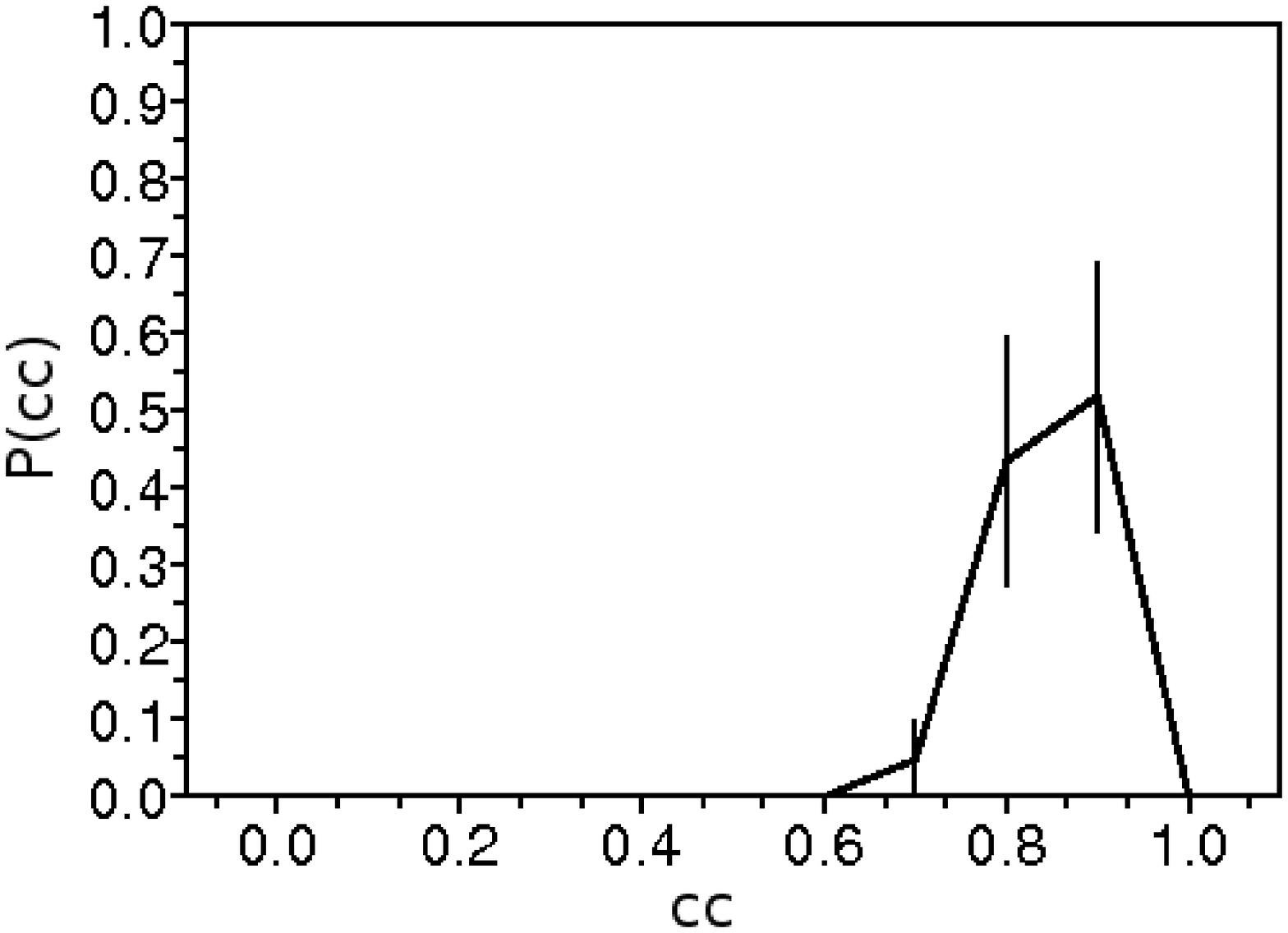}
    \includegraphics[scale=0.26]{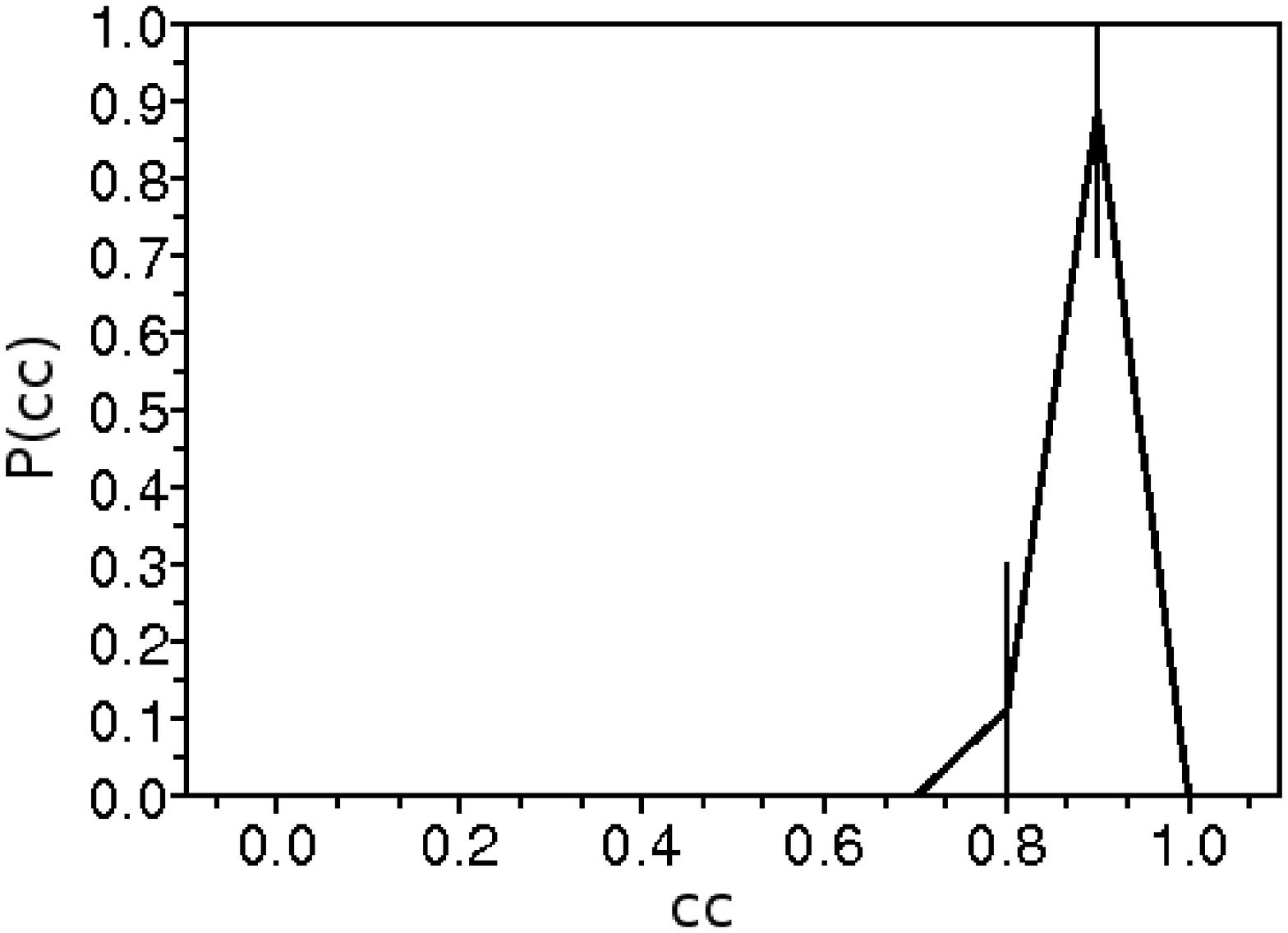}  \\
    (a) $\beta_{predator}=0.5$ \hspace{2cm} (b) $\beta_{predator}=0.125$ \hspace{2cm} (c) $\beta_{predator}=0.0625$
    \caption{Clustering coefficient distribution for
    $\Gamma_{predator}$ at three time steps: (i) $t=100000$, (ii)
    $t=300000$ and (iii) $t=500000$. We consider three attraction
    intensities between predators. } \label{fig:13}
  \end{center}
\end{figure*}

The clustering coefficients have nearly the same distribution in all
considered configurations for the predator network
(Fig.~\ref{fig:13}). The only difference is observed at the scale of
the distributions since smaller $\beta_{predator}$ implies more
movement and, consequently, faster connections between groups. The
normalized distributions become narrower and present characteristic
scales which move to the right over time, indicating that all groups
will become connected after a long period of time. The shapes of the
curves indicate that higher attraction between predators constrains
the movement and postpones creation of triangles between common
neighbours of a reference node. In case of weaker attraction,
predators are able to move between groups and consequently, the amount
of nodes with larger clustering coefficient becomes higher
(Fig.~\ref{fig:13}-c). Although the number of nodes with higher
clustering coefficient constantly increases over time, there are many
nodes with small values of $cc$ for larger $\beta_{predator}$
(Fig.~\ref{fig:13}-a).

\begin{figure*}
  \begin{center}
    \includegraphics[scale=0.26]{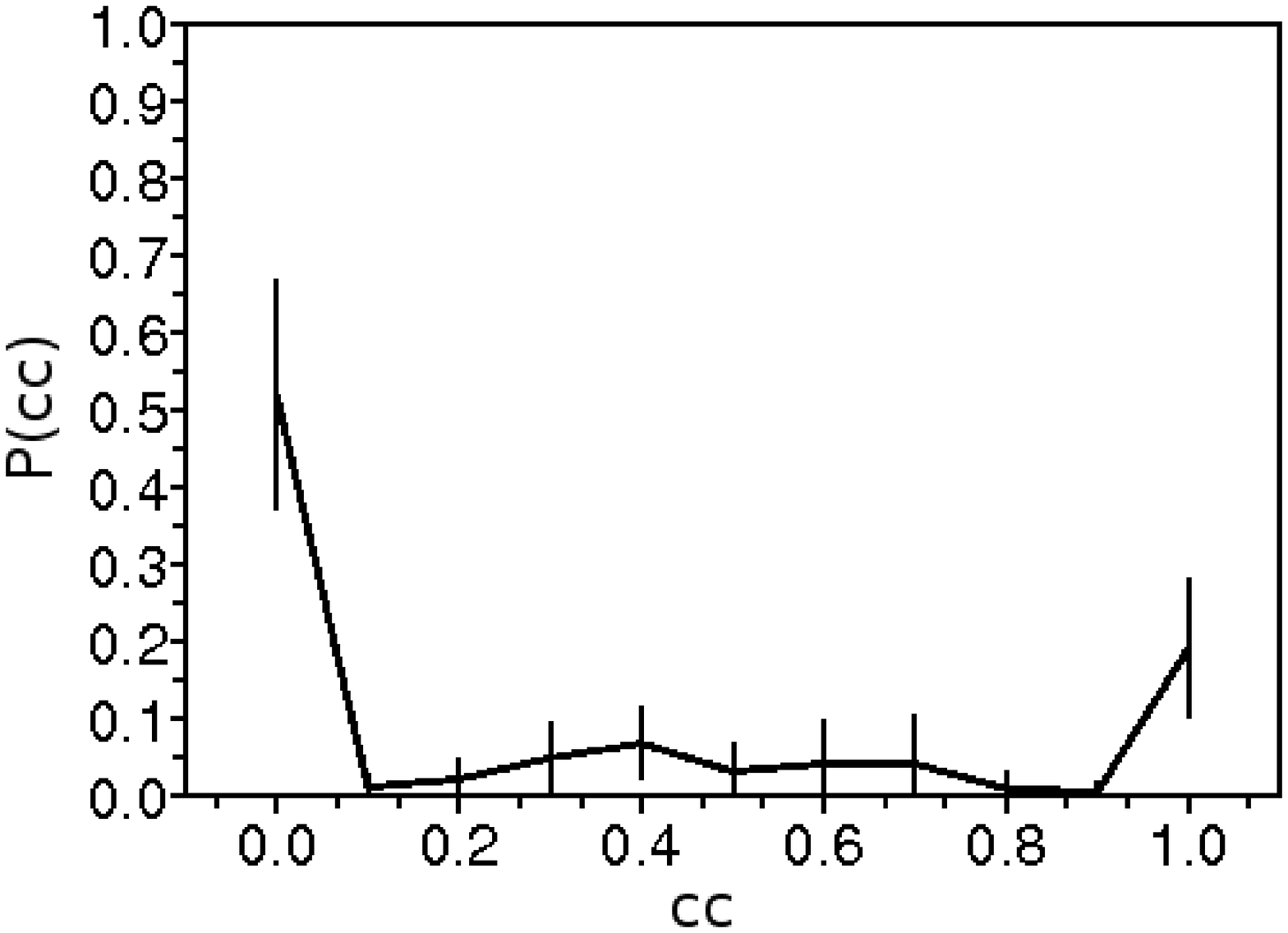}
    \includegraphics[scale=0.26]{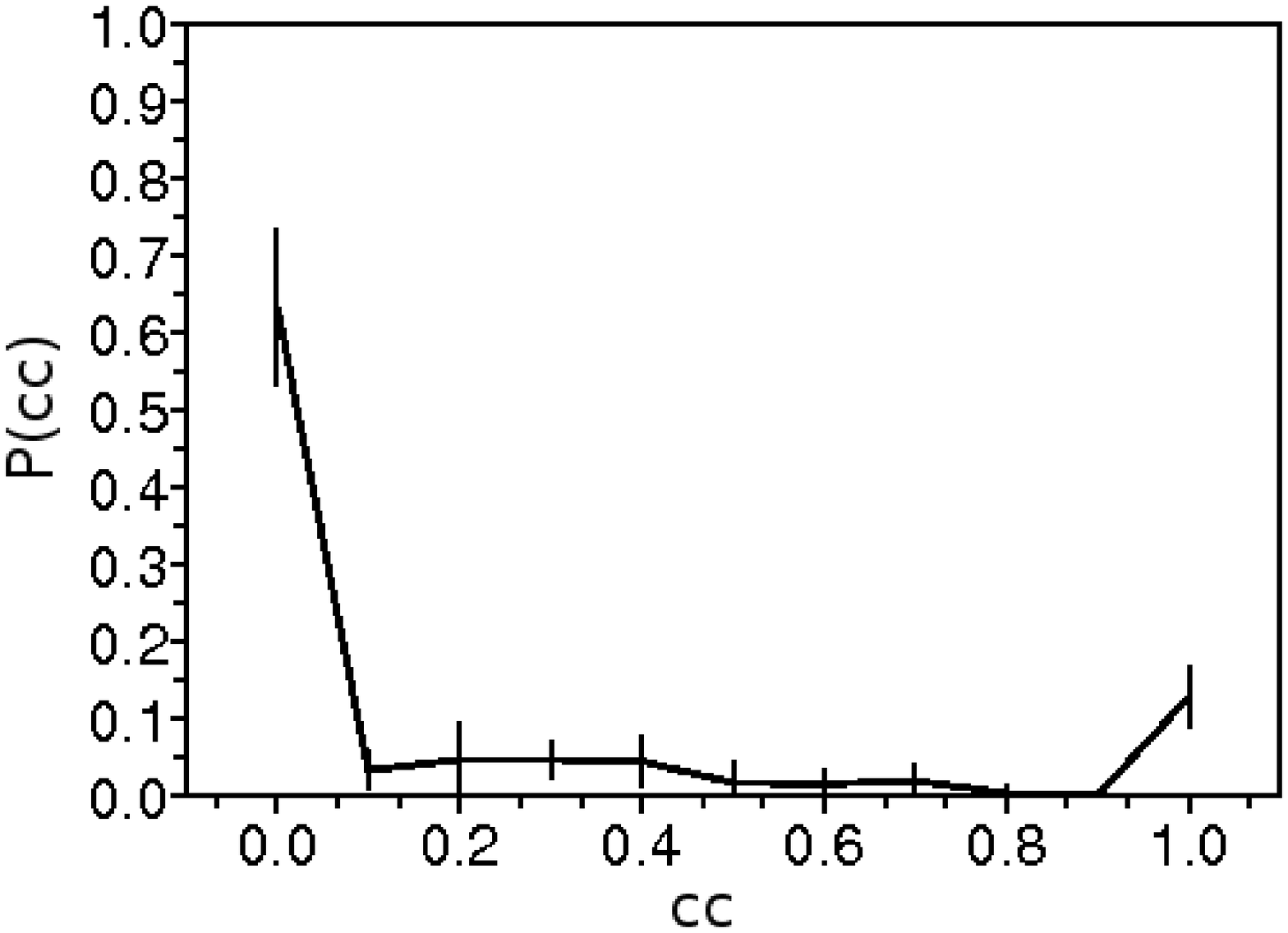}
    \includegraphics[scale=0.26]{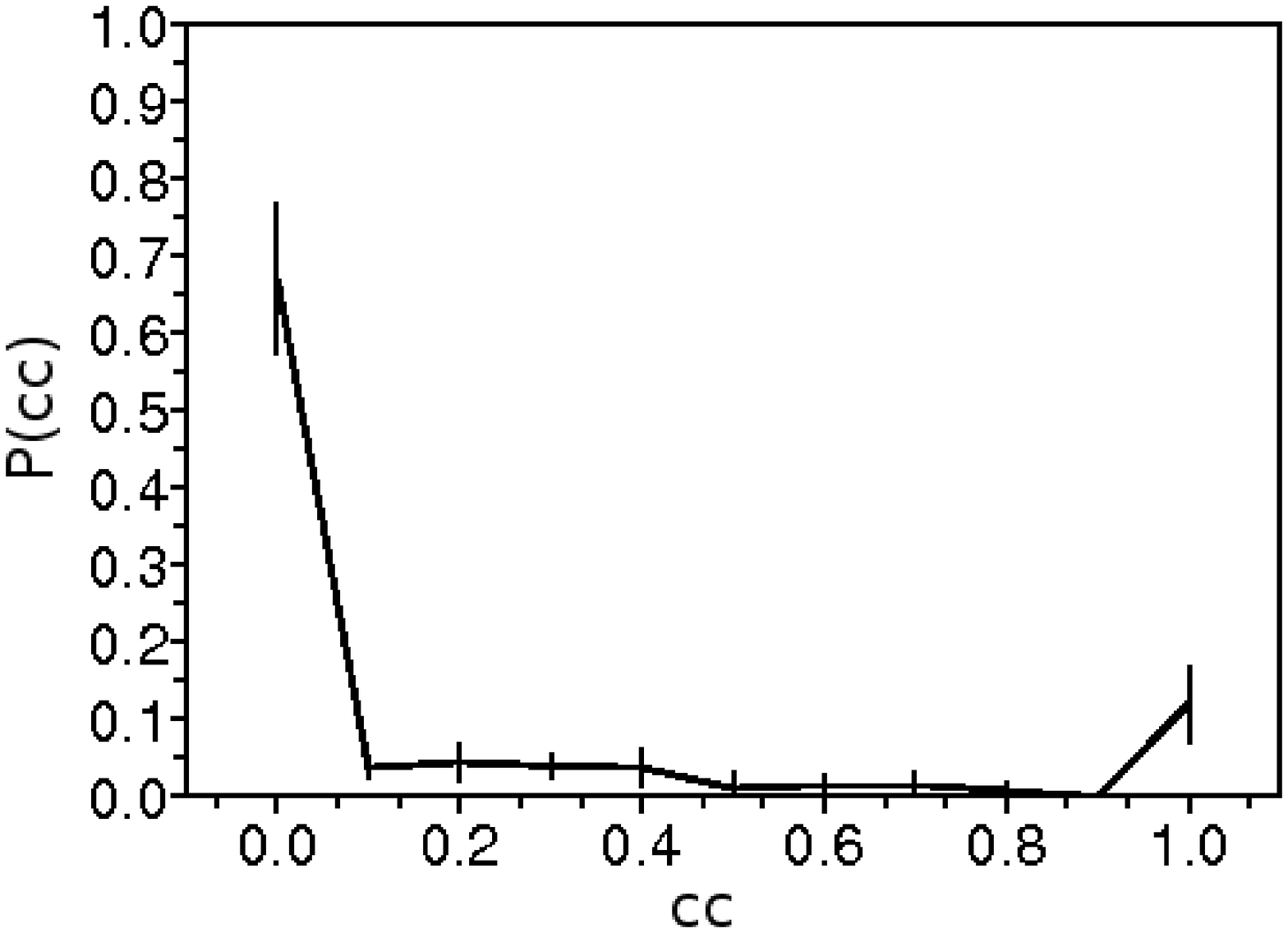}  \\
    (a) $\beta_{predator}=0.5$ \hspace{2cm} (b) $\beta_{predator}=0.125$ \hspace{2cm} (c) $\beta_{predator}=0.0625$
    \caption{Clustering coefficient distribution at $t=500000$ for
    $\Gamma_{prey}$ when three intensities of attraction between
    predators are considered. } \label{fig:14}
  \end{center}
\end{figure*}

In the preys network $\Gamma_{prey}$, all configurations resulted in a
clustering coefficient distribution with a clear division
(Fig.~\ref{fig:14}), which suggests the existence of two distinct
states in the system (nearly constant over time). A large number of
preys has small connectivity in their neighbourhood while a small
number presents a high level of local connectivity among its
neighbours. A small clustering coefficient is a consequence of two
effects: complete absence of movement or a high death rate, which
avoid triangle formation. Since the frequency of $cc$ null values
increases as we decrease the attraction between predators
(Fig.~\ref{fig:14}), we conclude that the main fact behind this effect
is exactly the death of preys.  As shown before, such deaths increase
as $\beta_{predator}$ decreases (Fig.~\ref{fig:06}-a). In case of
attraction between preys, the effect is the opposite, \textit{i.e.} the number
of dead preys and of nodes (predator groups) with null $cc$ increases
with $\beta_{prey}$. Since the configuration with $\beta_{prey}=0.5$
has regions with high concentration of preys, which are not completely
static and are not eliminated frequently too, the probability to
establish closed triangles increases and explains the higher frequency
observed at the maximum value of $cc$ ($cc=1$). The existence of
groups with maximum clustering coefficient indicates that some preys
stay a long time in the system and, although they are eliminated,
other preys keep some spatial groups united and permit triangle
formation (about $10\%$ of the nodes in $\Gamma_{prey}$). With
attraction between same species individuals, we observed a higher
frequency of intermediate clustering coefficient values. The existence
of clusters in both species resulted in extended life-time for the
preys, which allowed more movement and connections between nearby
spatial groups.

\section{Conclusions}

The current paper has focused on multiple networks systems involving
interacting species. Since the system was originally motivated by
ecology, we assumed that preys are eliminated when they move
sufficiently close to a predator, but new preys are randomly displaced
in the system in order to replace those which are
consumed. Spatio-temporal clusters emerge from the dynamics, which
involves movements of individual between clusters. Two complex
networks containing the same set of nodes are considered. In one
network, the weights represented the Euclidean distances between two
individuals. In the other network, the number of steps during which
two individuals were close enough are considered as weights. By
merging both networks, we obtained a third complex network whose nodes
represented spatial groups defined by the connected sub-networks.
Such a growth mechanism implied the connections to incorporate
information about the history of the individuals movement.

Several configurations, defined in terms of the intensity of the
attractions between preys and predators, were considered in our
simulations.  The increase in the attraction between same species
individuals generated dense clusters whose sizes and shapes were a
result of the same species attraction intensity as well as the other
species attraction features. Dense predator clusters constrained the
movement of single predators, while dense prey clusters generated
strong fields and attracted many predators. As a consequence, the rate
of eliminated preys was higher when the preys were organized into
clusters and predators had null attraction between them. We have
observed that group organization could benefit the species according
to the intensity of intra-species attraction.  On the other hand, some
configurations (with small attraction) evolved to states where the
group organization had few or no advantage to any of the two species.

By using the complex network theory, we observed that the average
degree of the resulting predator network increased up to a threshold
corresponding to the number of spatial groups, \textit{i.e.}, the
system evolved to a fully connected state. Since the average
clustering coefficient increased faster than the average degree, we
concluded that members exchanges between nearby spatial groups were
more frequent than between far away groups. The analysis at the last
time step showed that the system converged to an organized state. The
predator network presented a Gaussian-like strength and clustering
coefficient distributions in nearly all configurations while
approximately scale-free strength and a polarized clustering
coefficient distributions emerged in the case of the preys
network. The prey elimination mechanism was responsible for generating
the observed structure in the prey network. Since preys are eliminated
frequently, they are not able to move long enough throughout space and
establish connections between all clusters, as was the case with the
predators.

In order to further investigate the model, we propose the following
future developments: (i) study of the effect of individuals density
and other scale properties of the system, (ii) investigation of
clusters emergence when considering other attraction functions and
more species interacting together, (iii) analysis of collective
phenomena (\textit{e.g.}, epidemics and opinion formation as a result
of dynamical propagation of diseases and ideas from specific
individuals), and finally (iv) application of the same methodology in
empirical data, for example, to study the movement of animals in the
field.

\begin{acknowledgments}

LECR is grateful to CNPq for financial support. LFC is grateful to
CNPq (308231/03-1) and FAPESP (05/00587-5) for financial support.

\end{acknowledgments}

\bibliographystyle{unsrt}
\bibliography{predator_prey}

\begin{thebibliography}{10}

\bibitem{OltvaiPyramid02}
Zolt\'an~N. Oltvai and Albert-L\'aszl\'o Barab\'asi.
\newblock Life's complexity pyramid.
\newblock {\em Science}, 298(5594):763 -- 764, 2002.

\bibitem{CostaMultiple05}
Luciano da~Fontoura~Costa.
\newblock Socioeconomic development and stability: A complex network blueprint,
  2005.
\newblock arXiv:physics/0505008.

\bibitem{Erez05}
Tom Erez, Martin Hohnisch, and Sorin Solomon.
\newblock Statistical economics on multi-variable layered networks, 2005.
\newblock arXiv:physics/0406369.

\bibitem{Kurant06a}
Maciej Kurant and Patrick Thiran.
\newblock Layered complex networks.
\newblock {\em Physical Review Letters}, 96(138701), 2006.

\bibitem{Kurant06b}
Maciej Kurant and Patrick Thiran.
\newblock Extraction and analysis of trafﬁc and topologies of transportation
  networks.
\newblock {\em Physical Review E}, 74(036114), 2006.

\bibitem{Park06}
Juyong Park, Oscar Celma, Markus Koppenberger, Pedro Cano, and Javier~M.
  Buld\'u.
\newblock The social network of contemporary popular musicians, 2006.
\newblock arXiv:physics/0609229.

\bibitem{Rocha07}
Luis Enrique~Correa da~Rocha and Luciano da~Fontoura~Costa.
\newblock 2d pattern evolution constrained by complex network dynamics.
\newblock {\em New Journal of Physics}, 9(108), 2007.

\bibitem{Mingfeng:odor}
Mingfeng He, Pu~Li, and Changquan Ni.
\newblock Model odor-oriented predators and prey.
\newblock {\em International Journal of Modern Physics C}, 17(5):711 -- 720,
  2006.

\bibitem{Lotka_book}
Alfred~J. Lotka.
\newblock {\em Elements of physical biology}.
\newblock Williams and Wilkens, Baltimore, 1925.

\bibitem{Volterra_book}
Vito Volterra.
\newblock {\em Variazioni e fluttuazioni del numero d'individui in specie
  animali conviventi}, volume~2.
\newblock 1926.

\bibitem{Murray_book}
J.~D. Murray.
\newblock {\em Mathematical biology}.
\newblock Springer-Verlag, Berlin, 2005.

\bibitem{Waldau:book}
Nathalie Waldau, Peter Gattermann, Hermann Knoflacher, and Michael
  Schreckenberg.
\newblock {\em Pedestrian and Evacuation Dynamics}.
\newblock Springer Verlag, Berlin, 2005.

\bibitem{Sznajd:opinion}
Katarzyna Sznajd-Weron and JÓzef Sznajd.
\newblock Opinion evolution in closed community.
\newblock {\em International Journal of Modern Physics C}, 11(6):1157 -- 1165,
  2000.

\bibitem{Farkas:mexicanwave}
I.~Farkas, D.~Helbingt, and T.~Vicsek.
\newblock Mexican waves in an excitable medium.
\newblock {\em Nature}, 419:131 -- 132, 2002.

\bibitem{Barabasi:review}
R.~Albert and A.-L.Barab\'{a}si.
\newblock Statistical mechanics of complex networks.
\newblock {\em Reviews of Modern Physics}, 74:48 -- 98, 2002.

\bibitem{Newman:review}
M.~E.~J. Newman.
\newblock Structure and function of complex networks.
\newblock {\em SIAM Review}, 45(2):167 -- 256, 2003.

\bibitem{Dorog:review}
S.~N. Dorogovtsev and J.~F.~F. Mendes.
\newblock Evolution of networks.
\newblock {\em Advances in Physics}, 51:1079 -- 1187, 2002.

\bibitem{Costa:review}
L.~da~F. Costa, F.~A. Rodrigues, G.~Travieso, and P.~R. Villas~Boas.
\newblock Characterization of complex networks: A survey of measurements.
\newblock {\em Advances in Physics}, 56:167 -- 242, 2007.

\bibitem{Boccaletti:review}
S.~Boccaletti, V.~Latora, Y.~Moreno, M.~Chaves, and D.-U. Hwang.
\newblock Complex networks: structure and dynamics.
\newblock {\em Physics Reports}, 424:175 -- 308, 2006.

\bibitem{Barrat:weight}
M.~Pastor-Satorras~R. Barrat, A.~Barthelemy and A.~Vespignani.
\newblock The architecture of complex weighted networks.
\newblock {\em Proceedings of the National Academy of Sciences}, 101(11):3747
  -- 3752, 2004.

\bibitem{Motter:cascade}
A.~E. Motter and Y-C. Lai.
\newblock Cascade-based attacks on complex networks.
\newblock {\em Physical Review E}, 66:065102, 2002.

\bibitem{Satorras:epidemic}
Romualdo Pastor-Satorras and Alessandro Vespignani.
\newblock Epidemic dynamics and endemic states in complex networks.
\newblock {\em Physical Review E}, 63:066117, 2001.

\end{thebibliography}

\end{document}